\documentclass[useAMS,usenatbib]{mn2e}
\usepackage{graphicx}
\usepackage{amsmath}
\usepackage{textgreek}

\title[Modelling turbulent Comptonization]{A simple framework for modelling the dependence of bulk Comptonization by turbulence on accretion disc parameters}
\author[J. Kaufman, O. M. Blaes, and S. Hirose]{J. Kaufman$^1$\thanks{E-mail:
jason.kaufman09@gmail.com (JK); blaes@physics.ucsb.edu (OMB)}, O. M. Blaes$^1$, and S. Hirose$^2$\\
$^1$Department of Physics, University of California, Santa Barbara, CA 93106, USA\\
$^2$Department of Mathematical Science and Advanced Technology, JAMSTEC, Yokohama 236-0001, Japan}
\begin{document}

\date{Accepted ---. Received ---; in original form ---}

\pagerange{\pageref{firstpage}--\pageref{lastpage}} \pubyear{2018}

\maketitle

\label{firstpage}

\begin{abstract}
Warm Comptonization models for the soft X-ray excess in AGN do not self-consistently explain the relationship between the Comptonizing medium and the underlying accretion disc. Because of this, they cannot directly connect the fitted Comptonization temperatures and optical depths to accretion disc parameters. Since bulk velocities exceed thermal velocities in highly radiation pressure dominated discs, in these systems bulk Comptonization by turbulence may provide a physical basis in the disc itself for warm Comptonization models. We model the dependence of bulk Comptonization on fundamental accretion disc parameters, such as mass, luminosity, radius, spin, inner boundary condition, and $\alpha$. In addition to constraining warm Comptonization models, our model can help distinguish contributions from bulk Comptonization to the soft X-ray excess from those due to other physical mechanisms, such as absorption and reflection. By linking the time variability of bulk Comptonization to fluctuations in the disc vertical structure due to magnetorotational instability (MRI) turbulence, our results show that observations of the soft X-ray excess can be used to study disc turbulence in the radiation pressure dominated regime. Because our model connects bulk Comptonization to one dimensional vertical structure temperature profiles in a physically intuitive way, it will be useful for understanding this effect in future simulations run in new regimes. 

\end{abstract}


\section{Introduction}
\label{sec_intro}
The soft X-ray excess in active galactic nuclei (AGN) refers to the part of the spectrum below 1keV that lies above the power law fit to the hard (2-10keV) X-rays \citep{sin85,arn85}. The origin of this component is not well understood. In narrow-line Seyfert 1s (NLS1), which are high Eddington, comparatively low mass ($M \sim 10^6$) AGN, the disc color temperature is high enough that it contributes to the soft excess, but it cannot fully explain it (\citealt{don12}, hereafter D12). In other systems, the soft excess must entirely be due to some other physical mechanism.

The soft excess is often fitted with warm Comptonization models (D12; \citealt{mag98,jan01,dew07,jin09,meh11,pal16,pet17}). In this picture, photons from the disc are upscattered by a warm ($\sim 0.2$keV), optically thick medium. While these models have had considerable success fitting the data, they do not explain the physical origin of such a medium in some sources and therefore cannot predict its properties from the parameters of the underlying accretion disc. Models more tied to intrinsic disc physics include reflection models \citep{cru06,ros05} and absorption models \citep{gie04,sch07}.

In this work we explore the extent to which Comptonization by turbulent, bulk velocities in the disc itself, rather than thermal velocities, can provide a physical basis for warm Comptonization models. Since the ratio of turbulent velocities to thermal velocities for the alpha disc model \citep{sha73} is given by
\begin{align}
\label{eq_ratio_1}
\frac{\left\langle v_{\rm turb}^2\right\rangle}{\left\langle v_{\rm th}^2 \right\rangle}
\sim \alpha \left(m_e\over m_{\rm p}\right) \left({P_{\rm rad}\over P_{\rm gas}}\right),
\end{align}
we expect Comptonization by turbulence to be relevant in radiation pressure dominated systems. The ratio of radiation to gas pressure increases with mass and accretion rate, so we expect this effect to be most significant in high Eddington accretion discs around supermassive black holes, such as NLS1s. It is likely insignificant, therefore, in lower Eddington flows, even though the data show that in these flows the soft excess carries a more significant fraction of the power \citep{jin12,meh11,meh15}.

Preliminary work on bulk Comptonization in accretion discs was done by \cite{soc04} and \cite{soc10}. \cite{kau16} studied the physical principles that govern bulk Comptonization in detail. \cite{kau17} (hereafter, K17) used data from radiation magnetohydrodynamic (MHD) simulations to model bulk Comptonization for accretion discs with parameters typical of NLS1s. They characterized their results with temperatures and optical depths to make contact with warm Comptonization models of the soft excess. For a system with disc parameters similar to those fit by D12 to the source REJ1034+396, the temperatures, optical depths, and Compton $y$ parameters found by K17 broadly agree with those measured by D12.

This work simplifies and generalizes the bulk Comptonization model presented in K17 in order to develop greater physical insight into this process and explore a larger space of accretion disc parameters. In K17, bulk Comptonization is modeled by fitting the Comptonization temperature and optical depth parameters to spectra computed with Monte Carlo post-processing of simulations of the turbulence. Here we develop a procedure to infer these Comptonization parameters from the underlying disc vertical structure radiation MHD simulation data without computing spectra. The immediate benefit of this is that we can efficiently explore a larger space of accretion disc parameters. We can also find the time-averaged Comptonization parameters for a given simulation since without computing spectra we can now efficiently calculate Comptonization parameters for multiple timesteps.

More importantly, our model provides a physically intuitive framework for understanding bulk Comptonization. In particular, we show that the variation of this effect with disc parameters can be understood in terms of the vertical gas temperature and ``wave'' temperature (Section \ref{sec_overview}) profiles. This allows us to determine the dependence of bulk Comptonization on each accretion disc parameter separately without exhaustively exploring a multiparameter space. We can also probe how various physical effects, such as vertical radiation advection \citep{bla11,jia13,jia14}, may impact bulk Comptonization, as well as evaluate how robust our model's predictions are to changes in the disc vertical structure. In particular, although the specific bulk Comptonization parameters that we calculate here result from applying our model to limited, scaled data from the shearing box simulation 110304a (K17), we show that our principal findings regarding how bulk Comptonization scales with fundamental accretion disc parameters are likely to be robust to differences in the disc vertical structure seen in other simulations. Furthermore, understanding this framework should be useful for developing physical intuition in new situations in which some of our particular results may no longer hold, such as shearing box or global disc simulations run in radically different regimes.

The structure of this paper is as follows. In Section \ref{sec_modeling} we describe our model and show why it is effective. In Section \ref{sec_results} we apply our model to data from radiation MHD simulations. We show how the dependence of bulk Comptonization on shearing box parameters can be understood in terms of one dimensional temperature profiles (Section \ref{res_part_1}), and then proceed to examine its dependence on each accretion disc parameter individually (Section \ref{res_part_2}). We estimate bulk Comptonization for an entire disc as well by fixing the radius to the region of maximum luminosity (Section \ref{res_part_3}). We consider the effect of including radiation advection (Section \ref{res_advection}) and discuss the time variability of bulk Comptonization within a given simulation (Section \ref{res_timestep}). In Section \ref{sec_discussion} we discuss our results, and we summarize our findings in Section \ref{sec_Conclusion}.

\section{Efficiently modeling bulk Comptonization}
\label{sec_modeling}
\subsection{Overview}
\label{sec_overview}
Since this work simplifies and generalizes the bulk Comptonization model that we presented in K17, we begin by summarizing how they calculated the bulk Comptonization temperature and optical depth for a given system. At each radius in an accretion disc, they used radiation MHD stratified shearing box simulation data to calculate spectra both including and excluding velocities. Each spectral computation was performed by running a post-processing Monte Carlo simulation on a simulation data snapshot at a particular epoch in time. Since their data was limited, they developed a scheme to scale the original data to the accretion disc parameters of interest. At each radius in the disc, they used the Kompaneets equation to pass the spectrum computed without velocities through a Comptonizing medium with a given electron temperature and optical depth, and the resulting spectra were superposed to obtain the observed spectrum. Meanwhile, the spectra computed with velocities at each radius were superposed to obtain a different observed spectrum. The temperature and optical depth parameters (which are assumed to be the same at all radii) were adjusted until the two spectra match.

Here we develop a more efficient and physically revealing procedure for calculating the Comptonization temperature and optical depth. We focus on computing these parameters for each radius individually rather than for the whole disc at once, a choice that also allows us to study the dependence of bulk Comptonization on radius. This choice does not limit us to studying individual radii, since bulk Comptonization for a whole disc can be estimated by the Comptonization parameters at the radius where the luminosity is greatest (Section \ref{res_part_3}). One other difference from K17 is that we include only turbulent velocities, not shear velocities. This allows us to study the effects of turbulence alone. The preliminary results in K17 suggest that bulk Comptonization by shear is subdominant to bulk Comptonization by turbulence, but to rigorously calculate the effect of shear near the photosphere will require global simulations. Since the scalings for the shear velocities are nearly identical to the scalings for the turbulent velocities (K17), this omission should not affect our conclusions regarding the general dependence of bulk Comptonization on accretion disc parameters.

At a given radius, we define the Kompaneets parameters, consisting of the electron temperature $T_{\rm K}$ and optical depth $\tau_{\rm K}$, analogously to how the temperature and optical depth are defined for the whole disc in K17. In other words, spectra are calculated with and without velocities at a given radius, and the spectrum computed without velocities is passed through a Comptonizing medium using the Kompaneets equation. The temperature and optical depth of the medium are adjusted until the resulting spectrum matches the spectrum calculated with velocities. To scale simulation data to different accretion disc parameters we use the scheme developed in K17. 

We show that the Kompaneets parameters are approximated by what we will henceforth refer to as the Comptonization parameters. We define these parameters with an efficient and physically revealing procedure that we outline here and then discuss in greater detail in the following sections. First, we map the bulk velocity grid to a temperature grid by defining at each point a bulk ``wave'' temperature,
\begin{align}
\label{eq_wave_temp}
\frac{3}{2}k_{\rm B} T_{\rm w} = \frac{1}{4}m_{\rm e} \left\langle \left( \Delta {\bf v}\right)^2 \right\rangle_{\bf r}, 
\end{align}
where $\left\langle \left( \Delta {\bf v}\right)^2 \right\rangle_{\bf r}$ is the average square velocity difference between subsequent photon scatterings at ${\bf r}$. We note that if instead of applying equation (\ref{eq_wave_temp}) to the bulk velocities in a region we apply it to the thermal velocity distribution at a particular point, then since
\begin{align}
\left\langle \left( \Delta {\bf v}\right)^2\right\rangle &= \int \left({\bf v_2} - {\bf v_1} \right)^2 f\left({\bf v_1} \right)f\left({\bf v_2} \right) d {\bf v_1} d {\bf v_2} \\
&=2\left(\left\langle v^2 \right\rangle - \left\langle {\bf v} \right\rangle^2 \right) \\
&= 2\left\langle v^2 \right\rangle,
\end{align}
we find that
\begin{align}
\frac{3}{2}k_{\rm B} T_{\rm w} = \frac{1}{2}m_{\rm e} \left\langle v^2 \right\rangle_{\bf r}, 
\end{align}
as expected. We call this a ``wave" temperature and not a ``turbulent" temperature because it depends on the power spectrum of the turbulence and is less than the temperature that one usually associates with a turbulent velocity distribution, which is given by $\frac{3}{2}k_{\rm B} T = \frac{1}{2}m_{\rm e} \left\langle v^2 \right\rangle_{\bf r}$, analogous to a thermal temperature. We discuss $\left\langle \left( \Delta {\bf v} \right)^2 \right\rangle_{\bf r}$ in more detail in Section \ref{sec_step_map}. Next, we horizontally average all simulation variables, including the newly defined wave temperature, to obtain 1D profiles of the data. For the systems of interest, the wave temperature is negligible compared to the gas temperature at the effective photosphere, and it increases going outward so that near the scattering photosphere it may exceed it. We define the Comptonization optical depth $\tau_{\rm C}$ as the optical depth of the region in which the wave temperature is at least half the gas temperature, a region that we will henceforth refer to as the bulk Comptonization region. We define the Comptonization temperature $T_{\rm C}$ as a weighted average of the sum of the gas and wave temperatures in this region, given by
\begin{align}
\label{eq_T_C}
T_{\rm C} = \frac{\int_0^{\tau_{\rm C}} \left(T_{\rm g} + T_{\rm w}\right) \tau d \tau}{\int_0^{\tau_{\rm C}} \tau d \tau}.
\end{align}

In the next section we both describe in detail why this procedure approximates the Kompaneets parameters and demonstrate its effectiveness by comparing what it predicts with the results of actual Monte Carlo spectral calculations. 

\subsection{Physical justification for the bulk Comptonization model}
\label{sec_justification}
To justify our definition of the Comptonization parameters, $\tau_{\rm C}$ and $T_{\rm C}$, we start with the procedure that defines the Kompaneets parameters and then incrementally simplify it. In the following sections we detail each step of this process. Where appropriate we invoke an accretion disc parameter set for which $M = 2 \times 10^6 M_\odot$ and $L/L_{\rm Edd} = 5$ as a test case. The other parameters are given in Table \ref{table_disc_param}. The parameter $a$ is the black hole dimensionless spin parameter, $\Delta \epsilon$ is the change in efficiency for a non-zero torque inner boundary condition \citep{ago00}, $\alpha$ is the ratio of vertically integrated stress to vertically integrated pressure, and $\alpha_0$ is the value of $\alpha$ for the original simulation data. We list $\alpha/\alpha_0$ rather than $\alpha$ since it is the former that we can directly adjust with the scaling scheme from K17. Typically $\alpha_0 \sim 0.01$. We note that these parameters are nearly identical to those of the systems modeled in K17, and were originally chosen to correspond to those fit by D12 to the NLS1 REJ1034+396. The only parameter whose value differs from the value in K17 is $L/L_{\rm Edd}$, which we set here to $5$ rather than $2.5$. Since bulk Comptonization increases with $L/L_{\rm Edd}$, we choose a higher value here so that it is easier to see the effectiveness of our approximations in plots of actual spectra. After describing all steps, we demonstrate the effectiveness of the resulting procedure at six different radii for not only the $M = 2 \times 10^6 M_\odot$, $L/L_{\rm Edd} = 5$ parameter set but for two others as well. One is the same except that $L/L_{\rm Edd} = 2.5$, the original value in K17. For the other, $M = 2\times 10^8 M_\odot$ and $L/L_{\rm Edd} = 4.2$. All disc parameter sets are given in Table \ref{table_disc_param}. As in K17, all spectra in this section are computed with Monte Carlo post processing simulations \citep{dav09,poz83}, using data from the 140 orbits timestep of simulation 110304a. We discuss this simulation in more detail in Section \ref{sec_dependence_1}.

\begin{table}
\caption{Accretion disc parameter sets}
\begin{tabular}{lllll}
$M/M_\odot$ & $L/L_{\rm Edd}$ & $a$ & $\Delta \epsilon$ & $\alpha/\alpha_0$ \\
$2\times 10^6$ & $5$ & $0$ & $0$ & $2$ \\
$2\times 10^8$ & $4.2$ & $0$ & $0$ & $2$ \\
$2\times 10^6$ & $2.5$ & $0$ & $0$ & $2$ \\
\end{tabular}
\label{table_disc_param}
\end{table}

\subsubsection{Step 1 - Truncate the simulation data inside the effective photosphere and turn off emissivity above this surface}
\label{sec_step_trunc}
To begin, we observe that we can modify the defining procedure for calculating the Kompaneets parameters by using simulation data that is truncated inside the effective photosphere and turning off the emissivity everywhere except at this surface, without changing the resulting parameters. The effective photosphere is defined by using the Planck mean opacity. This phenomenon was demonstrated in the context of fitting Kompaneets parameters for an entire disc at once in K17, but it arises from the fact that it is true for individual radii.

\begin{figure}
\includegraphics[width = 84mm]{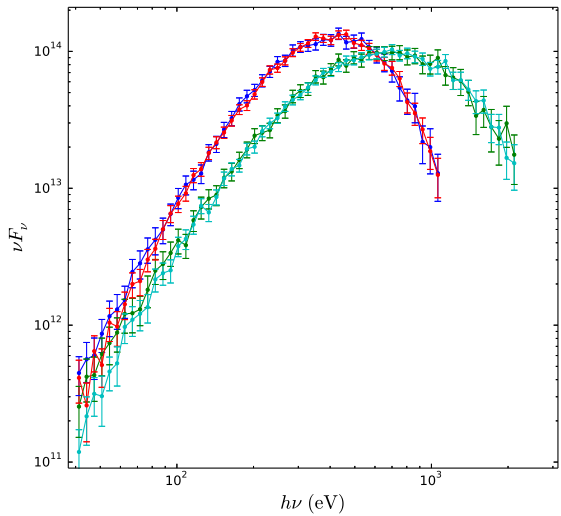}
\caption{Normalized accretion disc spectra at $r=14$ for the $M = 2 \times 10^6 M_\odot$, $L/L_{\rm Edd} = 5$ parameter set (Table \ref{table_disc_param}). The green and cyan curves are computed with velocities, and the blue and red curves are computed without velocities. For the cyan and red curves, the simulation data is truncated inside the effective photosphere and the emissivity is set to zero everywhere except at the effective photosphere.}
\label{fig_spectra_truncated_r14}
\end{figure}

For example, we calculate spectra both with and without velocities at $r = 14$ for the $M = 2 \times 10^6 M_\odot$, $L/L_{\rm Edd} = 5$ parameter set (Table \ref{table_disc_param}). We note that all numerical radii in this paper are in units of the gravitational radius $GM/c^2$ of the black hole. We also calculate spectra using simulation data that is truncated at the effective photosphere and in which the emissivity is set to zero everywhere except the base. All four resulting spectra are plotted in Figure \ref{fig_spectra_truncated_r14}. We see that except at very low energies, the two spectra calculated with velocities coincide and the two spectra calculated without velocities coincide.

Two reasons underlie this result. First, the emergent spectrum is dominated by photons originally emitted at or near the effective photosphere since free-free emission depends strongly on density. Second, bulk Comptonization is negligible except near the scattering photosphere (as we show in Section \ref{sec_step_map}) and so its effect does not depend on the precise effective optical depth at which photons are originally emitted. Because all systems in this work meet these conditions, this result is robust. Therefore, modifying the defining procedure for calculating the Kompaneets parameters in this way has a negligible effect on the outcome.

Calculating the temperature and optical depth with this modified method is the first step to simplifying the calculation and developing physical insight into the problem. Not only does this modified procedure run faster, but it shows that in order to model and understand bulk Comptonization we only need to understand the effect of bulk Comptonization on photons emitted at the effective photosphere.

\subsubsection{Step 2 - Map the velocity grid to a ``wave'' temperature grid}
\label{sec_step_map}
Next, we map the bulk velocity grid to a temperature grid by defining at each point a bulk ``wave'' temperature given by equation (\ref{eq_wave_temp}). This definition is motivated by the results of Section 4.2 in K16. There they showed that for a periodic box with statistically homogeneous turbulence and an escape probability, bulk Comptonization can be treated as thermal Comptonization by solving the Kompaneets equation with a temperature given by $\frac{3}{2}k_{\rm B} T_{\rm w} = \frac{1}{4}m_{\rm e} \left\langle \left( \Delta {\bf v} \right)^2 \right\rangle$, where $\left\langle \left( \Delta {\bf v}\right)^2 \right\rangle$ is the volume average of $\left\langle \left( \Delta {\bf v}\right)^2 \right\rangle_{\bf r}$, which is in turn given by K16 equation (56):
\begin{align}
\label{eq_def_delta_v}
\langle (\Delta {\bf v})^2 \rangle_{{\bf r}} = \int \left( \Delta {\bf v} (\Delta {\bf r},{\bf r})\right)^2 P_{\Delta {\bf r}}(\Delta {\bf r}) d^3 \Delta {\bf r}. 
\end{align}
Here, $\Delta {\bf v} (\Delta {\bf r},{\bf r})$ is the velocity difference between positions ${\bf r}$ and ${\bf r} + \Delta {\bf r}$, and $P_{\Delta {\bf r}}(\Delta {\bf r})$ is the probability density that a photon scattering at ${\bf r}$ subsequently scatters at ${\bf r} + \Delta {\bf r}$. Hence $\langle (\Delta {\bf v})^2 \rangle_{{\bf r}}$ is the average square velocity difference between subsequent photon scatterings at ${\bf r}$. In this work, therefore, equation (\ref{eq_wave_temp}) defines a wave temperature at each point instead of taking a volume average and defining it for an entire box.

To develop physical intuition into equation (\ref{eq_wave_temp}) it is important to understand the dependence of $\langle (\Delta {\bf v})^2 \rangle_{{\bf r}}$ on density. In the high density limit the velocity difference between subsequent photon scatterings is small. In particular, $\langle (\Delta {\bf v})^2 \rangle_{{\bf r}}$ is proportional to the square of the mean free path $\lambda_{\rm p}$ (K16) so $T_{\rm w}$ decreases significantly with increasing density. In the low density limit, on the other hand, $\langle (\Delta {\bf v})^2 \rangle_{{\bf r}}$ approaches $2 \left\langle v^2\right\rangle$ so that $\frac{3}{2}k_{\rm B} T_{\rm w}$ approaches $\frac{1}{2}m_{\rm e} \left\langle  v ^2 \right\rangle_{\bf r}$. We also define the bulk temperature, given by
\begin{align}
\frac{3}{2}k_{\rm B} T_{\rm bulk} = \frac{1}{2}m_{\rm e} v^2.
\end{align}
Applying equation (\ref{eq_def_delta_v}) directly to simulation data is somewhat problematic, so we discuss our implementation in detail in Appendix \ref{sec_implementation}.

\begin{figure}
\includegraphics[width = 84mm]{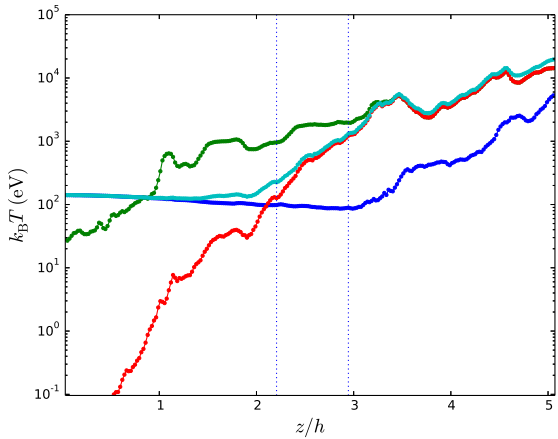}
\caption{Horizontally averaged profiles at $r=14$ for the $M = 2 \times 10^6 M_\odot$,  $L/L_{\rm Edd} = 5$ parameter set (Table \ref{table_disc_param}) for the gas temperature $T_{\rm g}$ (blue), bulk temperature $T_{\rm bulk}$ (green), wave temperature $T_{\rm w}$ (red), and sum of gas and wave temperatures (cyan). The dashed lines denote where $\tau_{\rm s} =$ $1$ and $\tau_{\rm s} =$ $10$.}
\label{fig_T_profiles_mdotfactor_3}
\end{figure}

For example, in Figure \ref{fig_T_profiles_mdotfactor_3} we plot the profile of the density weighted horizontal average of the wave temperature at $r=14$ for the $M = 2 \times 10^6 M_\odot$,  $L/L_{\rm Edd} = 5$ parameter set (Table \ref{table_disc_param}). We also plot the gas temperature, $T_{\rm g}$, the bulk temperature, $T_{\rm bulk}$, and the sum of the gas and wave temperatures. The dashed line on the right denotes the location of the scattering photosphere, which we define as the height at which the Thomson optical depth $\tau_{\rm s} =1$. We see that the wave temperature significantly increases with decreasing density (that is, moving rightward) as the photon mean free path grows, and is comparable to the bulk temperature only near the scattering photosphere.

\begin{figure*}
\begin{tabular}{ll}
\includegraphics[width = 84mm]{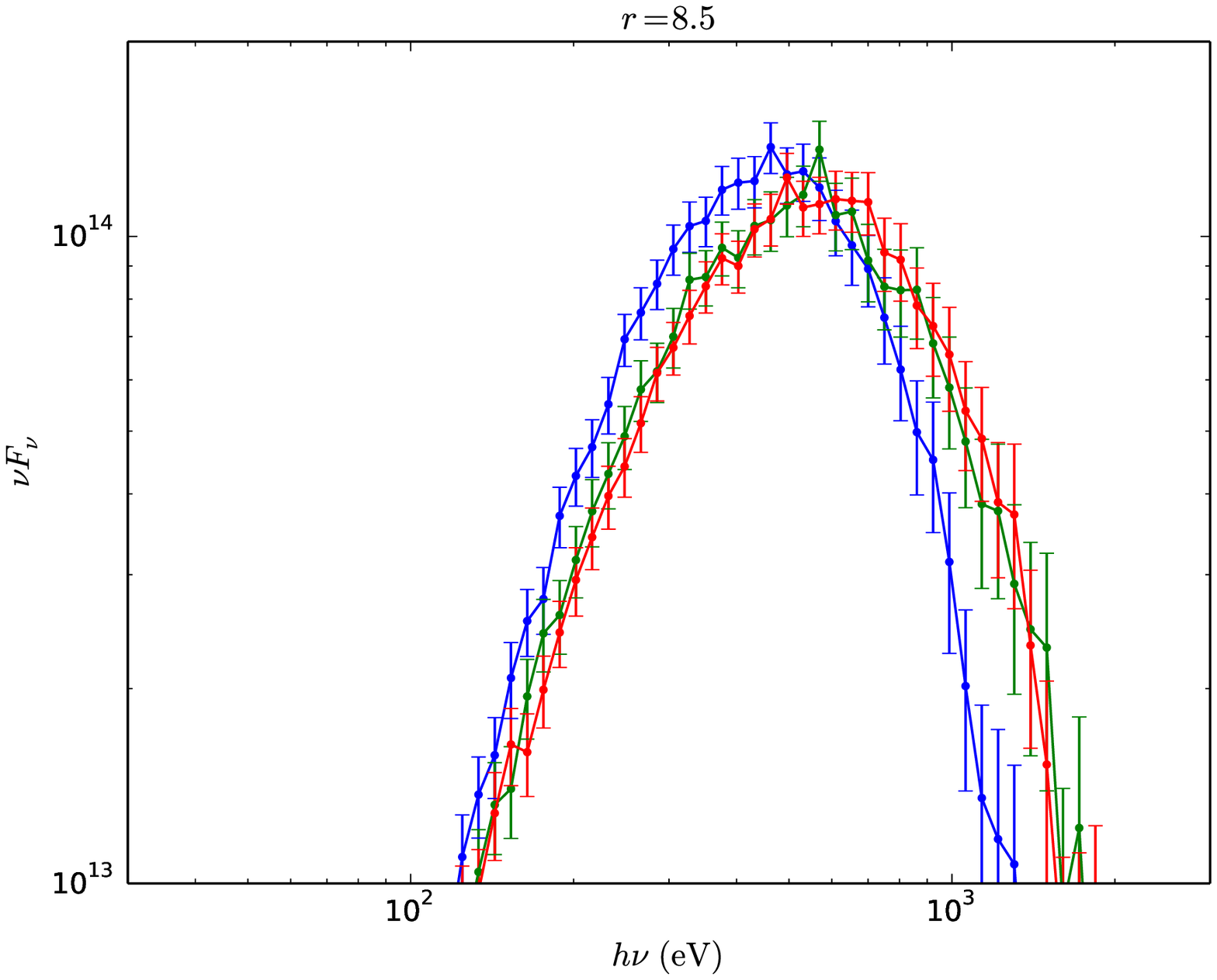} & \includegraphics[width = 84mm]{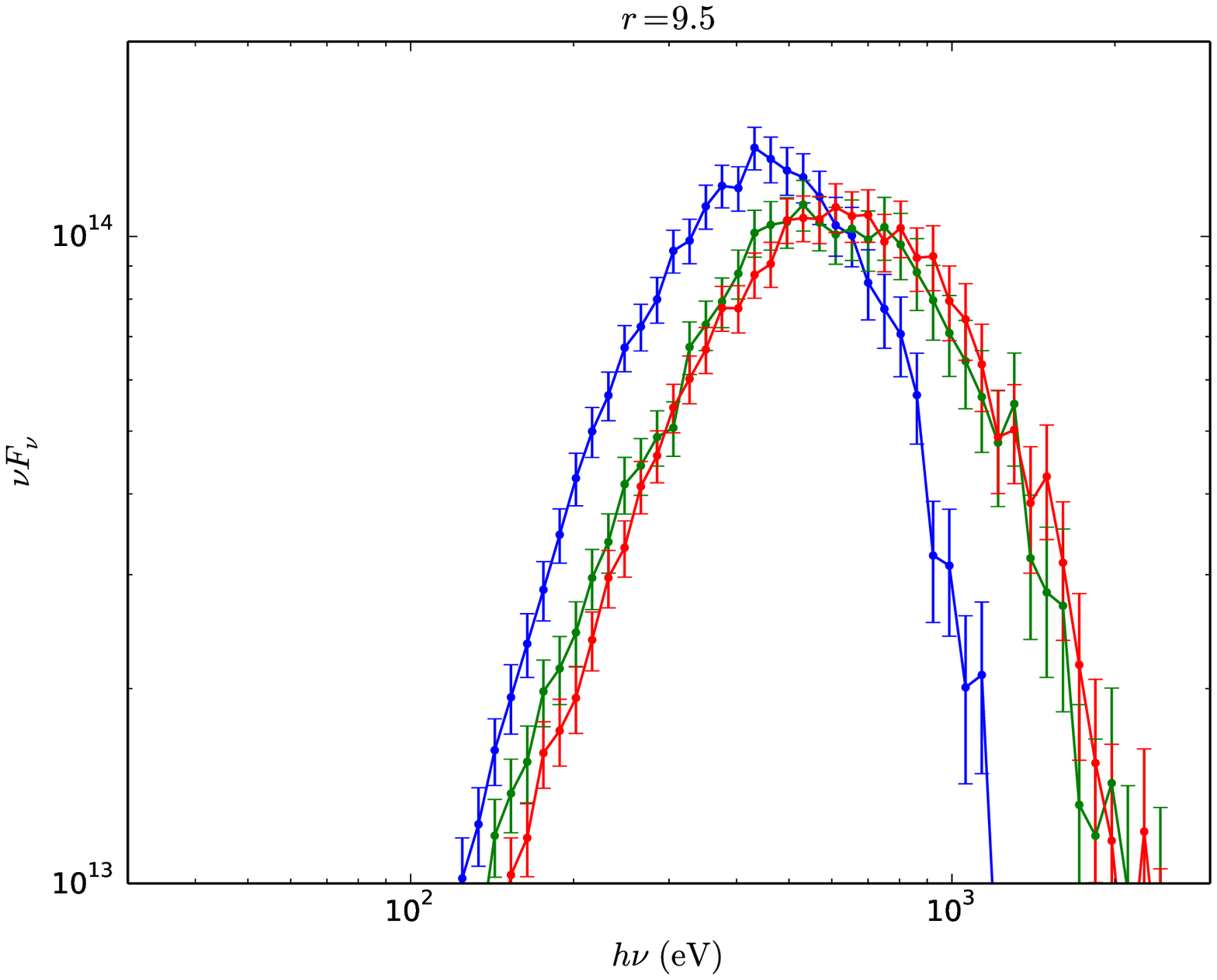} \\
\includegraphics[width = 84mm]{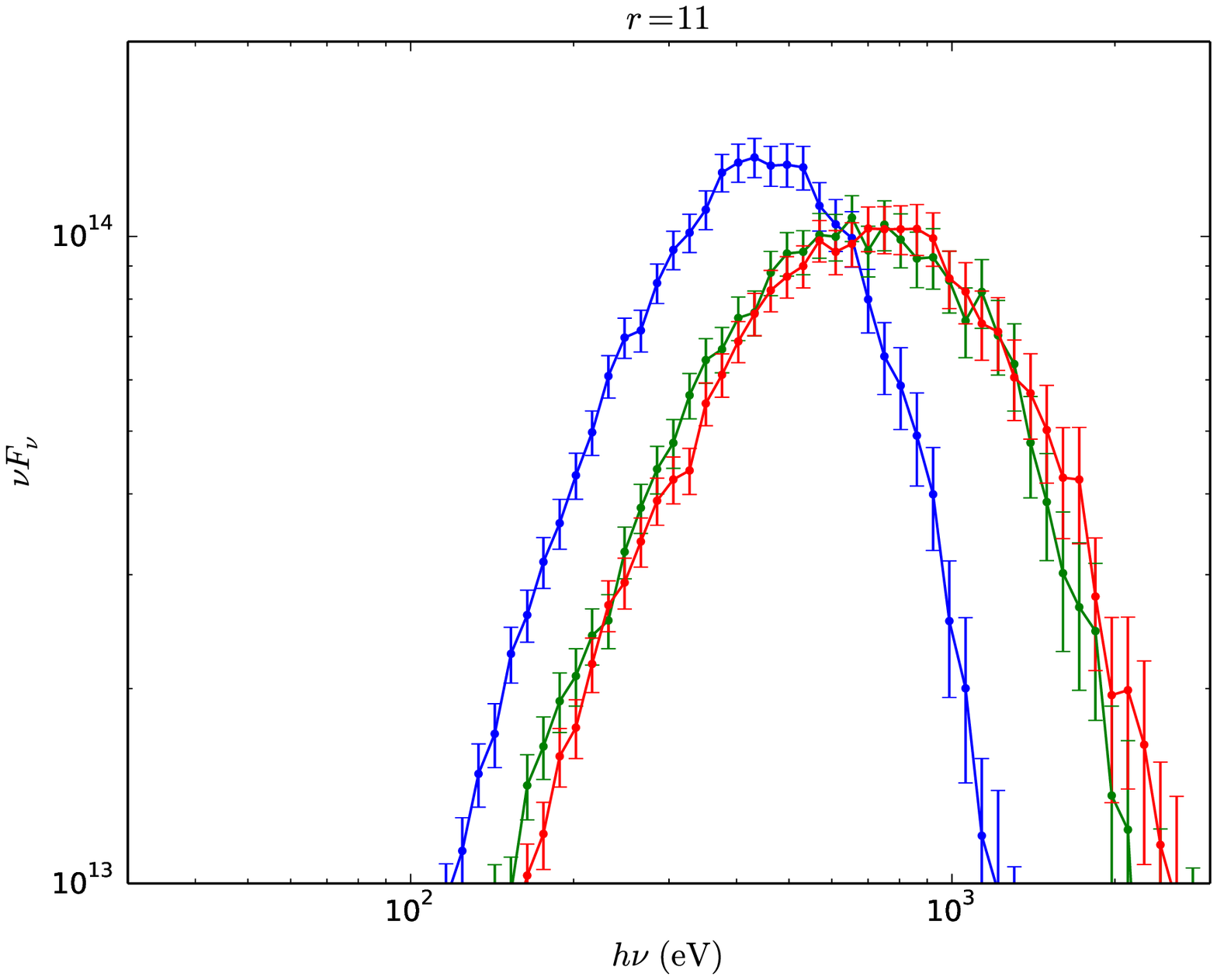} & \includegraphics[width = 84mm]{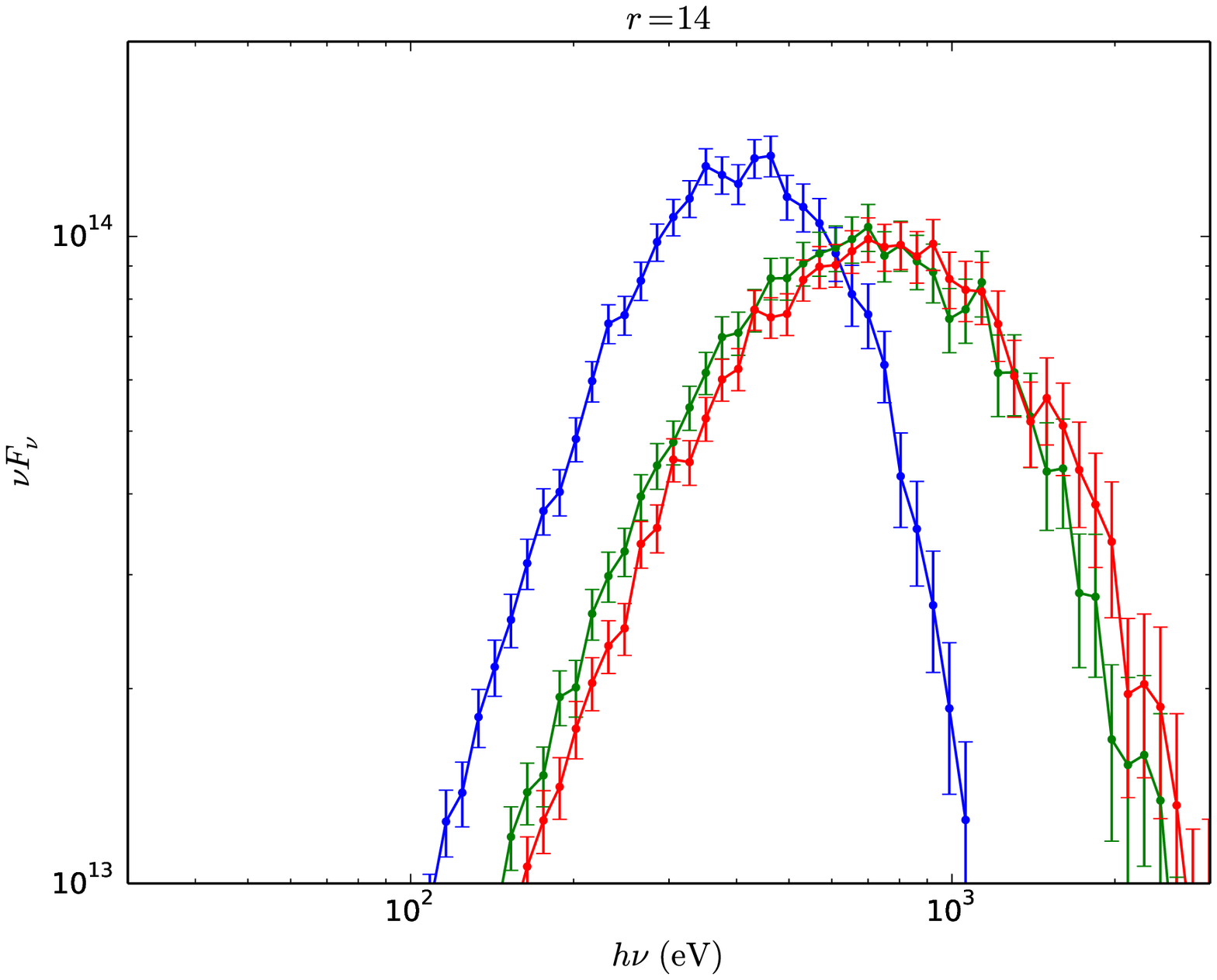} \\
\includegraphics[width = 84mm]{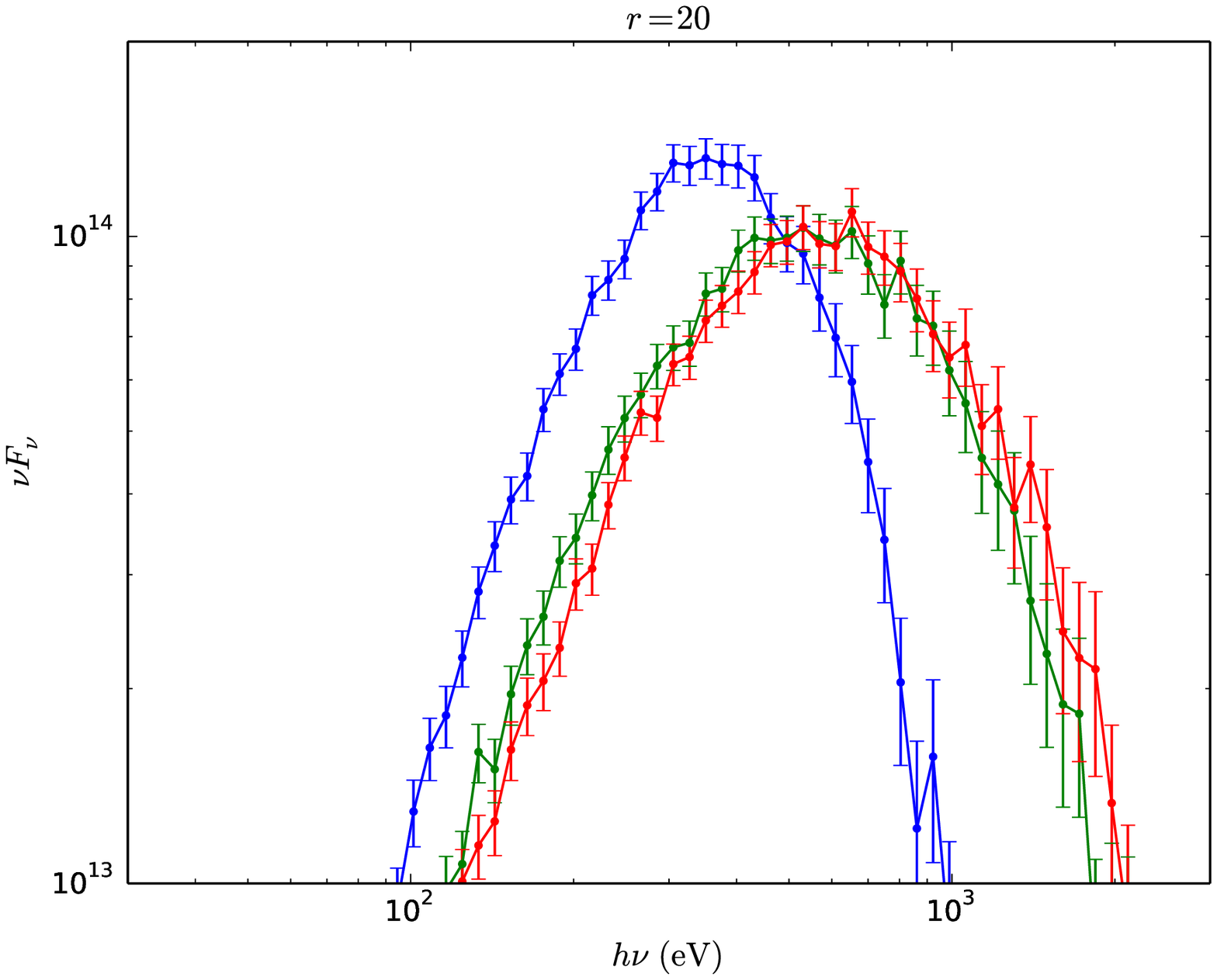} & \includegraphics[width = 84mm]{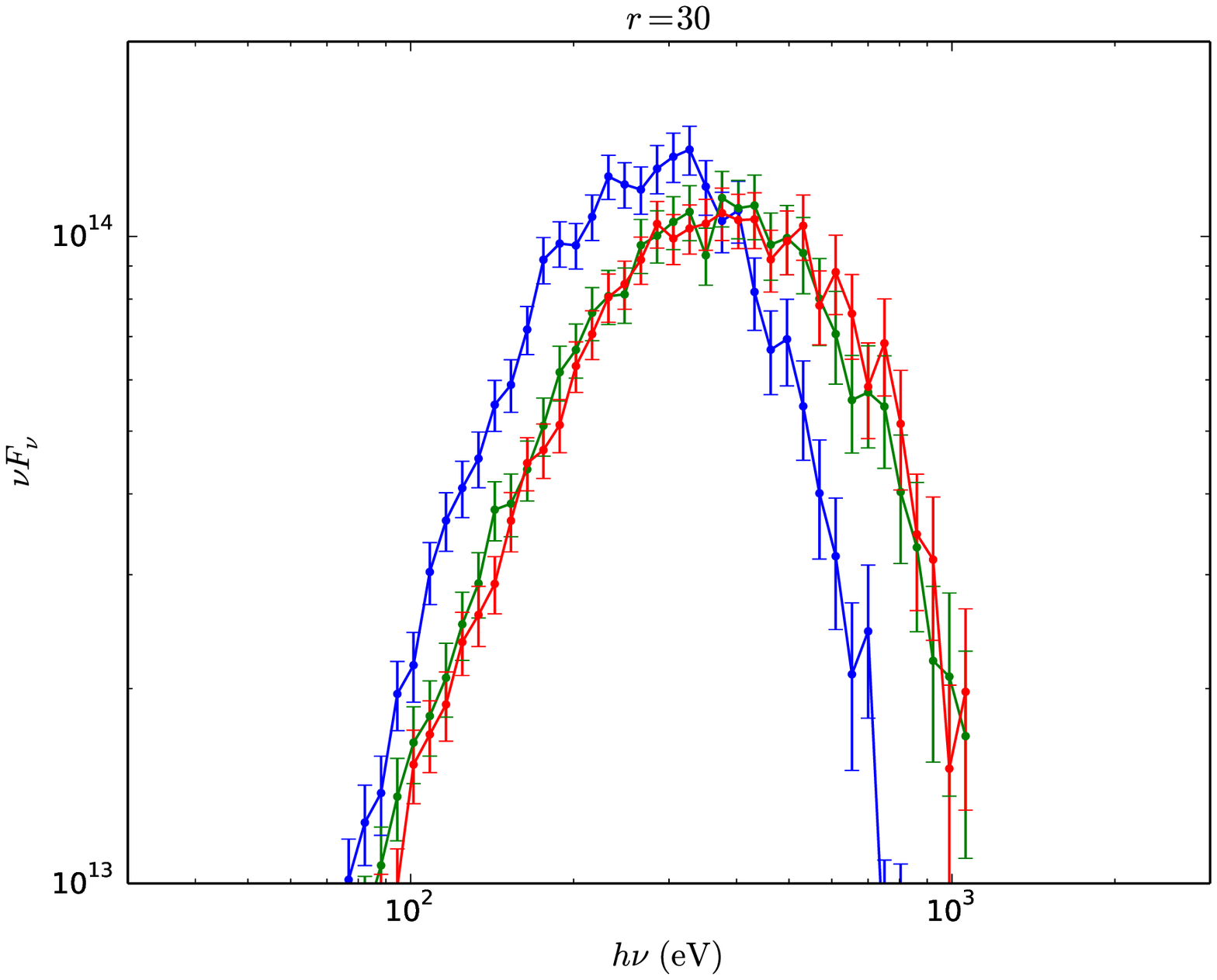} \\
\end{tabular}
\caption{Normalized accretion disc spectra at multiple radii for the $M = 2 \times 10^6 M_\odot$,  $L/L_{\rm Edd} = 5$ parameter set (Table \ref{table_disc_param}) computed with (green) and without (blue) the velocities. For the red curve, the velocities were not included but the wave temperatures were added to the gas temperatures.}
\label{fig_spectra_Twave}
\end{figure*}

We find that photon spectra computed with simulation data in which the velocities are turned off and the wave temperatures are added to the gas temperatures approximate photon spectra computed with the velocities turned on. For example, in Figure \ref{fig_spectra_Twave} we plot spectra at $r = 8.5$, $9.5$, $11$, $14$, $20$, and $30$ for the $M = 2 \times 10^6 M_\odot$,  $L/L_{\rm Edd} = 5$ parameter set (Table \ref{table_disc_param}), computed with and without velocities. We also plot spectra computed with data in which the velocities are turned off and the wave temperatures are added to the gas temperatures. We see that these spectra approximate the spectra computed with velocities. In other words, bulk Comptonization can be modeled by thermal Comptonization in which the temperature is given by equation (\ref{eq_wave_temp}).

\subsubsection{Step 3 - Horizontally average the simulation data}
\label{sec_step_1D}
Once bulk velocities are replaced by wave temperatures, it is straightforward to further simplify the problem by horizontally averaging the simulation data. In order that the effects of bulk Comptonization remain unchanged, the wave temperature data must be density averaged, not volume averaged.

\begin{figure}
\includegraphics[width = 84mm]{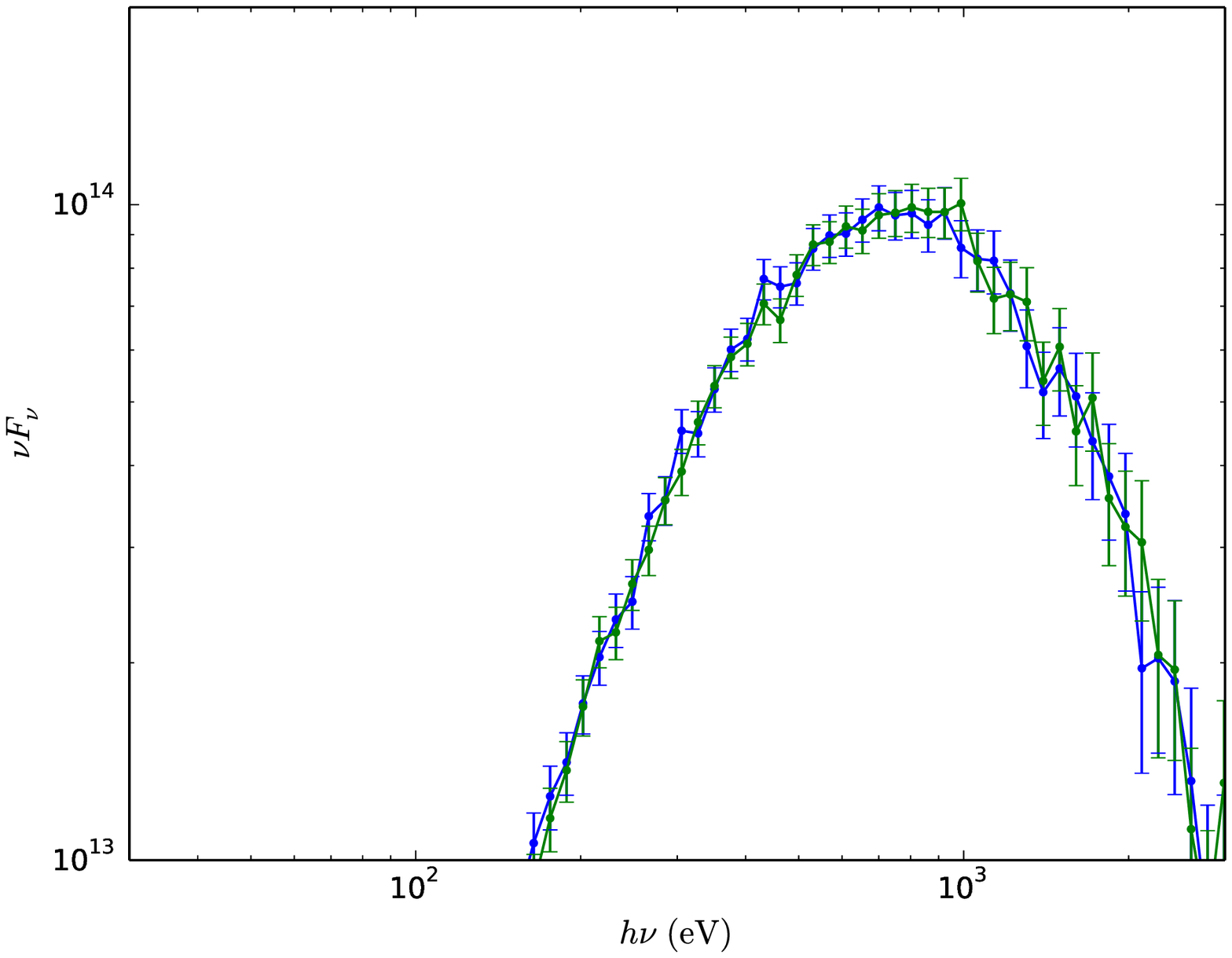}
\caption{Normalized accretion disc spectra at $r=14$ for the $M = 2 \times 10^6 M_\odot$,  $L/L_{\rm Edd} = 5$ parameter set (Table \ref{table_disc_param}) computed by omitting the velocities and instead adding the wave temperatures to the gas temperatures. The blue curve is computed with unaveraged data, and the green curve is computed with horizontally density weighted averaged data.}
\label{fig_spectra_Twave_avg_r14}
\end{figure}

\begin{figure}
\includegraphics[width = 84mm]{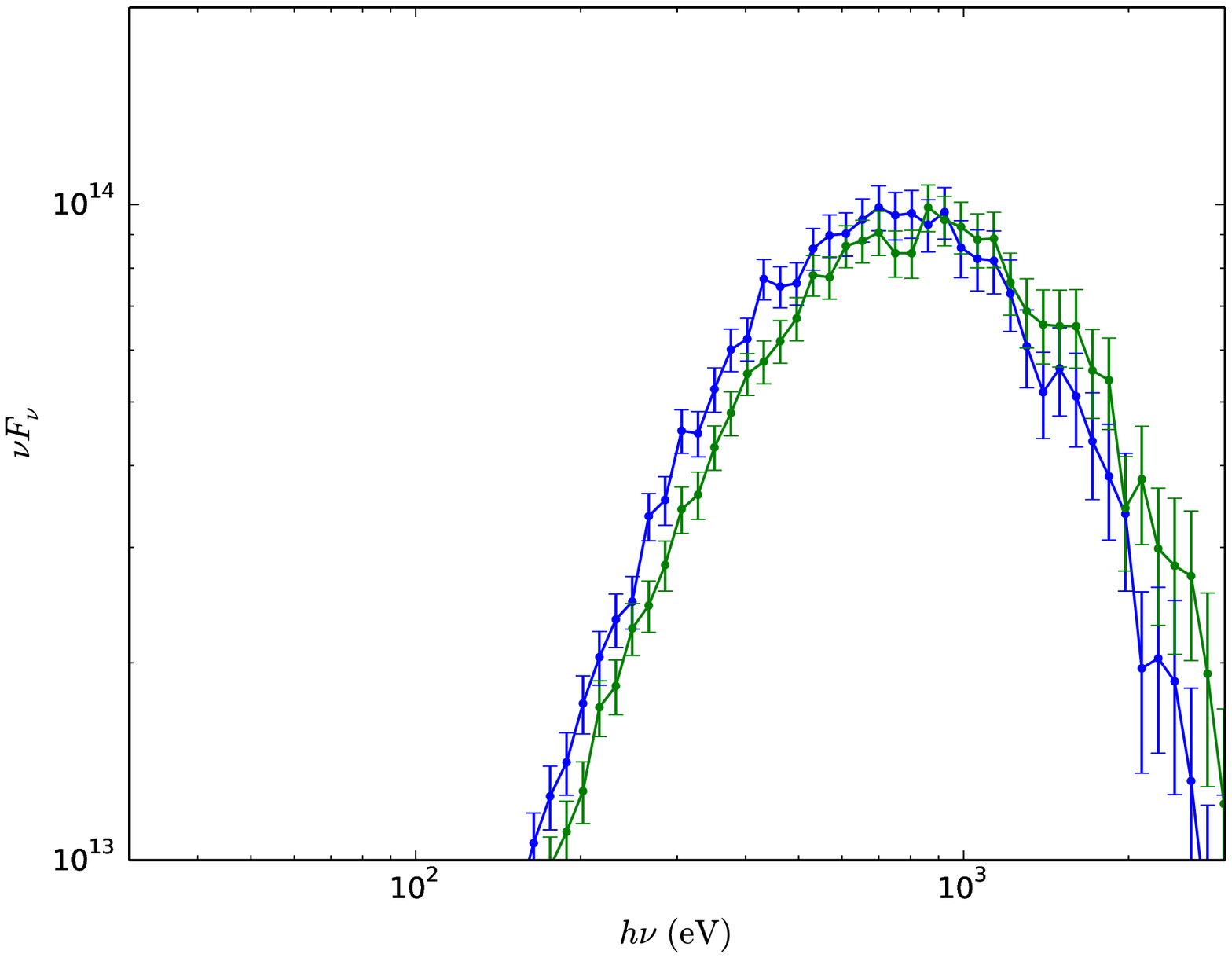}
\caption{Normalized accretion disc spectra at $r=14$ for the $M = 2 \times 10^6 M_\odot$,  $L/L_{\rm Edd} = 5$ parameter set (Table \ref{table_disc_param}) computed by omitting the velocities and instead adding the wave temperatures to the gas temperatures. The blue curve is computed with unaveraged data, and the green curve is computed with horizontally averaged data. For the latter, the wave temperature averages are volume weighted and the other variable averages are density weighted.}
\label{fig_spectra_Twave_vol_avg_r14}
\end{figure}

For example, we calculate the spectrum at $r=14$ for the $M = 2 \times 10^6 M_\odot$,  $L/L_{\rm Edd} = 5$ parameter set (Table \ref{table_disc_param}) using data in which the velocities are turned off and the wave temperatures are added to the gas temperatures, as described in Section \ref{sec_step_map}. We repeat this calculation using horizontally, density weighted averaged data and plot both spectra in Figure \ref{fig_spectra_Twave_avg_r14}. We see that the two spectra coincide. In Figure \ref{fig_spectra_Twave_vol_avg_r14} we again plot the spectrum calculated with the unaveraged data as well as a spectrum calculated with horizontally averaged data, except that this time the wave temperatures are computed with a simple spatial horizontal average instead. We see that simple spatially averaging the wave temperatures with no density weighting overestimates bulk Comptonization.

\begin{figure}
\includegraphics[width = 84mm]{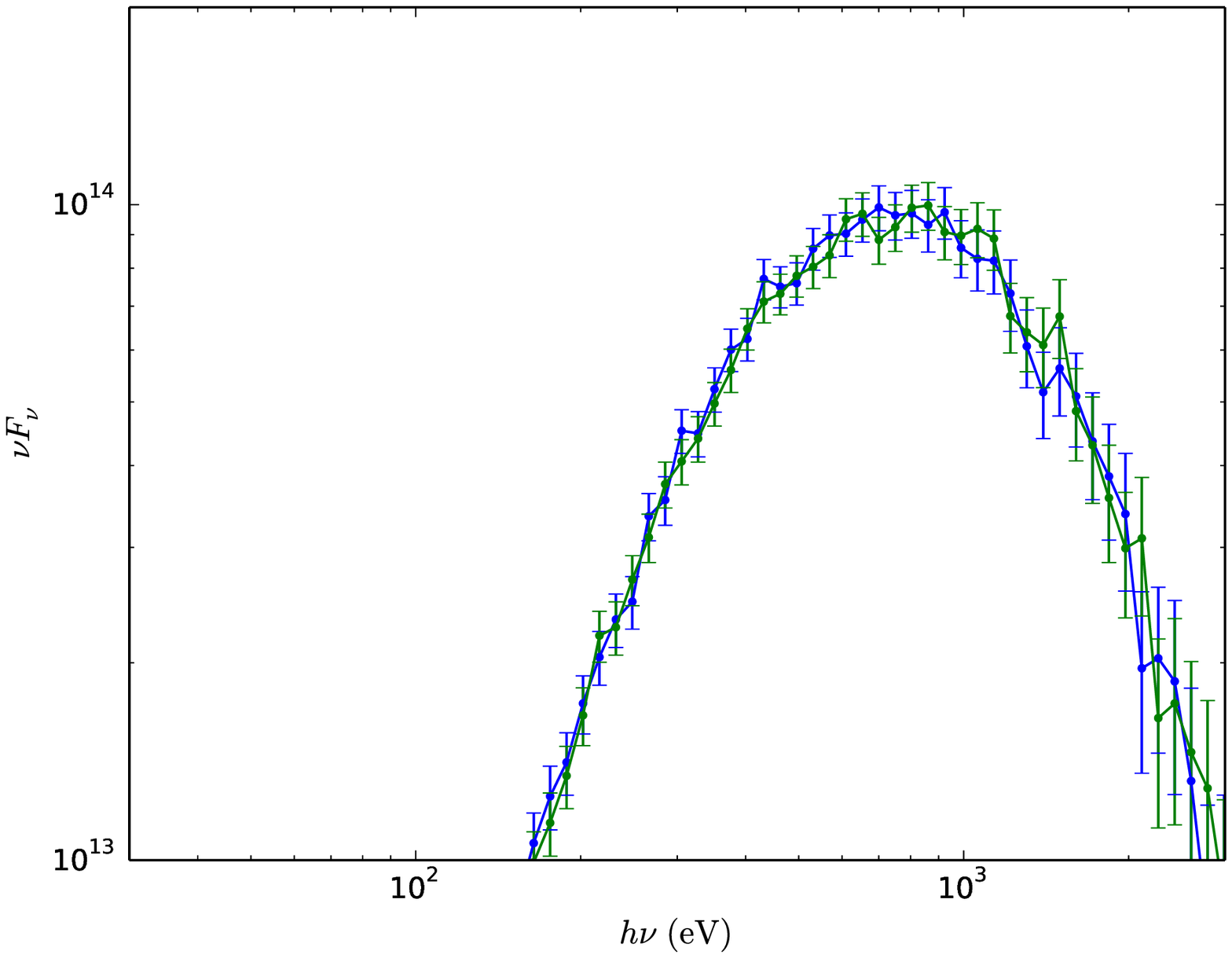}
\caption{Normalized accretion disc spectra at $r=14$ for the $M = 2 \times 10^6 M_\odot$,  $L/L_{\rm Edd} = 5$ parameter set (Table \ref{table_disc_param}) computed by omitting the velocities and instead adding the wave temperatures to the gas temperatures. The blue curve is computed with unaveraged data, and the green curve is computed with horizontally averaged data. For the latter, the gas temperature averages are volume weighted and the other variable averages are density weighted.}
\label{fig_spectra_Twave_avg_Tgas_vol_avg_r14}
\end{figure}

Density averaging improves the accuracy because the time photons spend in a region increases with the region's density. The reason that volume weighting overestimates bulk Comptonization is that the wave temperature is strongly correlated with density. As we discussed in Section \ref{sec_step_map}, the wave temperature decreases with density, and so volume averaging gives too much weight to regions where the wave temperature is larger. Because the gas temperature, on the other hand, is not strongly correlated with density, horizontally volume weighting the gas temperature has a negligible impact on the spectrum. For example, we calculate another spectrum with horizontally averaged data, except that this time the gas temperatures are computed with a simple spatial average instead. We plot the result alongside the spectrum calculated from the unaveraged data in Figure \ref{fig_spectra_Twave_avg_Tgas_vol_avg_r14}. We see that the two spectra coincide, which indicates that gas temperature inhomogeneities at a given height are not sufficiently correlated with density inhomogeneities to affect the spectrum.

The fact that we can use horizontally averaged quantities means we can map a three dimensional problem to a one dimensional problem, an important step to efficiently calculating and understanding bulk Comptonization. Plots of horizontally averaged quantities such as Figure \ref{fig_T_profiles_mdotfactor_3}, first introduced in Section \ref{sec_step_map}, will be of great use in the remainder of this work. 

\subsubsection{Step 4 - Solve the 1D inhomogeneous thermal Comptonization problem}
\label{sec_step_TC}
By this point, we have modified the original procedure for calculating the Kompaneets parameters by instead calculating spectra with and without adding the wave temperature profile to the gas temperature profile (Section \ref{sec_step_map}), using horizontally averaged data (Section \ref{sec_step_1D}) truncated inside the effective photosphere in which velocities are turned off and emission is zeroed everywhere except at the effective photosphere (Section \ref{sec_step_trunc}). The temperature and optical depth are adjusted until the spectra match.

To further simplify the problem, we first need to understand the effect of thermal Comptonization on photons emitted at the base of an inhomogenous one dimensional medium. We expect that if the optical depth is not too high, so that the average photon energy is always significantly below the local temperature (i.e. the photon spectrum does not saturate), then this process can always be well described by a homogeneous thermal Comptonization model. Since the number of scatterings is proportional to the square of the optical depth, the appropriate average scattering temperature should be given by
\begin{align}
\label{eq_T_1D}
T_{\rm 1D} = \frac{\int T \tau d \tau}{\int \tau d \tau}.
\end{align}
We test this description of 1D thermal Comptonization at $r = 8.5$, $9.5$, $11$, $14$, $20$, and $30$ for the $M = 2 \times 10^6 M_\odot$,  $L/L_{\rm Edd} = 5$ parameter set (Table \ref{table_disc_param}). At each radius, we add the wave temperature to the gas temperature and truncate the data inside $\tau_{\rm s} = 10$, an optical depth that is large enough to result in significant Comptonization but small enough to prevent the saturation of photon spectra for our purposes. We place a $50{\rm eV}$ Planck source at this location and calculate the emergent spectra. At each radius we also use the Kompaneets equation to pass the source through a homogeneous medium with temperature $T_{\rm 1D}$ and optical depth $\tau = 10$. We plot the resulting spectra for $r=14$ in Figure \ref{fig_Comp_1D}. Spectra at the other radii illustrate the same effect and are plotted in Appendix \ref{sec_figures} (Figure \ref{fig_Comp_1D_full}). The value of $T_{\rm 1D}$ at each radius is given in Table \ref{table_T_1D}. We see that the spectrum calculated using the Kompaneets equation coincides with the spectrum computed directly from the data, confirming that unsaturated, 1D inhomogeneous thermal Comptonization is well modeled by homogeneous thermal Comptonization, with the temperature given by $T_{\rm 1D}$.

\begin{table}
\caption{Values of $T_{\rm 1D}$ for vertical structure data truncated at $\tau_{\rm s} = 10$ at multiple radii for the $M = 2 \times 10^6 M_\odot$,  $L/L_{\rm Edd} = 5$ parameter set (Table \ref{table_disc_param}).}
\begin{tabular}{lllllll}
$r$                      & $8.5$ & $9.5$ & $11$  & $14$  & $20$  & $30$ \\
$T_{\rm 1D}$ $({\rm eV})$ & $226$ & $304$ & $361$ & $408$ & $344$  & $212$ 
\end{tabular}
\label{table_T_1D}
\end{table}

\begin{figure}
\includegraphics[width = 84mm]{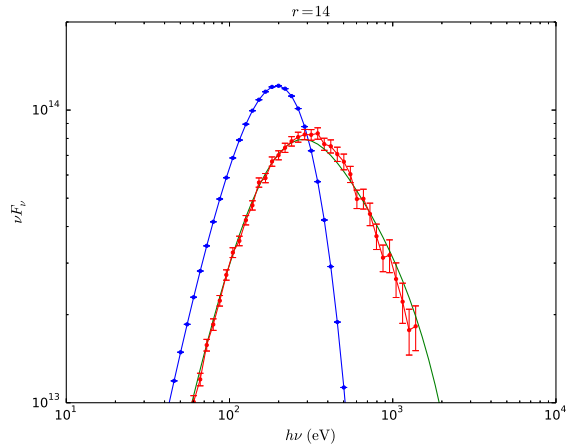}
\caption{Normalized spectrum (red) computed by passing a $50{\rm eV}$ Planck source (blue) through vertical structure data truncated at $\tau_{\rm s} = 10$ at $r = 14$ for the $M = 2 \times 10^6 M_\odot$,  $L/L_{\rm Edd} = 5$ parameter set (Table \ref{table_disc_param}). The velocities are zeroed and the wave temperatures are added to the gas temperatures. The green curve is calculated by using the Kompaneets equation to pass the $50{\rm eV}$ Planck source through a homogeneous medium with temperature  $T_{\rm 1D}$, given in Table \ref{table_T_1D}. Spectra at other radii are plotted in Appendix \ref{sec_figures} (Figure \ref{fig_Comp_1D_full}).}
\label{fig_Comp_1D}
\end{figure}

\subsubsection{Step 5 - Use the solution to the 1D thermal Comptonization problem to model bulk Comptonization}
\label{sec_step_model}
Armed with the results of Section \ref{sec_step_TC}, we return to the original problem. We begin by observing that typically in the region between the effective photosphere and the scattering photosphere (that is, where $\tau_{\rm eff} < 1$ and $\tau_{\rm s} > 1$), the gas temperature does not vary significantly. The wave temperature, on the other hand, changes rapidly with density. It is negligible compared to the gas temperature at the bottom of the effective photosphere and increases moving outward. Near the scattering photosphere it may exceed the gas temperature, depending on the parameters of the problem. The effect of adding the wave temperature profile to the gas temperature profile, therefore, is to take the spectrum that results from when there is no wave temperature and pass it through a Comptonizing medium of optical depth given by that of the region where the wave temperature is comparable to the gas temperature. We define this to be the region in which the wave temperature is at least half the gas temperature. We refer to it as the bulk Comptonization region and define the Comptonization optical depth parameter $\tau_{\rm C}$ to be its optical depth. We then define the associated Comptonization temperature parameter $T_{\rm C}$ by equation (\ref{eq_T_1D}), where $T$ is the sum of the gas and wave temperatures.

For example, we calculate spectra with and without velocities at $r = 8.5$, $9.5$, $11$, $14$, $20$, and $30$ for the $M = 2 \times 10^6 M_\odot$,  $L/L_{\rm Edd} = 5$ parameter set (Table \ref{table_disc_param}).  At each radius we also use the Kompaneets equation to pass the spectrum computed without velocities through a homogeneous medium with temperature $T_{\rm C}$ and optical depth $\tau_{\rm C}$. We plot the resulting spectra for $r=14$ in Figure \ref{fig_spectra_fit_set1}. Spectra at the other radii illustrate the same effect and are plotted in Appendix \ref{sec_figures} (Figure \ref{fig_spectra_fit_set1_full}). The temperature and optical depth parameters at all radii are given in Table \ref{table_comp_param_A}. We see that the spectrum computed with the Kompaneets equation approximates the spectrum computed with velocities, indicating that the Comptonization parameters $T_{\rm C}$ and $\tau_{\rm C}$ approximate the Kompaneets parameters, $T_{\rm K}$ and $\tau_{\rm K}$.

\begin{table}
\caption{Comptonization temperatures and optical depths at multiple radii for the $M = 2 \times 10^6 M_\odot$,  $L/L_{\rm Edd} = 5$ parameter set (Table \ref{table_disc_param}).}
\begin{tabular}{lllllll}
$r$                       & $8.5$ & $9.5$ & $11$  & $14$  & $20$  & $30$ \\
$T_{\rm C}$ $({\rm eV})$   & $210$ & $246$ & $251$ & $253$ & $203$ & $149$ \\
$\tau_{\rm C}$            & $11$ & $13$   & $16$  & $17$  & $18$  & $16$
\end{tabular}
\label{table_comp_param_A}
\end{table}

\begin{figure}
\includegraphics[width = 84mm]{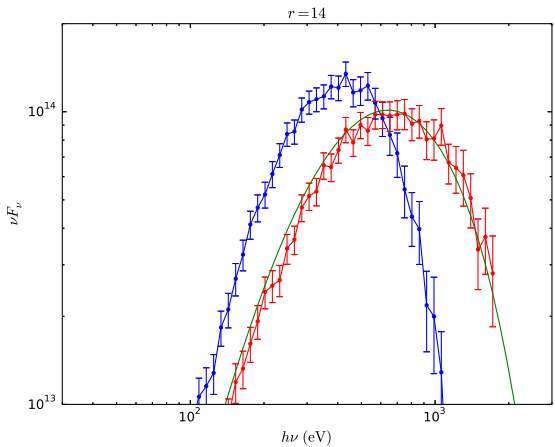}
\caption{Normalized spectra at $r=14$ for the $M = 2 \times 10^6 M_\odot$,  $L/L_{\rm Edd} = 5$ parameter set (Table \ref{table_disc_param}) computed with (red) and without (blue) velocities. The green curve is calculated by using the Kompaneets equation to pass the blue curve through a homogeneous Comptonizing medium with parameters $T_{\rm C}$ and $\tau_{\rm C}$, given in Table \ref{table_comp_param_A}. Spectra at other radii are plotted in Appendix \ref{sec_figures} (Figure \ref{fig_spectra_fit_set1_full}).}
\label{fig_spectra_fit_set1}
\end{figure}

\begin{figure}
\includegraphics[width = 84mm]{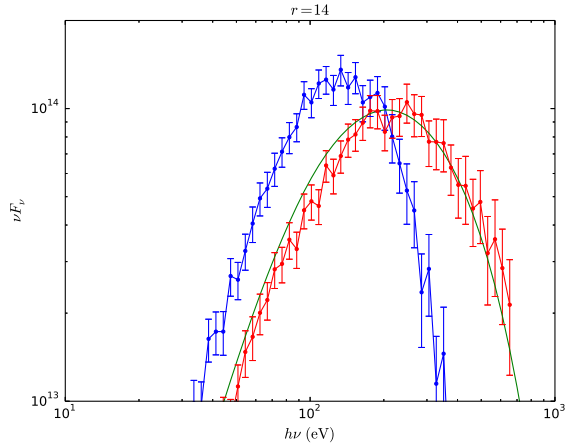}
\caption{Normalized spectra at $r=14$ for the $M = 2 \times 10^8 M_\odot$,  $L/L_{\rm Edd} = 5$ parameter set (Table \ref{table_disc_param}) computed with (red) and without (blue) velocities. The green curve is calculated by using the Kompaneets equation to pass the blue curve through a homogeneous Comptonizing medium with parameters $T_{\rm C}$ and $\tau_{\rm C}$, given in Table \ref{table_comp_param_B}. Spectra at other radii are plotted in Appendix \ref{sec_figures} (Figure \ref{fig_spectra_fit_set4_full}).}
\label{fig_spectra_fit_set4}
\end{figure}

To demonstrate that the effectiveness of the parameters $T_{\rm C}$ and $\tau_{\rm C}$ at describing bulk Comptonization is not limited to a narrow mass range, we modify the parameter set by choosing a significantly higher mass, $M/M_\odot = 2\times 10^8$ (Table \ref{table_disc_param}). We again calculate spectra at multiple radii, with and without velocities, and plot the results for $r=14$ in Figure \ref{fig_spectra_fit_set4}. In the same figure we plot the spectrum predicted by the parameters $T_{\rm C}$ and $\tau_{\rm C}$ for $r=14$, given in Table \ref{table_comp_param_B}, and see that the resulting spectrum well approximates the spectrum computed with velocities. Spectra at all radii are plotted in Appendix \ref{sec_figures} (Figure \ref{fig_spectra_fit_set4_full}).

\begin{table}
\caption{Comptonization temperatures and optical depths at multiple radii for the $M = 2 \times 10^8 M_\odot$,  $L/L_{\rm Edd} = 5$ parameter set (Table \ref{table_disc_param}).}
\begin{tabular}{lllllll}
$r$                       & $8.5$ & $9.5$ & $11$  & $14$  & $20$  & $30$ \\
$T_{\rm C}$ $({\rm eV})$   & $75$  & $80$  & $95$  & $90$  & $80$  & $52$ \\
$\tau_{\rm C}$            & $18$  & $24$  & $24$  & $27$  & $25$  & $26$
\end{tabular}
\label{table_comp_param_B}
\end{table}

\begin{figure}
\includegraphics[width = 84mm]{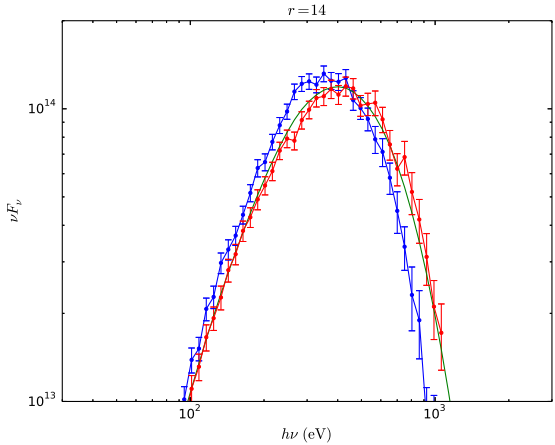}
\caption{Normalized spectra at $r=14$ for the $M = 2 \times 10^6 M_\odot$,  $L/L_{\rm Edd} = 2.5$ parameter set (Table \ref{table_disc_param}) computed with (red) and without (blue) velocities. The green curve is calculated by using the Kompaneets equation to pass the blue curve through a homogeneous Comptonizing medium with parameters $T_{\rm C}$ and $\tau_{\rm C}$, given in Table \ref{table_comp_param_C}. Spectra at other radii are plotted in Appendix \ref{sec_figures} (Figure \ref{fig_spectra_fit_set2_full}).}
\label{fig_spectra_fit_set2}
\end{figure}

Finally, we show that the Comptonization parameters $T_{\rm C}$ and $\tau_{\rm C}$ well describe bulk Comptonization for the $M = 2 \times 10^6 M_\odot$,  $L/L_{\rm Edd} = 2.5$ parameter set (Table \ref{table_disc_param}), whose value of $L/L_{\rm Edd}$ is the same as in K17. The corresponding spectra for $r=14$ are plotted in Figure \ref{fig_spectra_fit_set2}, and the Comptonization parameters are given in Table \ref{table_comp_param_C}. We see that the spectrum predicted by the parameters $T_{\rm C}$ and $\tau_{\rm C}$ well approximates the spectrum computed with velocities. Spectra at all radii are plotted in Appendix \ref{sec_figures} (Figure \ref{fig_spectra_fit_set2_full}).

\begin{table}
\caption{Comptonization temperatures and optical depths at multiple radii for the $M = 2 \times 10^6 M_\odot$,  $L/L_{\rm Edd} = 2.5$ parameter set (Table \ref{table_disc_param}).}
\begin{tabular}{lllllll}
$r$                       & $8.5$ & $9.5$ & $11$  & $14$  & $20$  & $30$ \\
$T_{\rm C}$ $({\rm eV})$   & $0$  & $160$  & $162$  & $159$  & $133$  & $100$ \\
$\tau_{\rm C}$            & $0$  & $6.7$  & $9.5$  & $11$  & $11$  & $8.0$
\end{tabular}
\label{table_comp_param_C}
\end{table}

\subsubsection{Step 6 - Qualify the bulk Comptonization model scope}
\label{sec_step_scope}
In Sections \ref{sec_step_trunc}-\ref{sec_step_model}, we justified each step of a process that results in a simplified, physically revealing model for bulk Comptonization by turbulence, and demonstrated the success of this model for multiple radii, masses, and accretion rates. We acknowledge that underlying each step are various assumptions, some of which may not hold over the entire range of accretion disc parameters of interest. This is a limitation only if the sole goal is to reproduce the Kompaneets temperature and optical depth as they are originally defined. But in this work our primary goal is rather to characterize bulk Comptonization in a physically revealing way so that we can easily map out its dependence on a wide range of disc parameters and understand how this dependence itself may change depending on the robustness of certain features in the disc vertical structure. Therefore, each step of this process should be viewed more as a search for parameters that are physically revealing and easily calculated rather than as an attempt to merely speed up the calculation of the Kompaneets parameters.

For example, we may find that in some regimes the calculated Comptonization region optical depth is sufficiently large that photon spectra saturate, which violates an assumption we made in Section \ref{sec_step_TC}. In this case, the Comptonization temperature and optical depth will probably differ somewhat from the Kompaneets parameters. But since they would still, by definition, tell us the optical depth of the region in which bulk Comptonization is significant as well as the weighted sum of the gas and wave temperatures in this region, they would still provide a useful characterization of bulk Comptonization.

\section{Results}
\label{sec_results}
\subsection{Overview}
\label{res_overview}
The independent variables in radiation MHD shearing box simulations are the surface density $\Sigma$, the vertical epicyclic frequency $\Omega_z$, and the strain rate $\partial_x v_y$. Since our simulation data is limited, we use the scheme developed in K17 to scale data from one set of independent variables to another. This scheme also allows for the variation of $\alpha$ \citep{sha73}, defined as the ratio of the vertically integrated total pressure to the vertically integrated total stress. 

In Section \ref{res_part_1} we calculate the dependence of bulk Comptonization  on the four shearing box parameters $\Sigma$, $\Omega_z$, $\partial_x v_y$, and $\alpha$. In particular, we show that these four parameters can in practice be reduced to two parameters, $\Sigma$ and $\alpha^3 \Omega_{\rm z}$, so that the dependence of bulk Comptonization on shearing box parameters can be illustrated in a single figure with multiple curves. In Section \ref{res_part_2} we show the dependence of bulk Comptonization on accretion disc mass, luminosity, radius, spin, and inner boundary condition. To do this, we examine the dependence of the shearing box parameters $\Sigma$ and $\Omega_z$ on these parameters. In Section \ref{res_part_3} we estimate bulk Comptonization for an entire disc by setting the radius equal to the value that contributes maximally to the luminosity. 

The scaling scheme from K17 assumes that the radiation energy flux is carried by radiation diffusion, but it does allow for variation in the opacity $\kappa$. In Section \ref{res_advection} we show that vertical radiation advection can be included indirectly by varying $\kappa$, and we examine the effect of this on bulk Comptonization. Finally, in Section \ref{res_timestep} we examine how bulk Comptonization is effected by time variability in the simulation data.

\subsection{Dependence of bulk Comptonization on shearing box parameters $\Sigma$, $\Omega_z$, $\partial_x v_y$, and $\alpha$}
\label{res_part_1}
\subsubsection{Reduction of four shearing box parameters to two}
\label{res_reduce}
\begin{figure}
\includegraphics[width = 84mm]{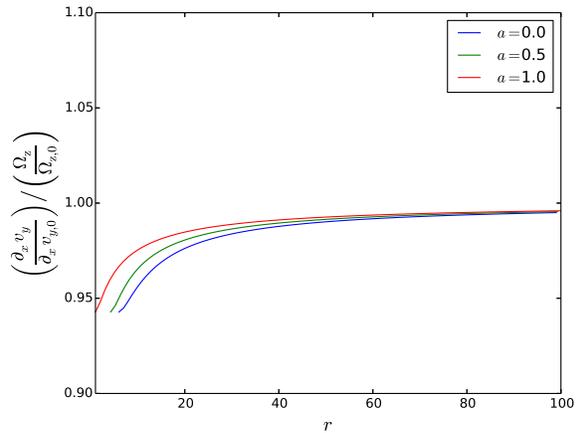}
\caption{Ratio of the strain rate $\partial_x v_y$ scaling to the vertical epicyclic frequency $\Omega_z$ scaling for Kerr discs, for different values of the spin parameter $a$. Note that the maximum deviation from unity is only $6\%$.}
\label{fig_omega_ratio}
\end{figure}

To simplify the problem, we observe that for Newtonian disc scalings
\begin{align}
\label{eq_strain_scaling}
\left(\frac{\partial_x v_y}{\partial_x v_{y,0}} \right) =\left(\frac{\Omega_{z}}{\Omega_{z,0}} \right)  = \left(\frac{M}{M_0}\right)^{-1} \left(\frac{r}{r_0} \right)^{-3/2}.
\end{align} 
For Kerr disc scalings, the strain rate scale factor is nearly equal to the vertical epicyclic frequency scale factor (K17). To show this, in Figure \ref{fig_omega_ratio} we plot the ratio of these quantities for multiple values of black hole spin. We see that at worst they agree to within $\sim 6\%$. For the purpose of understanding bulk Comptonization, then, we can set these factors equal to each other.

Next, we eliminate the dependence on $\alpha$ by proving the following statements for the bulk Comptonization parameters $T_{\rm C}$, $\tau_{\rm C}$, and $y_{\rm p, C}$ (the Compton $y$-parameter): For any constant $k$,
\begin{align}
\label{eq_tau}
\tau_{\rm C}\left(\Sigma,\Omega_z,k\alpha\right) = \tau_{\rm C}\left(\Sigma,k^3 \Omega_z,\alpha\right)
\end{align}
\begin{align}
\label{eq_T}
T_{\rm C}\left(\Sigma,\Omega_z,k\alpha\right) = \frac{1}{k}T_{\rm C}\left(\Sigma,k^3 \Omega_z,\alpha\right)
\end{align}
\begin{align}
\label{eq_yp}
y_{\rm p,C}\left(\Sigma,\Omega_z,k\alpha\right) = \frac{1}{k}y_{\rm p,C}\left(\Sigma,k^3 \Omega_z,\alpha\right).
\end{align}

To prove equation (\ref{eq_tau}), we first show that for fixed $\Sigma$ the wave temperature scales as the bulk temperature. The wave temperature depends not only on the bulk temperature but on the density via the ratio of each length scale in the turbulence $\lambda_{\rm w}$ to the photon mean free path $\lambda_{\rm p}$ (K16). The scalings for these parameters are $\lambda_{\rm w} \sim h$ and $\lambda_{\rm p} \sim \rho^{-1}$, where $h$ is the disc scale height. Since for fixed $\Sigma$ the scaling for the density is given by $\rho \sim h^{-1}$, varying only other parameters leaves the ratio $\lambda_{\rm w}/\lambda_{\rm p}$ unchanged and hence the wave temperature scales as the bulk temperature.

Next, we observe that the ratios of the turbulent velocities to the thermal velocities in the midplane and the photosphere are given by (K17)
\begin{align}
\frac{v_{\rm turb}^2}{v_{\rm th,c}^2} \sim \left(\alpha^3 \Omega_z \right)^{-1/4} \Sigma^{-2}
\end{align}
and
\begin{align}
\frac{v_{\rm turb}^2}{v_{\rm th,ph}^2} \sim \left(\alpha^3 \Omega_z \right)^{-1/4} \Sigma^{-7/4},
\end{align}
respectively. We see that if we vary the vertical epicyclic frequency inversely to $\alpha^3$, then the ratio of the turbulent kinetic energy to the thermal kinetic energy remains unchanged everywhere. Since for fixed $\Sigma$ the wave temperature is proportional to the gas temperature, it follows that the ratio of the wave temperature to the gas temperature is also everywhere unchanged. Then, since $\tau_{\rm C}$ is defined as the optical depth of the region in which the wave temperature is comparable to the gas temperature (Section \ref{sec_modeling}), under these circumstances it can change only if the overall density or the scale height changes. That is,
\begin{align}
\tau_{\rm C}\left(\Sigma,k^{-3}\Omega_z,k\alpha\right) = \tau_{\rm C}\left(\Sigma,\Omega_z,\alpha\right) \left(\frac{\rho_{\rm c}}{\rho_{\rm c,0}}\right)\left(\frac{h}{h_0}\right).
\end{align}
But $\rho_{\rm c}h \sim \Sigma$, and $\Sigma$ is held constant since it is an independent variable, so $\rho_{\rm c}$ varies inversely to $h$. In other words, for fixed $\Sigma$ the optical depth for any length scale is invariant. (Note that this is the exact same reason that the wave temperature scales as the bulk temperature.) It follows that
\begin{align}
\label{eq_tau_2}
\tau_{\rm C}\left(\Sigma,k^{-3}\Omega_z,k\alpha\right) = \tau_{\rm C}\left(\Sigma,\Omega_z,\alpha\right).
\end{align}
Equation (\ref{eq_tau}) follows directly from equation (\ref{eq_tau_2}).

To prove equation (\ref{eq_T}), we start with the turbulent and thermal velocity scalings individually (K17), rather than the ratio of scalings:
\begin{align}
\label{eq_v_turb}
v_{\rm turb}^2 \sim \alpha^{-1} \Sigma^{-2}
\end{align}
\begin{align}
\label{eq_v_th_c}
v_{\rm th,c}^2 \sim \left(\alpha^{-1} \Omega_z \right)^{1/4} 
\end{align}
\begin{align}
\label{eq_v_th_ph}
v_{\rm th,ph}^2 \sim \left(\alpha^{-1} \Omega_z \right)^{1/4} \Sigma^{-1/4}.
\end{align}
We observe that if we vary the vertical epicyclic frequency inversely to $\alpha^3$, as we just showed we must do in order to leave $\tau_{\rm C}$ unchanged,
then the scalings for the individual variables are
\begin{align}
v_{\rm turb}^2 \sim \alpha^{-1} \Sigma^{-2}
\end{align}
\begin{align}
v_{\rm th,c}^2 \sim \alpha^{-1}
\end{align}
\begin{align}
v_{\rm th,ph}^2 \sim \alpha^{-1} \Sigma^{-1/4}.
\end{align}
We see that all velocities scale inversely to $\alpha$. Since the Comptonization temperature is defined as a density weighted average of the sum of the wave and gas temperatures (Section \ref{sec_modeling}), it follows that
\begin{align}
\label{eq_T_2}
T_{\rm C}\left(\Sigma,k^{-3}\Omega_z,k\alpha\right) = \frac{1}{k}T_{\rm C}\left(\Sigma,\Omega_z,\alpha\right).
\end{align}
Equation (\ref{eq_T}) follows directly from equation (\ref{eq_T_2}). Finally, the definition of the Compton $y$-parameter is
\begin{align}
y_{\rm p} = \frac{4k_{\rm B} T}{m_{\rm e}c^2}N,
\end{align}
where $N$ is the average number of scatterings. For a plane parallel geometry with $\tau > 1$,\footnote{Compton scattering is negligible for $\tau < 1$ since $kT_{\rm C} \ll m_{\rm e}c^2$ for the systems we study here.} $N=1.6 \tau^2$, so
\begin{align}
y_{\rm p, C} = 1.6\left(\frac{4k_{\rm B} T_{\rm C}}{m_{\rm e}c^2} \right)\tau_{\rm C}^2,
\end{align}
and equation (\ref{eq_yp}) follows directly from equations (\ref{eq_tau}) and (\ref{eq_T}). 

Therefore, for the purpose of understanding bulk Comptonization, we can  regard $\Sigma$ and $\alpha^3 \Omega_z$ as the fundamental shearing box parameters and $\alpha T_{\rm C}$, $\tau_{\rm C}$, and $\alpha y_{\rm p,C}$ as the Comptonization parameters.

\subsubsection{Dependence of bulk Comptonization on $\Sigma$ and $\Omega_z$}
\label{sec_dependence_1}
\begin{figure}
\includegraphics[width = 84mm]{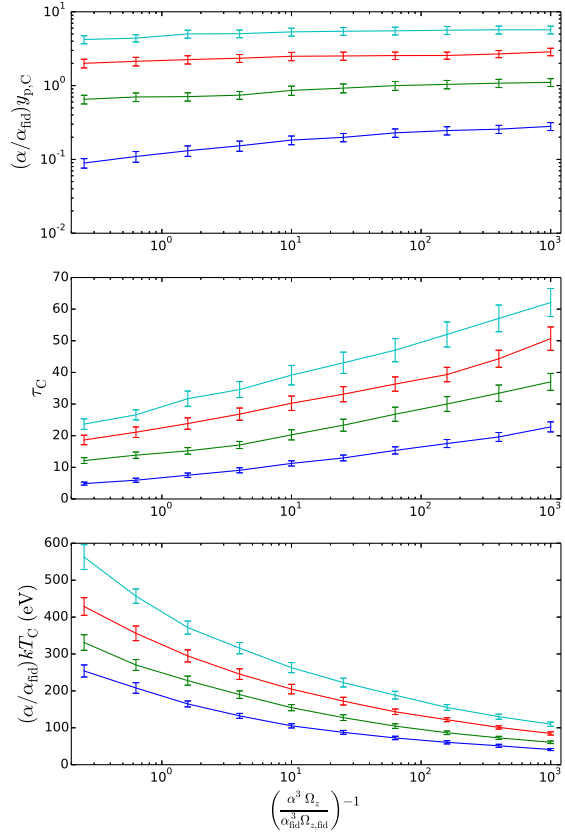}
\caption{Dependence of bulk Comptonization on shearing box parameters. The blue, green, red, and cyan curves correspond to $\left(\Sigma/\Sigma_{\rm fid} \right)^{-1}= $ $1$, $2$, $3.3$, and $5$, respectively.}
\label{fig_omega_sigma}
\end{figure}

The original data we use is from ZEUS simulation 110304a (K17). The shearing box parameters for this simulation are given in Table \ref{table_sim_param_1}. The time-averaged $\alpha$ parameter is $\alpha_0 = 0.01$ (which we do not list in Table \ref{table_sim_param_1} since it is not an independent variable). These correspond to an accretion disc annulus with parameters given in Table \ref{table_sim_param_2}. The opacities included are electron scattering and free-free. These should be good approximations for the opacities in AGN in the near and super-Eddington regimes of interest in this paper. At a given timestep, we scale the data to a range of values of $\Sigma$ and $\Omega_z$ with the scheme in K17, and then calculate the resulting Comptonization parameters $T_{\rm C}$, $\tau_{\rm C}$, and $y_{\rm p, C}$ with the procedure detailed in Section \ref{sec_modeling}. We repeat this for 21 timesteps spaced 10 orbital periods apart and plot the time-averaged results in Figure \ref{fig_omega_sigma}. The error at each point is estimated by dividing the sample standard deviation by the square root of the number of timesteps. For clarity we add the subscript ``fid" to the shearing box parameters of the fiducial system given in Table \ref{table_sim_param_3} to distinguish them from the original simulation parameters which we denote by the subscript ``0". The fiducial shearing box parameters correspond to $r=20$ for the $M = 2 \times 10^6 M_\odot$,  $L/L_{\rm Edd} = 2.5$ parameter set (Table \ref{table_disc_param}). Following K17, we chose these parameters to be similar to those fit to the NLS1 REJ1034+396 by D12.
\begin{table}
\caption{Original simulation shearing box parameters}
\begin{tabular}{lll}
Simulation & $\Omega_{\rm z,0}$ (s$^{-1}$) & $\Sigma_0$ (g cm$^{-2}$)       \\
110304a & $186.6$          & $2.5 \times 10^4$             \\
\end{tabular}
\label{table_sim_param_1}
\end{table}

\begin{table}
\caption{Accretion disc parameters corresponding to the original simulation shearing box parameters}
\begin{tabular}{llllll}
Simulation & $M/M_\odot$ & $r$ & $L/L_{\rm Edd}$ & $a$ & $\Delta \epsilon$ \\
110304a    & 6.62 & 30 & $\sim 1.7$ & 0 & 0 \\
\end{tabular}
\label{table_sim_param_2}
\end{table}

\begin{table}
\caption{Fiducial shearing box parameters, corresponding to $r=20$ for the $M = 2 \times 10^6 M_\odot$,  $L/L_{\rm Edd} = 2.5$ parameter set (Table \ref{table_disc_param}).}
\begin{tabular}{lll}
$\left( \Omega_{\rm z, fid}/\Omega_{\rm z,0} \right)^{-1}$ & $\left(\Sigma_{\rm fid}/\Sigma_0 \right)^{-1}$ & $\alpha_{\rm fid }/\alpha_0$       \\
$1.6 \times 10^{5}$ & $4.0$ & $2$ \\
\end{tabular}
\label{table_sim_param_3}
\end{table}

\begin{figure}
\includegraphics[width = 84mm]{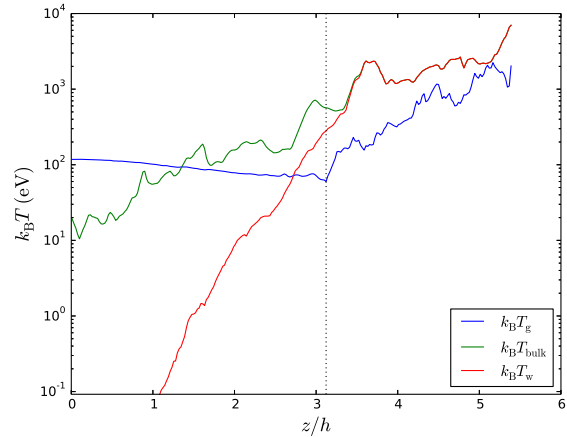}
\caption{Horizontally averaged temperature profiles for the fiducial shearing box parameters, given in Table \ref{table_sim_param_3}, at the $20$ orbits timestep. The dashed line denotes where $\tau_{\rm s} =$ $1$.}
\label{fig_T_profiles}
\end{figure}

We can intuitively understand these results by looking at equations (\ref{eq_v_turb})-(\ref{eq_v_th_ph}) and the horizontally averaged temperature profiles shown in Figure \ref{fig_T_profiles}. In particular, we show below why $\tau_{\rm C}$ and $T_{\rm C}$ strongly increase with increasing $\Sigma^{-1}$, while $\tau_{\rm C}$ increases weakly and $T_{\rm C}$ decreases weakly with increasing $\Omega_{\rm z}^{-1}$. It then follows that since $y_{\rm p,C}$ depends more strongly on $\tau_{\rm C}$ than on $T_{\rm C}$, $y_{\rm p,C}$ increases strongly with increasing $\Sigma^{-1}$
and weakly with increasing $\Omega_{\rm z}^{-1}$. Since $y_{\rm p}$ is generally used as a proxy for the overall magnitude of Comptonization, we conclude that bulk Comptonization increases strongly with increasing $\Sigma^{-1}$ and weakly with increasing $\Omega_{\rm z}^{-1}$. Because of this as well as the fact that $\Omega_{\rm z}^{-1} \sim M$ and $\Sigma^{-1} \sim L/L_{\rm Edd}$ (which we discuss in Section \ref{res_part_2}), we treat $\Omega_{\rm z}^{-1}$ and $\Sigma^{-1}$ as the fundamental parameters rather than $\Omega_z$ and $\Sigma$.

\paragraph*{Dependence of $\tau_{\rm C}$ on $\Omega_z^{-1}$}
\label{sec_tau_dependence}
We start by considering the dependence of the bulk Comptonization optical depth $\tau_{\rm C}$ on $\Omega_z^{-1}$. Equations (\ref{eq_v_turb})-(\ref{eq_v_th_ph}) show that the result of varying $\Omega_z^{-1}$ is to multiply the entire gas temperature profile by a constant and leave the bulk temperature profile unchanged. Since the wave temperature scales as the bulk temperature for fixed $\Sigma^{-1}$ (Section \ref{res_reduce}), the wave temperature profile is also unchanged. Increasing $\Omega_z^{-1}$, therefore, corresponds to moving the gas temperature profile downward in Figure \ref{fig_T_profiles}, increasing the optical depth of the region in which the wave temperature is comparable to the gas temperature, consistent with the results shown in Figure \ref{fig_omega_sigma}. 

\begin{figure}
\includegraphics[width = 84mm]{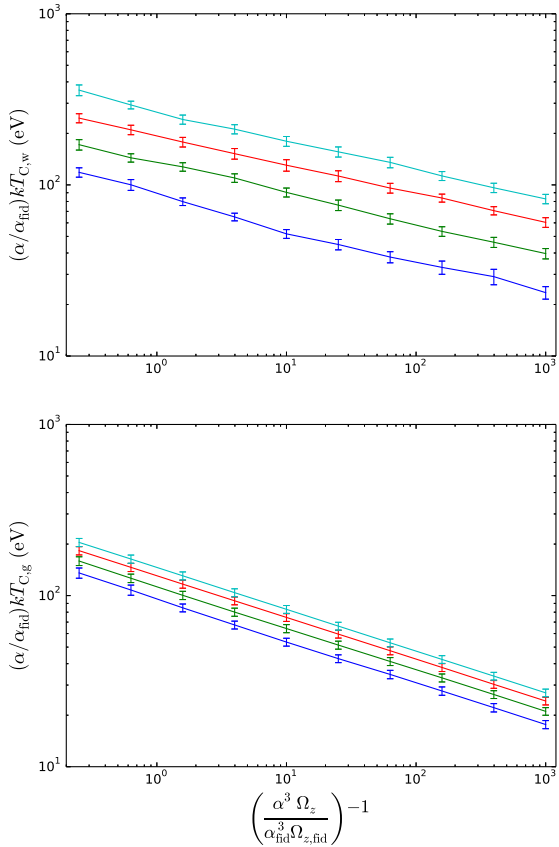}
\caption{Dependence of Comptonization gas and wave temperatures on shearing box parameters. The blue, green, red, and cyan curves correspond to $\left(\Sigma/\Sigma_{\rm fid} \right)^{-1}= $ $1$, $2$, $3.3$, and $5$, respectively.}
\label{fig_omega_sigma_separate}
\end{figure}

\paragraph*{Dependence of $T_{\rm C}$ on $\Omega_z^{-1}$}
To understand the dependence of the Comptonization temperature $T_{\rm C}$ on $\Omega_z^{-1}$, we first need to look at the bulk Comptonization region weighted average gas temperature $T_{\rm C,g}$ and wave temperature $T_{\rm C,w}$, individually. We plot the time-averaged dependence of these two parameters on $\Omega_z^{-1}$ in Figure \ref{fig_omega_sigma_separate}. As we already showed, increasing $\Omega_z^{-1}$ moves the gas temperature profile downward in Figure \ref{fig_T_profiles}. Since the gas temperature profile does not spatially vary significantly in this region, equations (\ref{eq_v_th_c}) - (\ref{eq_v_th_ph}) imply that $T_{\rm C,g}$ will be approximately proportional to $\Omega_z^{1/4}$. Typically the gas temperature profile is slightly decreasing at the lower boundary of this region so that the dependence is slightly shallower than $\Omega_z^{1/4}$, which is what we find in Figure \ref{fig_omega_sigma_separate}. Since increasing $\Omega_z^{-1}$ has no effect on the wave temperature profile, the only effect of increasing $\Omega_z^{-1}$ on $T_{\rm C,w}$ is to decrease the height of the lower boundary of the bulk Comptonization region. As it decreases, $T_{\rm C,w}$ also decreases since not only does the wave temperature profile decrease with increasing optical depth, but equation (\ref{eq_T_C}) gives greatest weight to $T_{\rm w}$ in the region where the optical depth is the largest. Therefore, $T_{\rm C,w}$ also decreases with increasing $\Omega_z^{-1}$ but less so than $T_{\rm C,g}$, as we see in Figure \ref{fig_omega_sigma_separate}. Since $T_{\rm C} =  T_{\rm C,g} + T_{\rm C,w}$, $T_{\rm C}$ also decreases with increasing $\Omega_z^{-1}$ at a rate slightly faster than $T_{\rm C,w}$ but slower than $T_{\rm C,g}$, as we see in Figure \ref{fig_omega_sigma}.

\paragraph*{Dependence of $\tau_{\rm C}$ on $\Sigma^{-1}$}
From equations (\ref{eq_v_turb})-(\ref{eq_v_th_ph}) we see that $v_{\rm turb}^2$ depends much more strongly on $\Sigma^{-1}$ than does $v_{\rm th}^2$, so decreasing $\Sigma^{-1}$ moves the bulk temperature profile downward in Figure \ref{fig_T_profiles} relative to the gas temperature profile. The wave temperature profile moves downward even more than the bulk temperature profile, since the wave temperature depends on the velocity difference between scatterings, which decreases with decreasing $\Sigma^{-1}$. Therefore, the Comptonization optical depth decreases with decreasing $\Sigma^{-1}$. We expect the dependence of $\tau_{\rm C}$ on $\Sigma^{-1}$ to be much stronger than its dependence on $\Omega_z^{-1}$ since $v_{\rm turb}^2 \sim \Sigma^{-2}$ whereas $v_{\rm th}^2 \sim \Omega_z^{1/4}$, which does not even take into account the fact that the wave temperature depends more strongly on $\Sigma^{-1}$ than does the bulk temperature. These conclusions are consistent with the results shown in Figure \ref{fig_omega_sigma}.

\paragraph*{Dependence of $T_{\rm C}$ on $\Sigma^{-1}$}
To understand the dependence of the Comptonization temperature $T_{\rm C}$ on $\Sigma^{-1}$, we first look at the dependence of $T_{\rm C,g}$ and $T_{\rm C,w}$, individually. If the gas and wave temperature profiles both decreased proportionally to the same power of $\Sigma^{-1}$, then both $T_{\rm C,g}$ and $T_{\rm C,w}$ would also decrease in proportion to this power of $\Sigma^{-1}$ because the size of the bulk Comptonization region would remain unchanged. But since the wave temperature profile decreases faster than the gas temperature profile, the effect on $T_{\rm C,g}$ and $T_{\rm C,w}$ also depends on other factors, such as the slopes of the gas and wave temperature profiles in the region. Since $v^2_{\rm turb} \sim \Sigma^{-2}$ gives a fairly strong dependence on density, we expect $T_{\rm C,w}$ to uniformly decrease with decreasing $\Sigma^{-1}$. But since $v^2_{\rm th,ph} \sim \Sigma^{-1/4}$ gives a very weak dependence on density, it is hard to see whether $T_{\rm C,g}$ will increase or decrease with decreasing $\Sigma^{-1}$. Either way, we expect the dependence of $T_{\rm C,g}$ on $\Sigma^{-1}$ to be weaker. Figure \ref{fig_omega_sigma_separate} confirms these expectations. Finally, since $T_{\rm C} =  T_{\rm C,g} + T_{\rm C,w}$, and since by definition the main contribution to $T_{\rm C}$ is from $T_{\rm C,w}$, we expect $T_{\rm C}$ to strongly increase with $\Sigma^{-1}$. Figure \ref{fig_omega_sigma} confirms this expectation.

\subsubsection{Dependence of bulk Comptonization on the Reynolds stress fraction}
\begin{figure}
\includegraphics[width = 84mm]{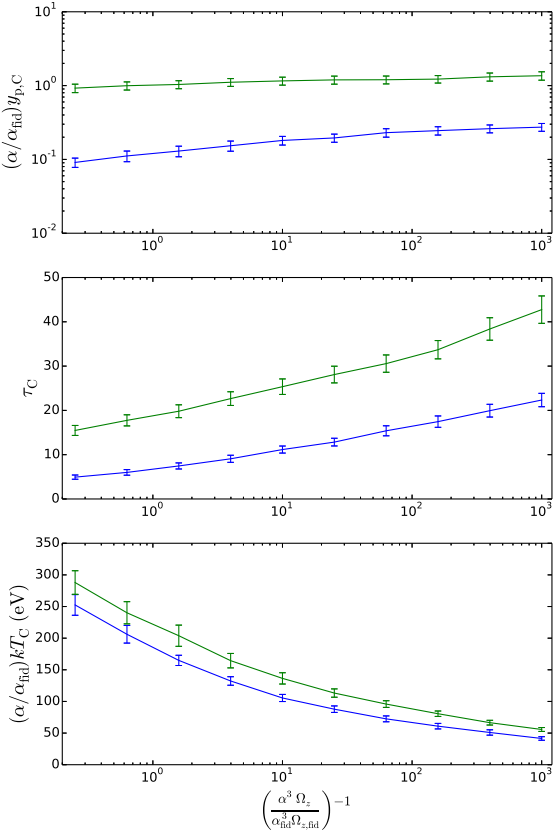}
\caption{Dependence of bulk Comptonization on shearing box parameters for $\beta / \beta_0 = $ $1$ (blue) and $4$ (green). For all curves, $\Sigma = \Sigma_{\rm fid}$.}
\label{fig_omega_beta}
\end{figure}

So far we have assumed that the $\beta$ parameter, defined as the ratio of the vertically integrated Reynolds stress to the vertically integrated total stress, is held constant. We note that this is not to be confused with the plasma $\beta$, which is the ratio of the plasma pressure to the magnetic pressure. In radiation MHD simulations it is typically found that $\beta \sim 0.2$. We now show how varying $\beta$ affects bulk Comptonization. The turbulent velocity scaling, equation (\ref{eq_v_turb}), becomes (K17)
\begin{align}
v_{\rm turb}^2 \sim \alpha^{-1} \beta \Sigma^{-2},
\end{align}
while the thermal velocity scalings, equations (\ref{eq_v_th_c})-(\ref{eq_v_th_ph}), remain unchanged. Since the dependence of $v_{\rm th,ph}$ on $\Sigma^{-1}$ is weak, we expect the dependence of bulk Comptonization on $\beta$ to be similar to its dependence on $\Sigma^{-2}$. In Figure \ref{fig_omega_beta} we plot the dependence of the bulk Comptonization parameters on $\Omega_{z}^{-1}$ for $\beta /\beta_0 = 1$ and $4$. As expected, we see the resulting curves are similar to those in Figure \ref{fig_omega_sigma} corresponding to $(\Sigma / \Sigma_{\rm fid})^{-1} = 1$ and $2$, respectively. We note that unlike the scaling for $\Sigma^{-2}$, the scaling for $\beta$ is restricted to a much narrower range since $\beta \leq 1$. In the rest of this work we suppress the dependence on $\beta$ for clarity.

\subsection{Dependence of bulk Comptonization on accretion disc parameters}
\label{res_part_2}
Now that we have analyzed in detail the dependence of bulk Comptonization on the shearing box parameters $\Omega_z^{-1}$ and $\Sigma^{-1}$, we proceed by relating these to the underlying accretion disc parameters. First we write $\Sigma^{-1}$ in terms of the local flux $F$ (K17):
\begin{align}
\label{eq_sigma_flux}
\Sigma^{-1} \sim \alpha \Omega_{\rm z}^{-1}F.
\end{align}
This says that for fixed $\Omega_z^{-1}$, $\Sigma^{-1}$ is simply proportional to the local flux. Next, we need the scalings for $\Omega_z^{-1}$ and $F$ as functions of the accretion disc parameters. For Newtonian discs, they are\footnote{K17 is missing the $\left(\frac{\eta + \Delta \epsilon}{\eta_0 + \Delta \epsilon_0} \right)$ factor in both the Newtonian and Kerr scalings for $\Sigma$.}
\begin{align}
\left(\frac{\Omega_z}{\Omega_{z,0}}\right) = \left(\frac{M}{M_0}\right)^{-1} \left(\frac{r}{r_0} \right)^{-3/2}
\end{align} 
and
\begin{align}
\label{eq_disc_flux_scaling_1}
\left(\frac{F}{F_0}\right) =& \left(\frac{M}{M_0}\right)^{-1}
\left(\frac{r}{r_0}\right)^{-3}
\left(\frac{\dot{m}}{\dot{m}_0}\right) \left(\frac{\eta + \Delta \epsilon}{\eta_0 + \Delta \epsilon_0} \right)^{-1} \notag \\
&\left(\frac{ 1-\sqrt{r_{\rm in}/r} + \left(\sqrt{r_{\rm in}/r}\right) r_{\rm in} \Delta \epsilon }{ 1-\sqrt{r_{\rm in,0}/r_0} + \left(\sqrt{r_{\rm in,0}/r_0}\right) r_{\rm in,0} \Delta \epsilon_0 }\right),
\end{align}
where $\eta$ is the efficiency assuming a no torque inner boundary condition, $\Delta \epsilon$ is the change in efficiency due to a non-zero torque inner boundary condition \citep{ago00}, $r_{\rm in}$ is the inner radius of the disc, and 
\begin{align}
\dot{m} = L/L_{\rm Edd}.
\end{align}
 For Kerr discs they are (\citealt{rif95}, K17)
\begin{align}
\left(\frac{\Omega_z}{\Omega_{z,0}}\right) = \left(\frac{M}{M_0}\right)^{-1} \left(\frac{r}{r_0} \right)^{-3/2} \left(\frac{C}{C_0}\right)^{1/2}\left(\frac{B}{B_0}\right)^{-1/2}
\end{align}
and
\begin{align}
\label{eq_disc_flux_scaling_2}
\left(\frac{F}{F_0}\right) =& \left(\frac{M}{M_0}\right)^{-1}
\left(\frac{r}{r_0}\right)^{-3}
\left(\frac{\dot{m}}{\dot{m}_0}\right) \left(\frac{\eta + \Delta \epsilon}{\eta_0 + \Delta \epsilon_0} \right)^{-1} \left(\frac{B}{B_0}\right)^{-1}
 \notag \\
&\left(\frac{r_{\rm in}^{3/2} B(r_{\rm in})^{1/2} \Delta \epsilon r^{-1/2} + D}{r_{{\rm in},0}^{3/2} B(r_{{\rm in},0})^{1/2} \Delta \epsilon_0 r_0^{-1/2} + D_0}\right),
\end{align}
where $B$, $C$, and $D$ are functions of $r$ and the spin parameter $a$, and go to unity for $r \gg r_{\rm in}$. In order that the scalings for both Newtonian and Kerr discs be functions of the same underlying parameters, for Newtonian discs we set $r_{\rm in}$ equal to the innermost stable circular orbit, which is in turn a function of the black hole spin parameter $a$.  

We note that since $\dot{m} = L/L_{\rm Edd}$, one should not think of $\dot{m}$ as the mass accretion rate. For example, for fixed mass $M$ and fixed $\dot{m}$, if we vary $\eta + \Delta \epsilon$ (by varying the spin or the inner boundary condition) then the luminosity is unchanged since $L = \dot{m}L_{\rm Edd}$ and $L_{\rm Edd}$ is proportional only to $M$ ($L_{\rm Edd} = 4 \pi G M m_{\rm p} c / \sigma_{\rm T}$, so all the other parameters are constants of physics). But the mass accretion rate is NOT unchanged since $\dot{M} = L/\left(\eta + \Delta\epsilon\right)c^2$. In other words, when varying other parameters (except for the mass) at fixed $\dot{m}$, one should think of this as varying the mass accretion rate at fixed luminosity.

We will find that it is helpful to reduce the above equations to the following simplified form. For the Newtonian scalings,
\begin{align}
\Omega_{\rm z}^{-1} \sim M r^{3/2}
\end{align}
and
\begin{align}
\Sigma^{-1} \sim& \alpha r^{-3/2} \left(L/L_{\rm Edd}\right) \left(\eta + \Delta \epsilon\right)^{-1} \notag \\
&\left(1-\sqrt{r_{\rm in}/r} + \left(\sqrt{r_{\rm in}/r}\right)r_{\rm in} \Delta \epsilon\right).
\end{align}
For the Kerr scalings,
\begin{align}
\Omega_{\rm z}^{-1} \sim M r^{3/2}C^{-1/2}B^{1/2}
\end{align}
and\footnote{The scaling for $\Sigma^{-1}$ differs slightly from that in K17 since here we have set the strain rate scaling equal to the scaling for the vertical epicyclic frequency, an excellent approximation for our purposes (Section \ref{res_part_1}).}
\begin{align}
\Sigma^{-1} &\sim \alpha r^{-3/2} \left(L/L_{\rm Edd}\right) \left(\eta + \Delta \epsilon\right)^{-1} B^{-1/2} C^{-1/2} \notag \\
&\left(r_{\rm in}^{3/2} B(r_{\rm in})^{1/2} \Delta \epsilon r^{-1/2} + D \right).
\end{align}
We now examine the dependence of the Comptonization parameters $T_{\rm C}$, $\tau_{\rm C}$, and $y_{\rm p,C}$ on the accretion disc parameters.

\paragraph*{Dependence on mass}
The dependence of the bulk Comptonization parameters on mass is straightforward. Since $\Omega_z^{-1}$ is directly proportional to mass and $\Sigma^{-1}$ is independent of mass, the dependence of the bulk Comptonization parameters on mass is identical to their dependence on $\Omega_z^{-1}$. That is, $T_{\rm C}$ decreases weakly, $\tau_{\rm C}$ increases weakly, and $y_{\rm p,C}$ increases weakly with increasing mass. Furthermore, we can immediately regard the $\Omega_z^{-1}$ axis in Figures \ref{fig_omega_sigma} and \ref{fig_omega_sigma_separate} as the mass axis.

\paragraph*{Dependence on luminosity}
The dependence of the bulk Comptonization parameters on $L/L_{\rm Edd}$ is straightforward. Since $\Sigma^{-1}$ is directly proportional to $L/L_{\rm Edd}$ and $\Omega_z^{-1}$ is independent of $L/L_{\rm Edd}$ for both Newtonian and Kerr discs, the dependence of the bulk Comptonization parameters on $L/L_{\rm Edd}$ is identical to their dependence on $\Sigma^{-1}$. That is, $T_{\rm C}$, $\tau_{\rm C}$, and $y_{\rm p,C}$ all increase strongly with increasing $L/L_{\rm Edd}$. Furthermore, we can immediately regard the curves corresponding to different values of $\Sigma^{-1}$ in Figures \ref{fig_omega_sigma} and \ref{fig_omega_sigma_separate} as corresponding to different values of $L/L_{\rm Edd}$. In Section \ref{sec_Conclusion} we reproduce the plots from Figure \ref{fig_omega_sigma} with the independent variables relabeled in order to summarize the dependence of bulk Comptonization on mass and luminosity.

\begin{figure}
\includegraphics[width = 84mm]{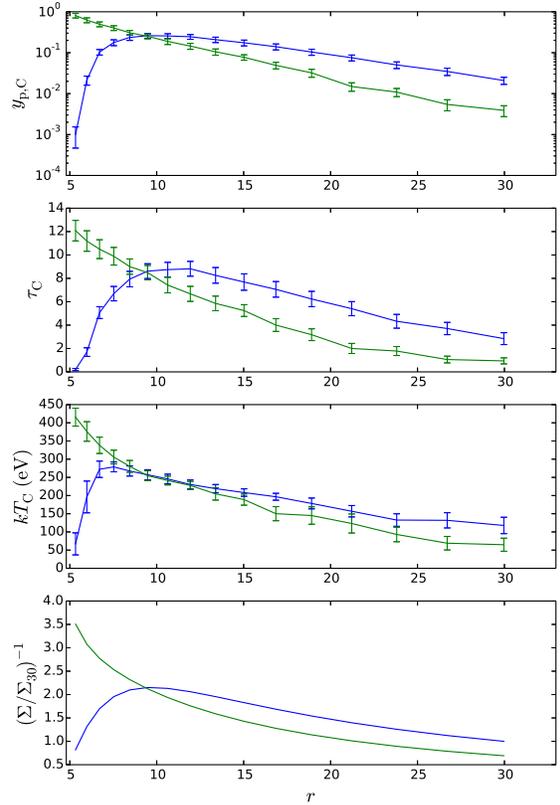}
\caption{Dependence of bulk Comptonization and $\Sigma^{-1}$ on $r$ for  $\Delta \epsilon = 0$ (blue) and $\Delta \epsilon = 0.05$ (green). The parameter $\Sigma_{30}$ denotes the surface density at $r = 30$ for  $\Delta \epsilon = 0$. The values of the parameters held constant are $M/M_\odot = 2 \times 10^{6}$, $L/L_{\rm Edd} = 2.5$, $a = 0.5$, and $\alpha/\alpha_0 = 2$.}
\label{fig_r_deltaep}
\end{figure}

\paragraph*{Dependence on radius}
Both $\Omega_z^{-1}$ and $\Sigma^{-1}$ depend on $r$. But since all bulk Comptonization parameters depend strongly on $\Sigma^{-1}$ and weakly on $\Omega_z^{-1}$, their dependence on $r$ is almost entirely explained by the dependence of $\Sigma^{-1}$ on $r$. For $r \gg r_{\rm in}$, $\Sigma^{-1}$ and hence the bulk Comptonization parameters increase with decreasing $r$ for both Newtonian and Kerr discs. For $\Delta \epsilon \ll 1$, $\Sigma^{-1}$ eventually begins to decrease as $r$ approaches $r_{\rm in}$, after which the bulk Comptonization parameters begin to decrease with decreasing $r$. The precise value of $r$ below which the bulk Comptonization parameters begin to decrease differs slightly from the value of $r$ at which $\Sigma^{-1}$ begins to decrease because the bulk Comptonization parameters also depend on $r$ through $\Omega_z^{-1}$, albeit weakly. For example, in Figure \ref{fig_r_deltaep} we plot the dependence of $\Sigma^{-1}$ and bulk Comptonization on $r$ for $\Delta \epsilon = 0$. We see that the dependence of bulk Comptonization on $r$ is well predicted by the variation in $\Sigma^{-1}$.

If, on the other hand, $\Delta \epsilon$ is large enough, then both $\Sigma^{-1}$ and the bulk Comptonization parameters monotonically increase with decreasing $r$, just as they do for $r \gg r_{\rm in}$. This holds true for both Newtonian and Kerr discs. For example, in Figure \ref{fig_r_deltaep} we also plot the dependence of $\Sigma^{-1}$ and the Comptonization parameters on $r$ for $\Delta \epsilon = 0.05$. We see that both $\Sigma^{-1}$ and the bulk Comptonization parameters uniformly increase with decreasing $r$.

\begin{figure}
\includegraphics[width = 84mm]{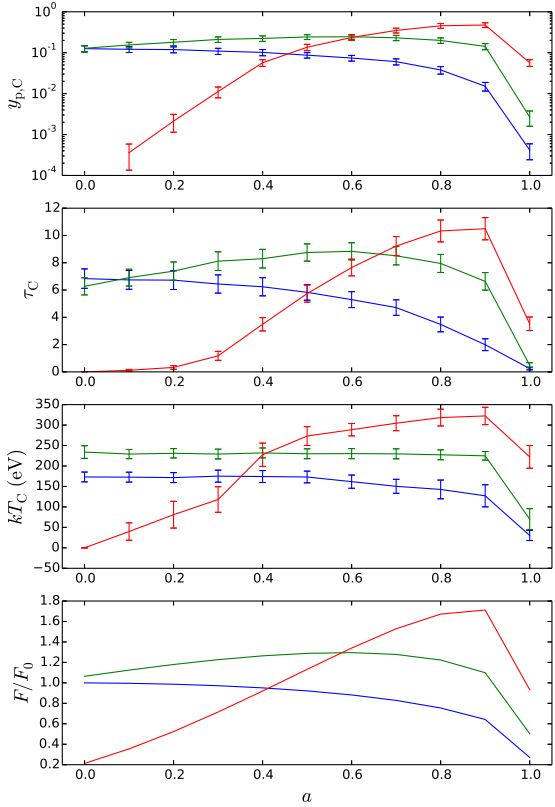}
\caption{Dependence of flux and bulk Comptonization on $a$ for $r=20$ (blue), $r=12$ (green), and $r=7$ (red). The parameter $F_{0}$ denotes the flux at $a = 0$ for $r=20$. The values of the parameters held constant are $M/M_\odot = 2 \times 10^{6}$, $L/L_{\rm Edd} = 2.5$, $\Delta \epsilon = 0$, and $\alpha/\alpha_0 = 2$.}
\label{fig_a_r20_r12_r7}
\end{figure}

\paragraph*{Dependence on spin}
For Newtonian discs $\Omega_{\rm z}^{-1}$ is independent of the spin parameter $a$, and for Kerr discs $\Omega_z^{-1}$ depends on $a$ only for $r$ very close to $r_{\rm in}$ through the functions $C$ and $B$. But since all bulk Comptonization parameters depend strongly on $\Sigma^{-1}$ and weakly on $\Omega_z^{-1}$, their dependence on $a$ is almost entirely explained by the dependence of $\Sigma^{-1}$ on $a$. For $r \gg r_{\rm in}$, the dependence of $\Sigma^{-1}$ on $a$ is given by
\begin{align}
\Sigma^{-1} &\sim \alpha r^{-3/2} \left(L/L_{\rm Edd}\right) \left(\eta + \Delta \epsilon\right)^{-1},
\end{align}
where for both Newtonian and Kerr discs $\eta$ is a monotonically increasing (albeit different) function of $a$. We see that $\Sigma^{-1}$ decreases with increasing $a$. The reason for this is straightforward. We recall that for fixed $\Omega_z^{-1}$, $\Sigma^{-1}$ is proportional to the flux. As the spin and efficiency increase, the flux increases in the inner radii so for fixed luminosity $L/L_{\rm Edd}$ the flux must decrease at large radii. The bulk Comptonization parameters therefore decrease at large radii with increasing spin. For example, in Figure \ref{fig_a_r20_r12_r7} we plot the dependence of flux and bulk Comptonization on spin for $r=20$, $12$, and $7$. We see that for $r=20$, both flux and bulk Comptonization decrease with increasing spin.

For $r$ sufficiently close to $r_{\rm in}$, on the other hand, since the flux increases with spin, so do the bulk Comptonization parameters. For example, in Figure \ref{fig_a_r20_r12_r7} we see that for $r=7$ flux and bulk Comptonization increase with spin until $a \approx 1$. This is expected because as $a$ approaches $1$, $r_{\rm in}$ approaches $1$ and so $r=7$ is no longer close to $r_{\rm in}$.

For an intermediate value of $r$ (at which flux does not monotonically increase or decrease with spin), the dependence of the bulk Comptonization parameters on spin can still be understood by simply plotting the flux as a function of spin. For example, in Figure \ref{fig_a_r20_r12_r7} we see that for $r=12$ the dependence of bulk Comptonization on spin tracks the variation in flux.

\paragraph*{Dependence on inner boundary condition}
The dependence of bulk Comptonization on the inner boundary condition is very similar to the dependence on spin. Since $\Sigma^{-1}$ is proportional to the flux for fixed $\Omega_z^{-1}$, the dependence of bulk Comptonization on the inner boundary condition follows the variation in the flux. The inner boundary is parameterized in terms of $\Delta \epsilon$, the change in efficiency due to a non-zero inner torque. Since increasing $\Delta \epsilon$ increases the flux in the inner radii, bulk Comptonization increases with increasing $\Delta \epsilon$ in this region. At large radii, increasing $\Delta \epsilon$ at fixed luminosity decreases the flux so that bulk Comptonization also decreases.

\begin{figure}
\includegraphics[width = 84mm]{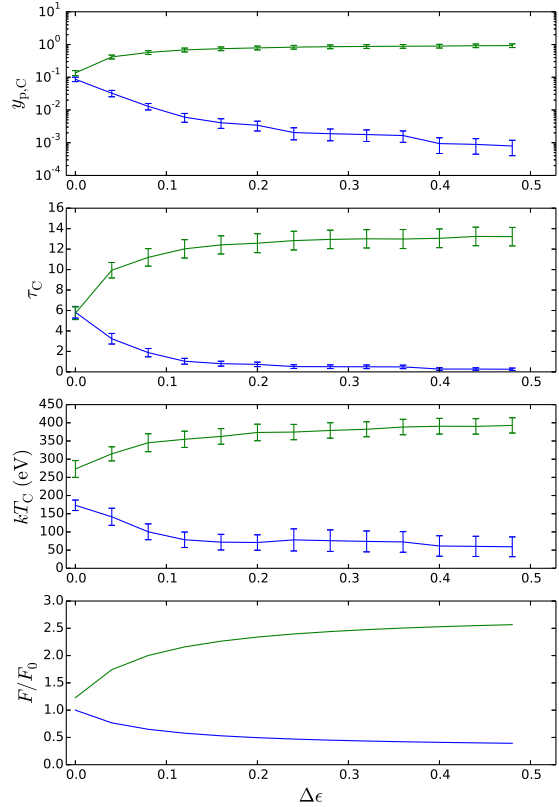}
\caption{Dependence of flux and bulk Comptonization on $\Delta \epsilon$ for $r = 20$ (blue) and $r = 7$ (green). The parameter $F_{0}$ denotes the flux at $\Delta \epsilon = 0$ for $r = 20$. The values of the parameters held constant are $M/M_\odot = 2 \times 10^{6}$, $L/L_{\rm Edd} = 2.5$, $a=0.5$, and $\alpha/\alpha_0 = 2$.}
\label{fig_deltaep_r20_r7}
\end{figure}

For example, in Figure \ref{fig_deltaep_r20_r7} we plot the dependence of flux and bulk Comptonization on $\Delta \epsilon$ at large ($r=20$) and small ($r=7$) radii. In both cases we see that bulk Comptonization follows the variation in the flux.

\paragraph*{Dependence on $\alpha$}
For fixed $\Sigma^{-1}$ and $\Omega_z^{-1}$, the variation of bulk Comptonization with $\alpha$, given by equations (\ref{eq_tau})-(\ref{eq_yp}), is reflected in Figure \ref{fig_omega_sigma}. For $\alpha = \alpha_{\rm fid}$, these plots are uncomplicated. Multiplying $\alpha$ by a constant $k > 1$ translates each curve for $\tau_{\rm C}$ to the right on a log scale. For $T_{\rm C}$ and $y_{\rm p, C}$, multiplying $\alpha$ by $k$ not only translates each curve to the right but also multiplies each curve by $1/k$. Since we plot $y_{\rm p, C}$ on a log scale, for this variable multiplying $\alpha$ by $k$ is equivalent to moving each curve to the right and downward.

\begin{figure}
\includegraphics[width = 84mm]{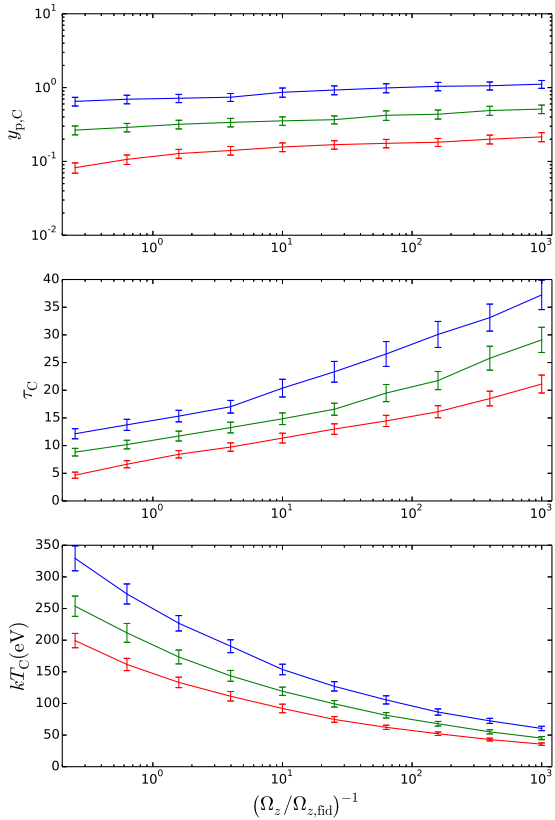}
\caption{Dependence of bulk Comptonization on $\Omega_z^{-1}$ for $\alpha/\alpha_{\rm fid}$ = $1$ (blue), $2$ (green), and $4$ (red). The surface density is $\left(\Sigma / \Sigma_{\rm fid} \right)^{-1} = 2$.}
\label{fig_omega_alpha}
\end{figure}

Alternatively, one can think of multiplying $\alpha$ by a constant $k > 1$ as moving leftward along each curve for $\tau_{\rm C}$. For a given value of $T_{\rm C}$ and $y_{\rm p, C}$, multiplying $\alpha$ by $k$ is equivalent to not only moving leftward but also dividing the resultant value by $k$. To develop physical intuition, we plot the dependence of bulk Comptonization on $\Omega_z^{-1}$ for multiple values of $\alpha$ in Figure \ref{fig_omega_alpha}. We see that bulk Comptonization overall decreases moderately with increasing $\alpha$.

\begin{figure}
\includegraphics[width = 84mm]{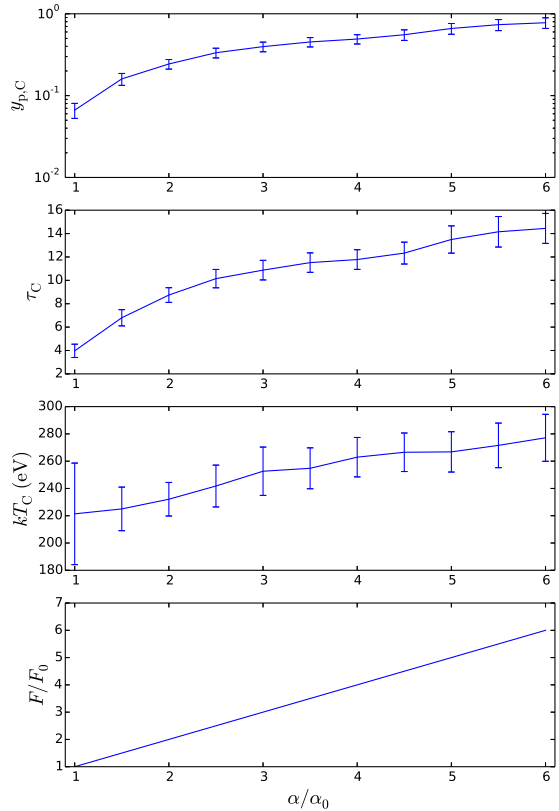}
\caption{Dependence of flux and bulk Comptonization on $\alpha$. The parameter $F_{0}$ denotes the flux for $\alpha/\alpha_0 = 1$. The values of the parameters held constant are $M/M_\odot = 2 \times 10^{6}$, $L/L_{\rm Edd} = 2.5$, $a=0.5$, and $r = r_{\rm max} = 11.8$.}
\label{fig_alpha_rmax}
\end{figure}

But in an accretion disc, we see that $\Sigma^{-1}$ itself is directly proportional to $\alpha$ via the flux. The effect of varying $\alpha$ as an accretion disc parameter, then, affects bulk Comptonization primarily by varying $\Sigma^{-1}$. Since bulk Comptonization increases strongly with $\Sigma^{-1}$, increasing $\alpha$ generally increases bulk Comptonization. For example, in Figure \ref{fig_alpha_rmax} we plot the dependence of the flux and bulk Comptonization on $\alpha$. We see that bulk Comptonization increases with increasing $\alpha$ for fixed accretion disc parameters, unlike for fixed shearing box parameters.

\subsection{Dependence of bulk Comptonization on accretion disc parameters at the radius of maximum luminosity}
\label{res_part_3}
To estimate the magnitude of bulk Comptonization for an entire accretion disc, we calculate the bulk Comptonization parameters at the radius $r_{\rm max}$ where the local luminosity is maximized. The luminosity at $r$ is given by
\begin{align}
L \sim F(2\pi)r dr \sim r^2 F,
\end{align}
and $r_{\rm max}$ is determined by maximizing this function with respect to $r$. Since $r_{\rm max}$ is a function only of the spin $a$ and the inner boundary condition parameter $\Delta \epsilon$, the dependence of bulk Comptonization on mass, luminosity, and $\alpha$ at this radius is the same as for fixed $r$, described in the previous section. The dependence on $a$ and $\Delta \epsilon$, however, is different.

\begin{figure}
\includegraphics[width = 84mm]{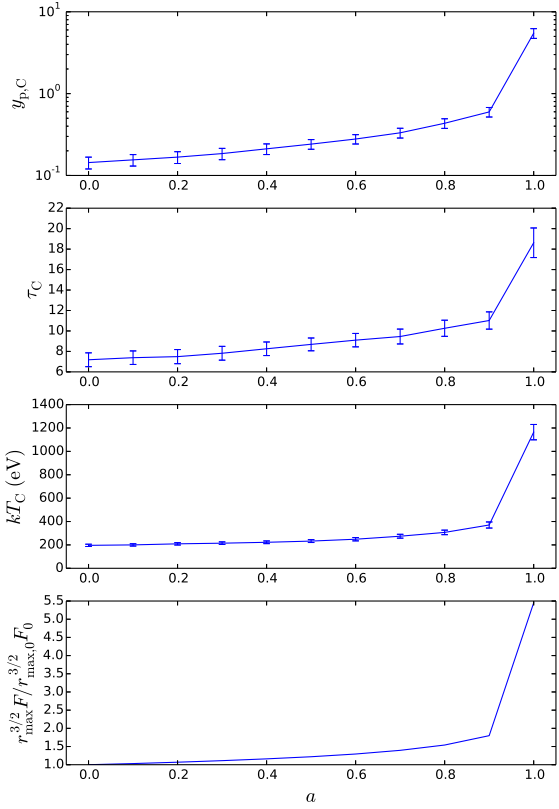}
\caption{Dependence of $\Sigma^{-1}$ and bulk Comptonization on $a$ at the radius where the luminosity is greatest. The subscript zero denotes the value at $a = 0$. The values of the parameters held constant are $M/M_\odot = 2 \times 10^{6}$, $L/L_{\rm Edd} = 2.5$, $\Delta \epsilon = 0$, and $\alpha/\alpha_0 = 2$.}
\label{fig_a_rmax}
\end{figure}

\paragraph*{Dependence on spin}
We attempt to determine the dependence of bulk Comptonization on spin by analyzing how the flux depends on spin, as we did earlier. But in this case we have to be careful. Since $r_{\rm max}$ depends on $a$, $r$ is not held constant. Given that
\begin{align}
\Sigma^{-1} \sim \alpha \Omega_{\rm z}^{-1}F \sim \alpha M r^{3/2} F,
\end{align}
we see that what really matters is the dependence of $r_{\rm max}^{3/2} F$ on $a$. Fortunately, since the dependence of $r^{3/2} F$ on $r$ is qualitatively similar to that of $F$, this does not change our physical intuition. As the spin parameter $a$ increases, $r_{\rm max}$ decreases, and as the region of maximum luminosity becomes smaller, we expect that the flux at each point in this region must increase in order for the total luminosity $L/L_{\rm Edd}$ to remain the same. In Figure \ref{fig_a_rmax} we plot the dependence of $r^{3/2} F$ and bulk Comptonization on $a$ at $r_{\rm max}$. We see that $r^{3/2} F$ and the bulk Comptonization parameters increase with spin, in agreement with our expectations.

\begin{figure}
\includegraphics[width = 84mm]{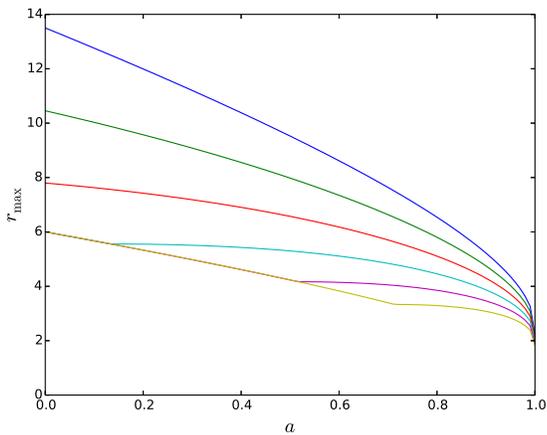}
\caption{Dependence of $r_{\rm max}$ on $a$ for $\Delta \epsilon$ = $0$ (blue), $0.02$ (green), $0.04$ (red), $0.06$ (cyan), $0.08$ (magenta), and $0.1$ (yellow). }
\label{fig_rmax_a}
\end{figure}

We note that for $\Delta \epsilon > 0$, it may be the case that $r_{\rm max} = r_{\rm in}$ rather than a value of $r$ at which $dL/dr = 0$. Since $r_{\rm in}$ decreases with $a$, however, this does not effect our conclusions. For example, in Figure \ref{fig_rmax_a} we plot the dependence of $r_{\rm max}$ on $a$ for multiple values of $\Delta \epsilon$. We see that for sufficiently large $\Delta \epsilon$, $r_{\rm max}$ tracks $r_{\rm in}$ until $a$ is large enough that the value of $r$ at which $dL/dr$ equals zero is greater than $r_{\rm in}$. For very large $\Delta \epsilon$, $r_{\rm max}$ tracks $r_{\rm in}$ for almost all values of $a$.

\paragraph*{Dependence on inner boundary condition}
The dependence on the inner boundary condition parameter $\Delta \epsilon$ is similar to the dependence on spin. As $\Delta \epsilon$ increases, $r_{\rm max}$ decreases, and as the region of maximum luminosity becomes smaller, we expect that the flux at each point in this region must increase in order for the total luminosity $L/L_{\rm Edd}$ to remain the same. Even once $\Delta \epsilon$ is sufficiently large that $r_{\rm max} = r_{\rm in}$, we expect the flux at $r_{\rm in}$ to continue to increase since the increase in efficiency parameterized by $\Delta \epsilon$ should result in an increase in flux at all radii near $r_{\rm in}$.

\begin{figure}
\includegraphics[width = 84mm]{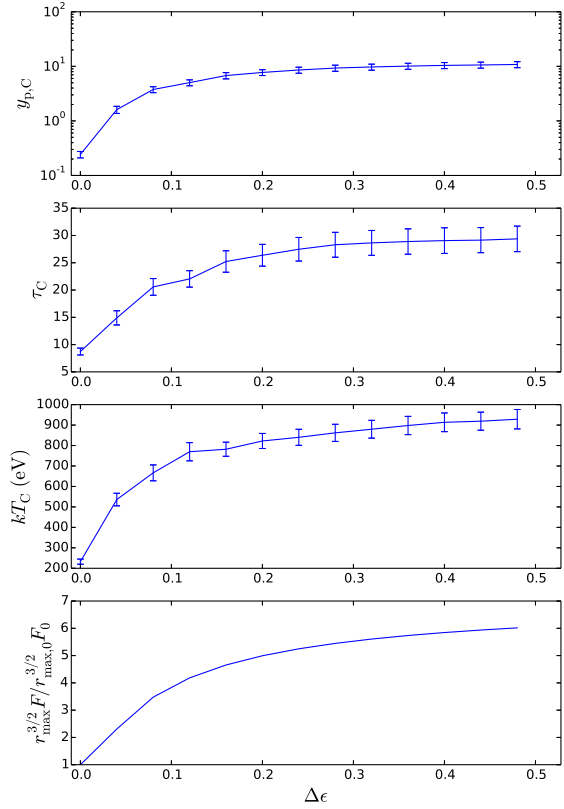}
\caption{Dependence of $\Sigma^{-1}$ and bulk Comptonization on $\Delta \epsilon$ at the radius where the luminosity is greatest. The subscript zero denotes the value at $\Delta \epsilon = 0$. The values of the parameters held constant are $M/M_\odot = 2 \times 10^{6}$, $L/L_{\rm Edd} = 2.5$, $a = 0.5$, and $\alpha/\alpha_0 = 2$.}
\label{fig_deltaep_rmax}
\end{figure}

\begin{figure}
\includegraphics[width = 84mm]{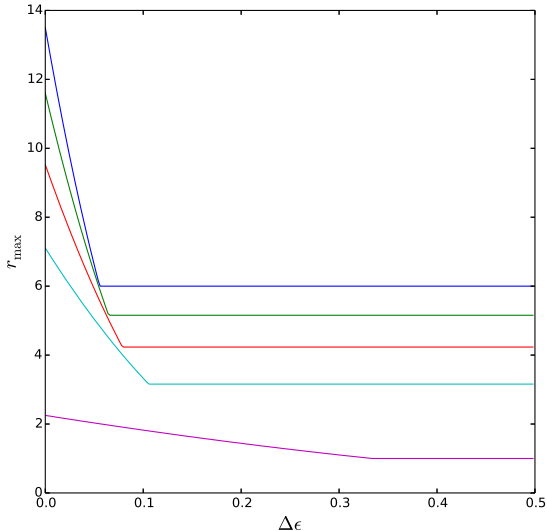}
\caption{Dependence of $r_{\rm max}$ on $\Delta \epsilon$ for $a$ = $0$ (blue), $0.25$ (green), $0.5$ (red), $0.75$ (cyan), and $1$ (magenta).}
\label{fig_rmax_deltaep}
\end{figure}

In Figure \ref{fig_deltaep_rmax} we plot the dependence of $r^{3/2} F$ and bulk Comptonization on $\Delta \epsilon$ at $r_{\rm max}$. We see that $r^{3/2} F$ and the bulk Comptonization parameters increase with $\Delta \epsilon$, in agreement with our expectations. For $\Delta \epsilon$ slightly greater than zero, flux increases both because the flux at all inner radii increases with $\Delta \epsilon$ and because $r_{\rm max}$ itself decreases towards smaller radii where the flux is larger. For larger values of $\Delta \epsilon$, $r_{\rm max}$ is fixed to $r_{\rm in}$, so the flux increases only due to the first effect and therefore increases at a slower rate. In Figure \ref{fig_rmax_deltaep} we plot the dependence of $r_{\rm max}$ on $\Delta \epsilon$ for multiple values of $a$. We see that for each curve $r_{\rm max}$ decreases until $r_{\rm max} = r_{\rm in}$.

\subsection{Effect of vertical radiation advection on bulk Comptonization}
\label{res_advection}
The scheme from K17 that we use to scale simulation data assumes that the flux is carried by radiation diffusion. Since we also see substantial vertical radiation advection in some radiation MHD simulations \citep{bla11}, we attempt here to incorporate this process into our analysis of bulk Comptonization. Vertical radiation advection has both a direct and indirect impact on bulk Comptonization. The direct effect is to transport photons through the bulk Comptonization region faster so that they scatter fewer times, reducing bulk Comptonization. But the direct effect is typically negligible since vertical advection is significant only deep inside the photosphere \citep{bla11}, outside the bulk Comptonization region. This is physically intuitive since radiation advection assists radiation diffusion in transporting photons out of the disc in order to maintain thermal equilibrium. Vertical advection is therefore most significant deep inside the photosphere where the photon diffusion time is comparatively large.

The indirect effect of vertical radiation advection on bulk Comptonization, on the other hand, is to modify the underlying vertical structure gas and wave temperature profiles, which in turn either increases or decreases bulk Comptonization depending on whether shearing box parameters or the accretion disc parameters are held constant. In order to study this effect we need to incorporate vertical advection into the scaling scheme. One way to do this is to rederive the shearing box scalings without assuming that the flux is carried by radiation diffusion. We take this approach in Appendix \ref{sec_scalings}. Although this gives physical insight and is necessary to implement a specific model of advection, it is unnecessarily complex for our purpose here. Instead, we begin by simply adding an advection term $F_{\rm a}$ to the radiation diffusion equation (K17), which gives
\begin{align}
F = \frac{2cP_{\rm c}}{\kappa \Sigma} + F_{\rm a}.
\end{align}
We observe that the only effect of adding $F_{\rm a}$ at fixed surface density $\Sigma$ is to increase the total flux. Since this is also the equation that introduces the opacity parameter $\kappa$ into the scaling scheme, it follows that the effect of adding $F_{\rm a}$ is the same as decreasing $\kappa$, as far as our scaling scheme is concerned. Conveniently, we see that the scheme in K17 already allows for scaling with respect to $\kappa$, even though for clarity we have suppressed the dependence on this parameter until now. We therefore proceed to determine the effect of advection on bulk Comptonization by varying $\kappa$.

\subsubsection{Effect of advection for fixed shearing box parameters}
\begin{figure}
\includegraphics[width = 84mm]{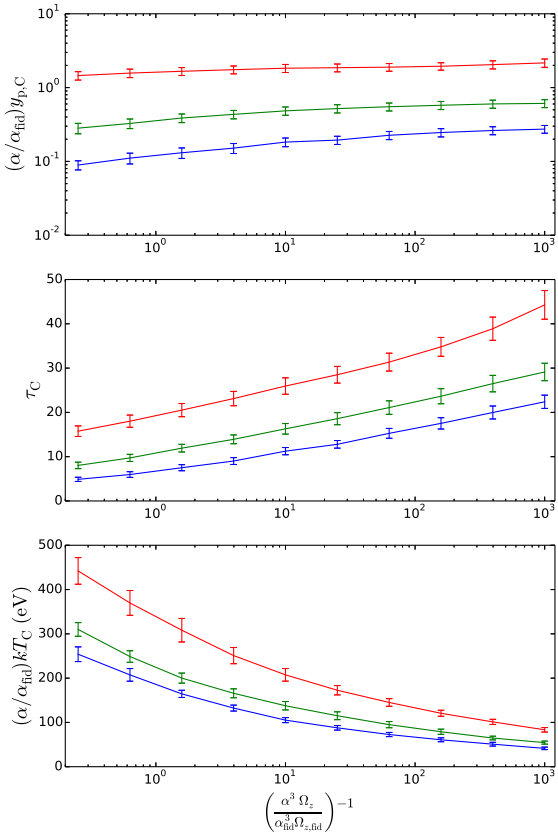}
\caption{Dependence of bulk Comptonization on shearing box parameters for $\kappa = $ $1$ (blue), $0.75$ (green), and $0.5$ (red). For all curves, $\Sigma = \Sigma_{\rm fid}$.}
\label{fig_omega_kappa}
\end{figure}

To determine the effect of including vertical radiation advection on bulk Comptonization for fixed shearing box parameters, we need the turbulent and thermal velocity scalings with $\kappa$ included (K17):
\begin{align}
\label{eq_v_turb_kappa}
v_{\rm turb}^2 \sim \kappa^{-2} \alpha^{-1} \Sigma^{-2}
\end{align}
\begin{align}
\label{eq_v_th_c_kappa}
v_{\rm th,c}^2 \sim  \kappa^{-1/4} \left(\alpha^{-1} \Omega_z \right)^{1/4} 
\end{align}
\begin{align}
\label{eq_v_th_ph_kappa}
v_{\rm th,ph}^2 \sim  \kappa^{-1/2} \left(\alpha^{-1} \Omega_z \right)^{1/4} \Sigma^{-1/4}.
\end{align}
We see that decreasing $\kappa$ primarily affects the turbulent velocity magnitude. In particular, it moves the bulk and wave temperature profiles upward in Figure \ref{fig_T_profiles}, increasing the size of the bulk Comptonization region. As advection increases at fixed $\Sigma$, therefore, we expect $\tau_{\rm C}$, $T_{\rm C}$, and $y_{\rm p,C}$ to increase. For example, in Figure \ref{fig_omega_kappa} we plot the dependence of bulk Comptonization on $\Omega_z^{-1}$ for multiple values of $\kappa$. We see that this result is consistent with our expectations.

\subsubsection{Effect of advection for fixed accretion disc parameters}
\begin{figure}
\includegraphics[width = 84mm]{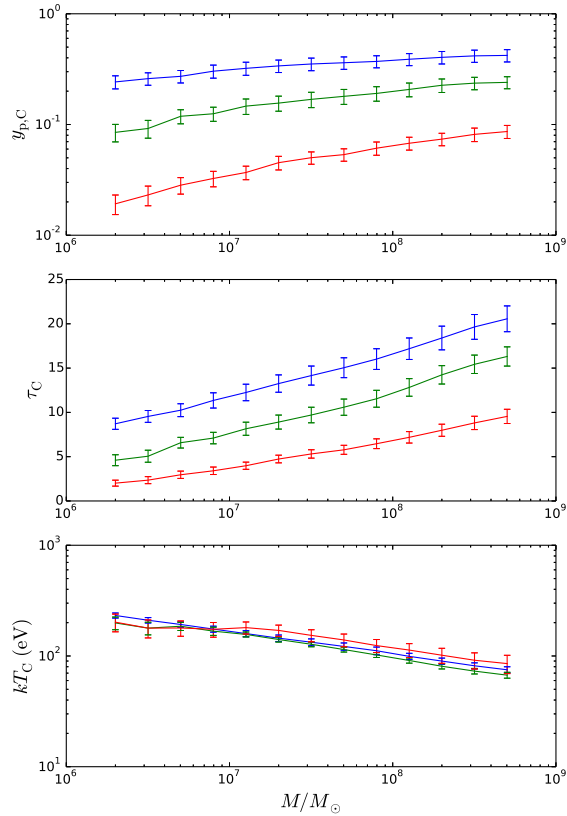}
\caption{Dependence of bulk Comptonization on mass for $\kappa = $ $1$ (blue), $0.75$ (green), and $0.5$ (red). The values of the parameters held constant are $M/M_\odot = 2 \times 10^{6}$, $L/L_{\rm Edd} = 2.5$, $r = r_{\rm max} = 11.8$, $a = 0.5$, and $\alpha/\alpha_0 = 2$.}
\label{fig_m_kappa_rmax}
\end{figure}

To determine the effect of including vertical radiation advection on bulk Comptonization for fixed accretion disc parameters, we write $\Sigma^{-1}$ in terms of the local flux $F$, this time allowing for the variation in $\kappa$ (K17):
\begin{align}
\Sigma^{-1} \sim \alpha \kappa^{2} \Omega_{\rm z}^{-1}F.
\end{align}
Since neither $\Omega_z^{-1}$ nor $F$ depends on $\kappa$, combining this with equations (\ref{eq_v_turb_kappa})-(\ref{eq_v_th_ph_kappa}) results in the following dependence on $\kappa$:
\begin{align}
\label{eq_v_turb_kappa_2}
v_{\rm turb}^2 \sim \kappa^{2} 
\end{align}
\begin{align}
\label{eq_v_th_c_kappa_2}
v_{\rm th,c}^2 \sim  \kappa^{-1/4} 
\end{align}
\begin{align}
\label{eq_v_th_ph_kappa_2}
v_{\rm th,ph}^2 \sim 1.
\end{align}
We see that decreasing $\kappa$ primarily affects the turbulent velocity magnitude, but in the opposite direction to the one in the previous section where the shearing box parameters are fixed. It moves the bulk and wave temperature profiles downward in Figure \ref{fig_T_profiles}, decreasing the size of the bulk Comptonization region. As advection increases, therefore, we expect $\tau_{\rm C}$, $T_{\rm C}$, and $y_{\rm p,C}$ to decrease. For example, in Figure \ref{fig_m_kappa_rmax} we plot the dependence of bulk Comptonization on $\Omega_z^{-1}$ for multiple values of $\kappa$. We see that this result is consistent with our expectations.

\subsection{Time variability of bulk Comptonization}
\label{res_timestep}
We now explore the effect of the time variability of the vertical structure on bulk Comptonization. We stress that the specific numerical results of this section should not be directly compared to observations of real discs for two primary reasons. First, variability is an inherently global phenomenon which shearing box simulations therefore cannot effectively capture. Second, shearing box simulations with narrow box widths have been found to overestimate variability at a particular radius in the disc, so even not taking into account global phenomena we would expect this analysis to overestimate the variability of bulk Comptonization. The purpose of this section, therefore, is only to demonstrate how the time variability of bulk Comptonization depends on the time variability of the vertical structure profiles. To model the latter will require global disc simulations.

\begin{figure}
\includegraphics[width = 84mm]{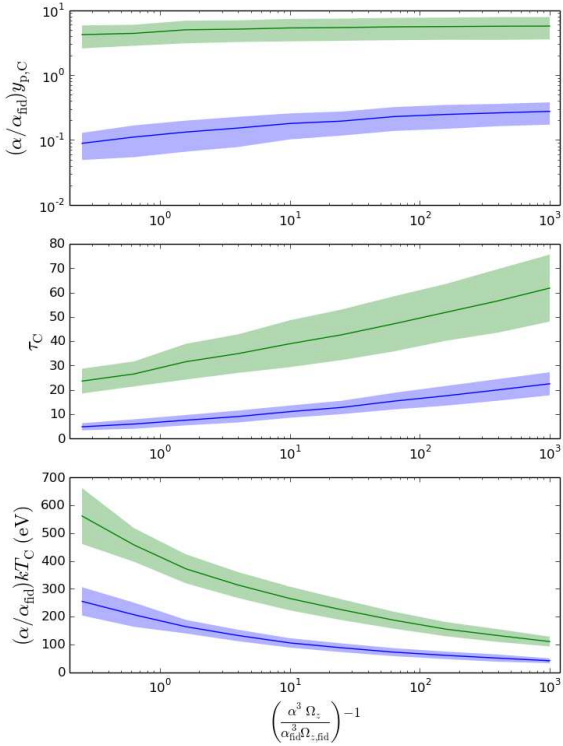}
\caption{Dependence of bulk Comptonization on shearing box parameters for $\left(\Sigma/\Sigma_{\rm fid} \right)^{-1} = $ $1$ (blue) and $5$ (green). The shaded region corresponds to points within $0.675 \sigma$ (i.e. $50 \%$ of the data for a Gaussian distribution) in the time variability.}
\label{fig_omega_sigma_range}
\end{figure}

\begin{figure}
\includegraphics[width = 84mm]{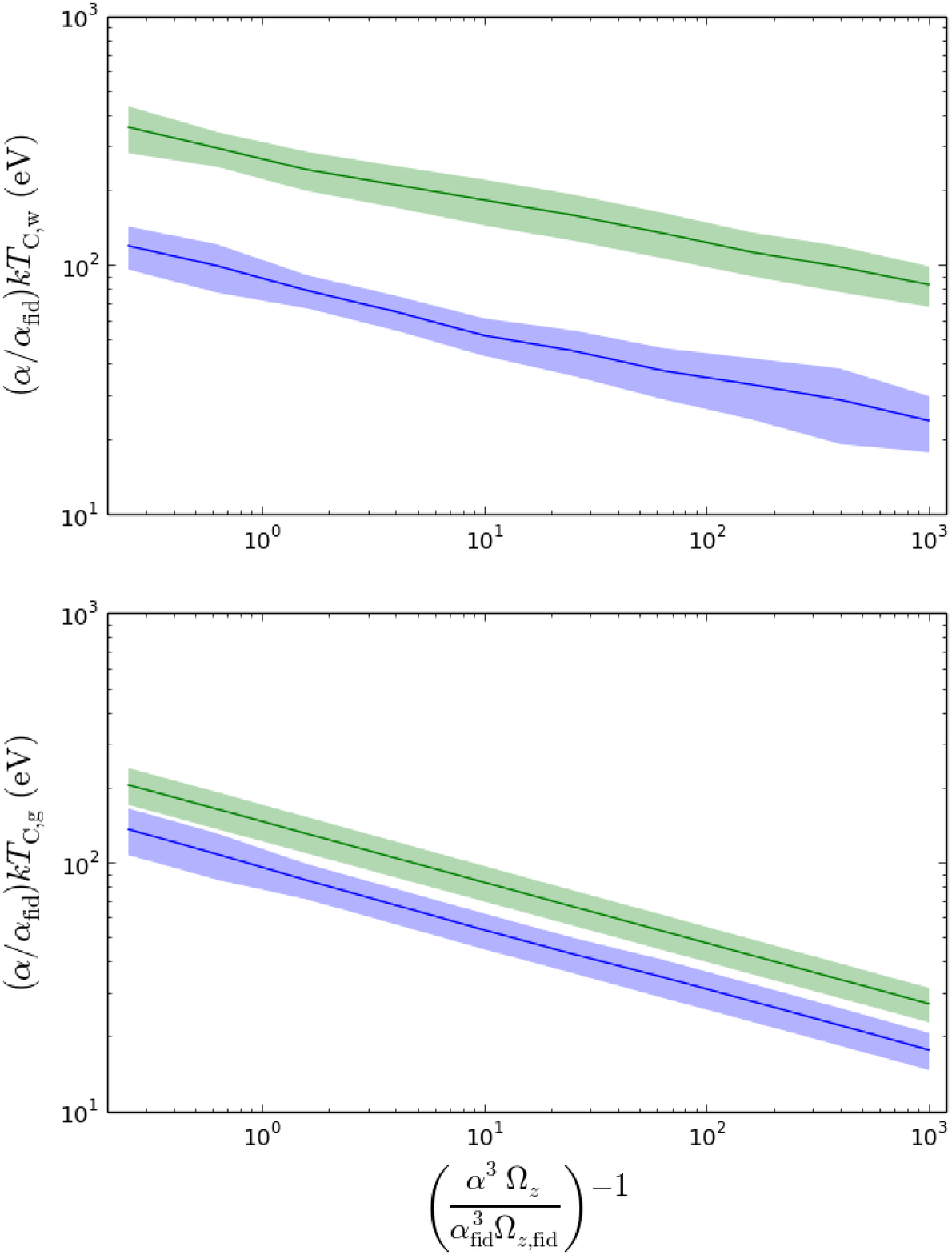}
\caption{Dependence of the Comptonization gas and wave temperatures on shearing box parameters, separately, for $\left(\Sigma/\Sigma_{\rm fid} \right)^{-1}= $ $1$ (blue) and $5$ (green). The shaded region corresponds to points within $0.675 \sigma$ (i.e. $50 \%$ of the data for a Gaussian distribution) in the time variability.}
\label{fig_omega_sigma_separate_range}
\end{figure}

\begin{figure}
\includegraphics[width = 84mm]{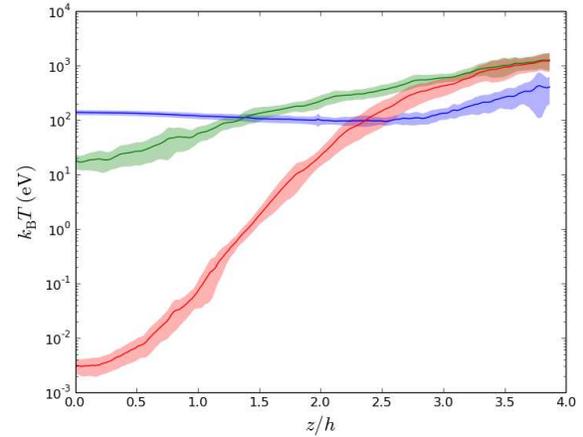}
\caption{Horizontally and time averaged gas (blue), bulk (green), and wave (red) temperature profiles for the fiducial shearing box parameters (Table \ref{table_sim_param_3}). The shaded region corresponds to points within $0.675 \sigma$ (i.e. $50 \%$ of the data for a Gaussian distribution) in the time variability.}
\label{fig_T_profiles_range}
\end{figure}

\begin{figure}
\includegraphics[width = 84mm]{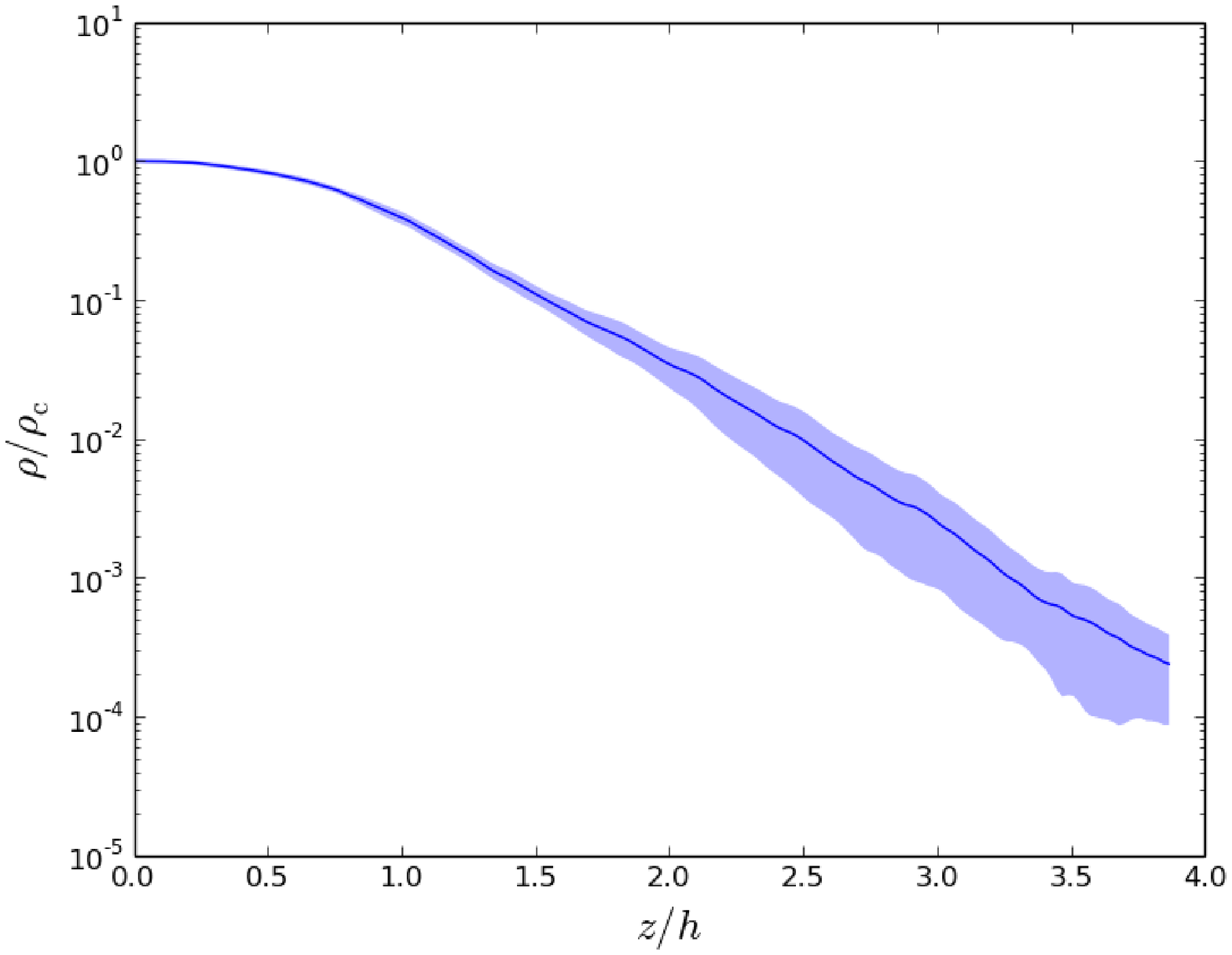}
\caption{Horizontally and time averaged density profile for the fiducial shearing box parameters (Table \ref{table_sim_param_3}). The shaded region corresponds to points within $0.675 \sigma$ (i.e. $50 \%$ of the data for a Gaussian distribution) in the time variability.}
\label{fig_rho_profile_range}
\end{figure}

We plot the standard deviation of the bulk Comptonization parameters over the $21$ equally spaced timesteps in Figure \ref{fig_omega_sigma_range}. We also plot the standard deviation of the Comptonization gas and wave temperatures individually in Figure \ref{fig_omega_sigma_separate_range}. In order to understand these results, we plot the standard deviations of the time-averaged temperature and density profiles in Figures \ref{fig_T_profiles_range} and \ref{fig_rho_profile_range}, respectively.

\begin{figure}
\includegraphics[width = 84mm]{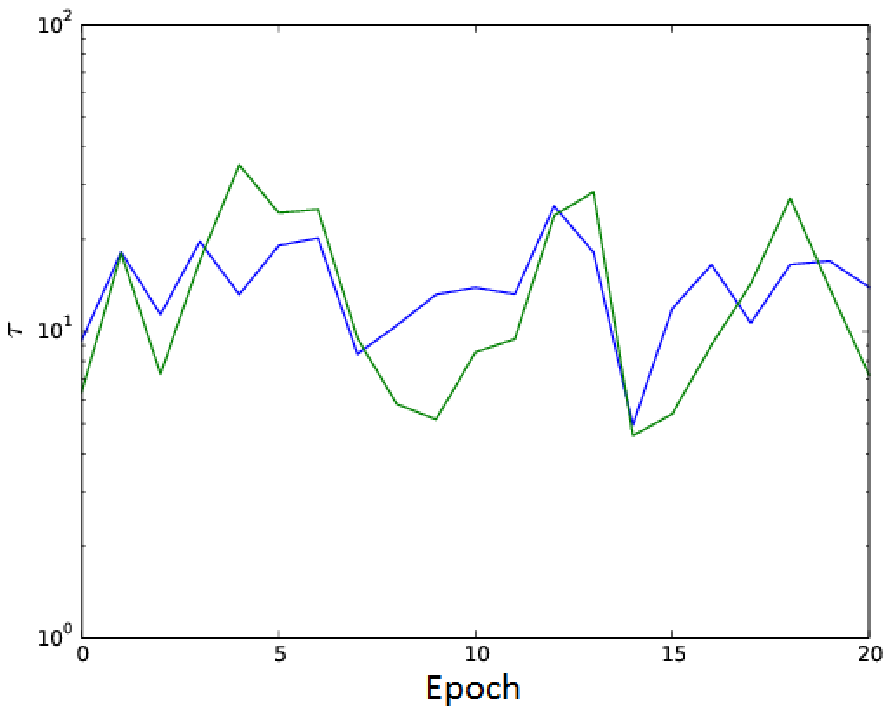}
\caption{Time variability of the Comptonization optical depth $\tau_{\rm C}$ (blue) and the (normalized) optical depth of a nearly identical region whose physical size is defined to be constant (green). The surface density here is $\left(\Sigma/\Sigma_{\rm fid} \right)^{-1}= 2$ and the other parameters are the fiducial shearing box parameters (Table \ref{table_sim_param_3}). The timesteps are spaced $10$ orbital periods apart, and each orbital period is $5535$ seconds.}
\label{fig_tau_time_variability}
\end{figure}

The time variation of the density profile alone may cause significant variations in the bulk Comptonization optical depth. For example, in Figure \ref{fig_tau_time_variability} at each timestep we plot the Comptonization optical depth $\tau_{\rm C}$ and the optical depth for a region of fixed size. The bottom of this region is taken to be the point at which the time-averaged gas and wave temperature profiles intersect in Figure \ref{fig_T_profiles_range}. Variations in the optical depth of this region result in variations in $\tau_{\rm C}$ that are caused by changes in the density profile alone, rather than changes in the size of the region itself. We see in Figure \ref{fig_tau_time_variability} that the overall variance of the optical depth of the region of fixed size is similar to the variance in $\tau_{\rm C}$. But we also see that the two quantities are only weakly correlated, which means that the variation in the bulk Comptonization optical depth must be due to other factors as well. For example, increasing the density may indirectly \emph{decrease} the bulk Comptonization optical depth by reducing the wave temperature (see Section \ref{sec_step_map}) and therefore the size of the bulk Comptonization region.

\begin{figure}
\includegraphics[width = 84mm]{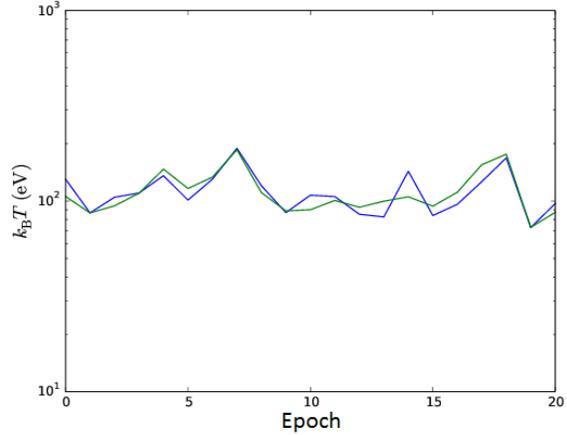}
\caption{Time variability of the Comptonization gas (blue) and wave (green) temperatures, normalized to the average Comptonization gas temperature. The surface density here is $\left(\Sigma/\Sigma_{\rm fid} \right)^{-1}= 2$ and the other parameters are the fiducial shearing box parameters (Table \ref{table_sim_param_3}). The timesteps are spaced $10$ orbital periods apart, and each orbital period is $5535$ seconds.}
\label{fig_T_time_variability}
\end{figure}

\begin{figure}
\includegraphics[width = 84mm]{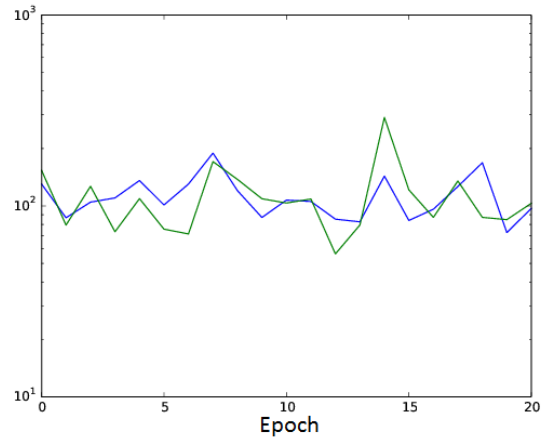}
\caption{Time variability of the Comptonization gas temperature $T_{\rm g, C}$ (blue) and inverse optical depth $\tau_{\rm C}^{-1}$ (green), normalized to the average Comptonization gas temperature. The surface density here is $\left(\Sigma/\Sigma_{\rm fid} \right)^{-1}= 2$ and the other parameters are the fiducial shearing box parameters (Table \ref{table_sim_param_3}). The timesteps are spaced $10$ orbital periods apart, and each orbital period is $5535$ seconds.}
\label{fig_T_tau_time_variability}
\end{figure}

Because the spatial variation of the gas temperature profile is so small in the bulk Comptonization region, we expect that its time variation should correlate with the time variation of the bulk Comptonization parameters in a predictable way. We expect that increasing the gas temperature decreases the size of the bulk Comptonization region, thereby increasing the Comptonization wave temperature and decreasing the Comptonization optical depth. In Figure \ref{fig_T_time_variability} at each timestep we plot the Comptonization gas temperature and the Comptonization wave temperature. We see that the two temperatures are strongly correlated in the direction we expect. In Figure \ref{fig_T_tau_time_variability} we plot the Comptonization gas temperature and optical depth. In this case, the correlation is also in the direction we expect, but it is weaker since density variations (among other factors) also play a significant role in determining the Comptonization optical depth.

\begin{figure}
\includegraphics[width = 84mm]{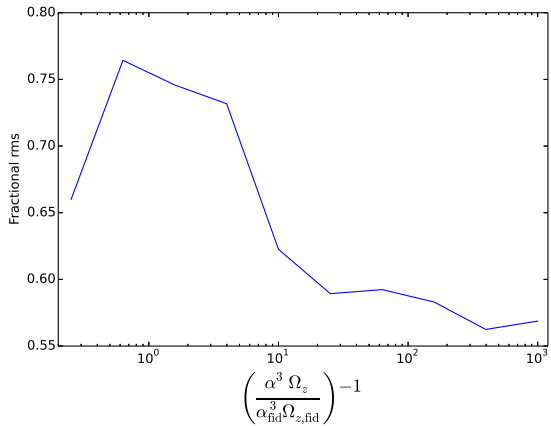}
\caption{Dependence of fractional rms on shearing box parameters for $\left(\Sigma/\Sigma_{\rm fid} \right)^{-1} = 1$.}
\label{fig_fractional_rms}
\end{figure}

We can also estimate the variability of the luminosity powered by bulk Comptonization according to the shearing box simulation data. The fraction of the luminosity powered by bulk Comptonization is just the total fractional photon energy change, which for unsaturated spectra is approximately
\begin{align}
\frac{\Delta \epsilon}{\epsilon} \approx e^{y_{\rm p,C}} - 1.
\end{align}
For $y_{\rm p,C} > 1$ we must check that spectra is unsaturated. For $y_{\rm p,C} \ll 1$ spectra is always unsaturated, and in addition the fractional energy change simplifies to 
\begin{align}
\frac{\Delta \epsilon}{\epsilon} \approx y_{\rm p,C}.
\end{align}
The variability of the luminosity powered by bulk Comptonization is therefore characterized by the fractional rms (root mean square) $y_{\rm p,C}$, which is the standard deviation divided by the mean of $y_{\rm p,C}$,
\begin{align}
\text{fractional rms} = \frac{\sigma_{y_{\rm p,C}}}{\left \langle y_{\rm p,C} \right \rangle}.
\end{align}
We plot the fractional rms for $\left(\Sigma/\Sigma_{\rm fid} \right)^{-1} = 1$ in Figure \ref{fig_fractional_rms}, and see that it is consistent with Figure \ref{fig_omega_sigma_range}. In particular, as the values on the $x$ axis increase from $10^0$ to $4 \times 10^2$, we see that $\sigma_{y_{\rm p,C}}$ decreases while $\left \langle y_{\rm p,C} \right \rangle$ increases so that the fractional rms increases. Since $\Omega_z^{-1} \propto M$ (Section \ref{res_part_2}), the fractional rms seems to vary insubstantially with mass. We note that the fiducial shearing box parameters (Table \ref{table_sim_param_3}) correspond to $r=20$ for the $M = 2 \times 10^6 M_\odot$,  $L/L_{\rm Edd} = 2.5$ parameter set (Table \ref{table_disc_param}). Since $r \approx 20$ in the region of the disc that contributes most to the luminosity for this parameter set (i.e. for $a=0$, $\Delta \epsilon = 0$), Figure \ref{fig_fractional_rms} also characterizes the variability of the luminosity powered by bulk Comptonization for the entire accretion disc.

\begin{figure}
\includegraphics[width = 84mm]{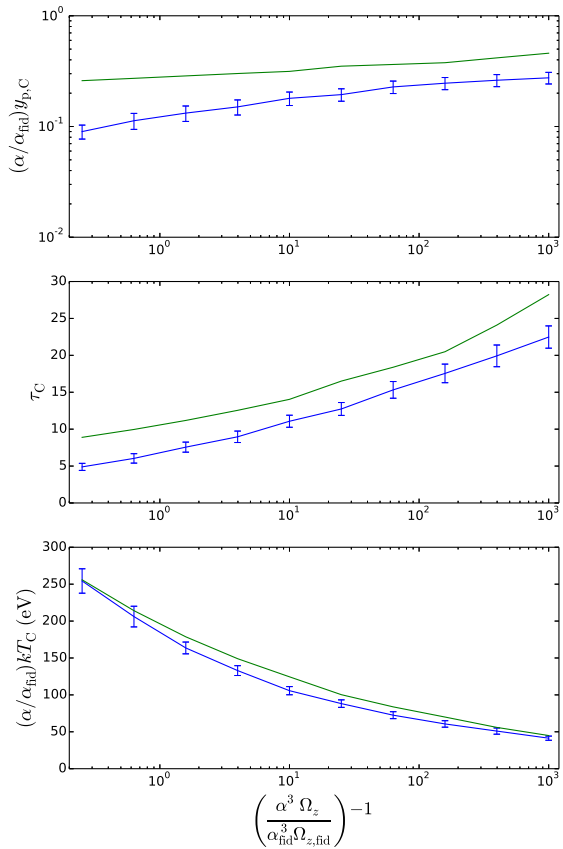}
\caption{Dependence of bulk Comptonization on the shearing box parameters, calculated by either time averaging the Comptonization parameters (blue) or time averaging the vertical structure profiles (green). The surface density is $\Sigma/\Sigma_{\rm fid}$ = 1.}
\label{fig_omega_sigma_avg2}
\end{figure}

We note that the time-averaged bulk Comptonizaton parameters are not equal to the bulk Comptonization parameters computed from the time-averaged temperature profiles. For example, in Figure \ref{fig_omega_sigma_avg2} we plot the bulk Comptonization parameters computed with the time-averaged data. To do this, we first time average the gas and wave temperature profiles and then compute the bulk Comptonization parameters. In the same figure we also plot the time-averaged parameters, originally plotted in Figure \ref{fig_omega_sigma}. We see that the parameters computed from the time-averaged profiles significantly overestimate the time-averaged parameters. This result is important because it means that the time-averaged profiles, while often useful, should not directly be used to model bulk Comptonization.

This work is based on only $21$ simulation snapshots spaced $10$ orbital periods apart, but we note that by applying our model to a complete set of simulation data one could also calculate how the power spectra of the vertical structure profiles affect that of the fraction of the luminosity powered by bulk Comptonization and other bulk Comptonization parameters.

\section{Discussion}
\label{sec_discussion}

\subsection{Comparison of results with previous work and observations}
\label{sec_discussion_obs}
Aside from the fact that our model implements a simplified version of the procedure used in K17 to calculate the bulk Comptonization parameters, our work differs from K17 in two important ways: In K17 bulk Comptonization is modelled for an entire accretion disc at once rather than at each radius individually, and shear velocities are included in addition to turbulent velocities. Because of these differences, the bulk Comptonization parameters found in that work depart slightly from those found here. But the dependence of bulk Comptonization on accretion disc parameters detailed in Sections \ref{res_part_2} and \ref{res_part_3} is consistent with the results of K17. In particular, the bulk Comptonization $y$ parameter for the overall disc increases with $\alpha$ and mass, while the bulk Comptonization temperature decreases with increasing mass. And within a given disc, bulk Comptonization is greatest at intermediate radii where the flux is also near maximal.

To make contact with observations, K17 modelled bulk Comptonization for a few systems with accretion disc parameters similar to those fit by D12 to REJ1034+396, a narrow-line Seyfert 1 (NLS1) with $L/L_{\rm Edd} = 2.4$. The bulk Comptonization parameters found by K17 broadly agree with those fit by D12. This agreement suggests that the large soft X-ray excess seen in REJ1034+396 may at least in part be due to bulk Comptonization. By generalizing the results of K17, our work provides a physical basis for more widely connecting warm Comptonization models of the soft excess to underlying accretion disc parameters.

\subsection{The importance of the disc inner boundary condition and implications for black hole X-ray binaries}
\label{section_inner}
An important consequence of our results is that bulk Comptonization is strongly dependent on the disc inner boundary condition parameter, $\Delta \epsilon$. Before proceeding, however, we provide context for the range of $\Delta \epsilon$ since it is not a widely used parameter and we need to have a sense of what it means for it to be large. To start, we observe that since the efficiency for a no torque inner boundary condition, zero spin system is $\eta = 0.057$, any value of $\Delta \epsilon > 0.01$ is relatively large. Even for spin $a = 0.9$, the efficiency with no inner torque is $\eta = 0.16$, so $\Delta \epsilon = 0.1$ corresponds to a substantial physical change.

Another way to understand the effect of $\Delta \epsilon$ is to examine how it affects the disc scalings presented in K17. We see that this parameter arises in the equations for the flux scalings, reproduced in Section \ref{res_part_2}, equations (\ref{eq_disc_flux_scaling_1}) and (\ref{eq_disc_flux_scaling_2}). To understand why $\Delta \epsilon$ appears here, we observe that for $\Delta \epsilon = 0$ in both equations the purpose of the final factor is to ensure that the flux goes to zero as $r$ approaches $r_{\rm in}$ rather than continue to increase as $r^{-3}$. It follows that we can regard $\Delta \epsilon$ large to the extent that it reverses the effects of this factor. For example, for the Newtonian scalings we see that setting $\Delta \epsilon = 1/r_{\rm in}$ removes the dependence on $r$ of this term altogether so that $F \sim r^{-3}$. For zero spin, $r_{\rm in} = 6$ so we should regard $\Delta \epsilon = 0.17$ as very large. For $a=0.9$, $r_{\rm in} = 2.32$, so the critical value of $\Delta \epsilon$ is 0.43. Therefore, values of $\Delta \epsilon$ anywhere from 0.1 to 0.4 should be viewed as very large, depending on the spin parameter $a$. The Kerr scalings lead to similar conclusions.

As $\Delta \epsilon$ approaches infinity the flux scaling asymptotes to a fixed value rather than continuing to increase. At $r = r_{\rm in}$, we see from the Newtonian scalings that this limit is reached when $\Delta \epsilon \sim 1$. Beyond this point, therefore, bulk Comptonization hardly varies at all with $\Delta \epsilon$. The reason for this is that $\Delta \epsilon$ changes the distribution of flux throughout the disc at a fixed overall luminosity $L/L_{\rm Edd}$. For $\Delta \epsilon \gg 1$, the distribution at the inner radii is fixed and the flux distribution continues to change only for $r \gg r_{\rm in}$.

\begin{figure}
\includegraphics[width = 84mm]{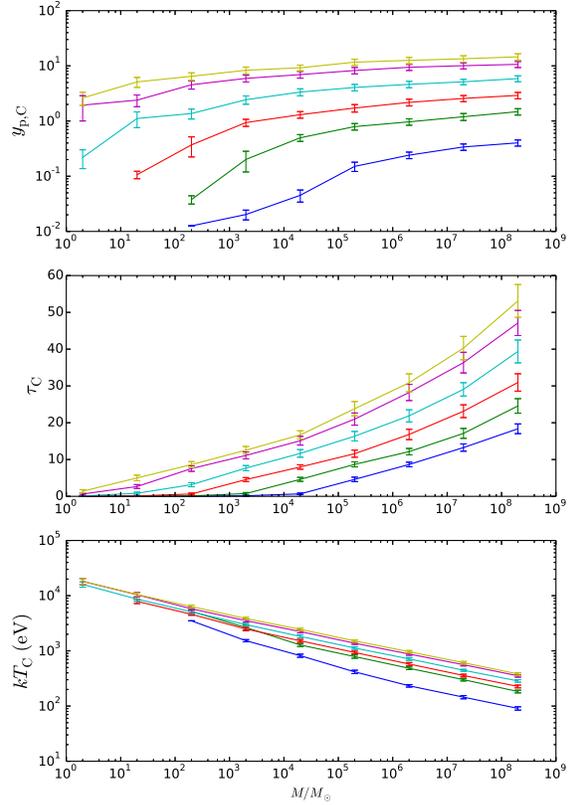}
\caption{Dependence of bulk Comptonization on mass for $\Delta \epsilon = $ $0$ (blue), $0.03$ (green), $0.05$ (red), $0.1$ (cyan), $0.3$ (magenta), and $1$ (yellow). The values of the parameters held constant are $L/L_{\rm Edd} = 2.5$, $a = 0.5$, $r = r_{\rm max}$, and $\alpha/\alpha_0 = 2$.}
\label{fig_m_deltaep_a0.5}
\end{figure}

In Figure \ref{fig_m_deltaep_a0.5} we plot the dependence of bulk Comptonization on mass for several values of $\Delta \epsilon$ for a system with moderate ($a=0.5$) spin. We see that all bulk Comptonization parameters strongly increase with increasing $\Delta \epsilon$. Note that this dependence holds only for the region where the disc is brightest, not radii for which $r \gg r_{\rm in}$ (see Sections \ref{res_part_2} and \ref{res_part_3}).

An important implication of this result is that bulk Comptonization is likely insignificant in black hole X-ray binaries unless the luminosity greatly exceeds Eddington or $\Delta \epsilon$ is large (i.e. $\Delta \epsilon > 0.1$). In Figure \ref{fig_m_deltaep_a0.5}, for which $L/L_{\rm Edd} = 2.5$, for example, we see that for $\Delta \epsilon = 0$ bulk Comptonization is non-existent for $M < 10^5 M_{\odot}$. For $M = 10 M_{\odot}$, we see a significant effect only for $\Delta \epsilon > 0.1$.

\subsection{Robustness of bulk Comptonization results to variations in the disc vertical structure}
In Section \ref{sec_dependence_1} we showed that the dependence of bulk Comptonization on $\Sigma^{-1}$ and $\Omega_z^{-1}$, plotted in Figure \ref{fig_omega_sigma}, can be understood in terms of the shearing box temperature profiles, plotted in Figure \ref{fig_T_profiles}. Since these profiles correspond to scaled data from a single radiation MHD simulation, we need to examine the extent to which our results are robust to changes in the disc vertical structure that may occur in shearing box simulations run in different regimes or global simulations. Certainly the exact values of the bulk Comptonization parameters are sensitive to such changes (Section \ref{res_timestep}), but we now show that the overall dependence on $\Sigma^{-1}$ and $\Omega_z^{-1}$ (and therefore on the accretion disc parameters) is more robust.

We first consider the dependence of bulk Comptonization on $\Omega_z^{-1}$. In Section \ref{sec_dependence_1}, using the profiles shown in Figure \ref{fig_T_profiles}, we showed that since only the gas temperature profile varies with $\Omega_z^{-1}$, the Comptonization temperature decreases and the Comptonization optical depth increases with increasing $\Omega_z^{-1}$. There is considerable uncertainty in the shape of the gas temperature profile outside the scattering photosphere, but fortunately the contribution of this region to bulk Comptonization is negligible since the bulk Comptonization temperature is weighted by the optical depth factor $\tau d \tau$. The greatest uncertainty in this analysis, therefore, is the bulk velocity field, which determines the shape of the wave temperature profile. But since the wave temperature is defined to strongly decrease with increasing density (Section \ref{sec_step_map}), we expect that even for significantly different velocity fields the wave temperature profile will increase near the scattering photosphere and that the resulting dependence of the Comptonization parameters on $\Omega_z^{-1}$ will be unchanged. Since our conclusions in Section \ref{sec_dependence_1} regarding the dependence of bulk Comptonization on $\Sigma^{-1}$ also rely primarily on the fact that the wave temperature profile strongly increases near the photosphere, we also expect them to be robust to changes in the vertical structure.

In addition to the above concerns, we must also check that as the size of the bulk Comptonization region increases it remains outside the effective photosphere. Otherwise, only part of the bulk Comptonization region will contribute to bulk Comptonization (since photons are emitted at the effective photosphere). However, this condition is likely always satisfied since the size of the bulk Comptonization region increases most significantly as $\Sigma^{-1}$ increases, which simultaneously moves the effective photosphere inward. In particular, for our data we find that the vertical structure becomes effectively thin well before the bulk Comptonization region optical depth is more than a small fraction of the optical depth of the half-thickness of the disc.

\subsection{Limitations to the scaling scheme parameter range}
We now make note of a subtlety that limits the applicability of the scaling scheme from K17: The scheme can scale data to lower surface densities, but not higher ones. To understand why, we examine how the gas temperature profile scales with decreasing $\Sigma$. First we note that since the gas temperature profile below the photosphere is significantly different from the profile above it, the regions must be scaled separately and then joined together. Next, we observe that the photosphere is not defined to be at a set number of scale heights $h$ away from the midplane but rather at the point at which the scattering optical depth is unity. As a result, as the surface density $\Sigma$ decreases the photosphere moves inward in $z/h$.  Since $h$ is the fundamental length scale for variations in the vertical structure and since fewer scale heights of data are needed outside the photosphere, fewer grid cells of data are needed to fill the region outside the photosphere of the scaled disc. This truncated data is scaled appropriately and then joined to the scaled data from above the photosphere. We see, therefore, that scaling to smaller surface densities requires deleting grid cells from the original simulation data. By the same reasoning, this scheme cannot scale to larger surface densities since it would require data from more grid cells than already exist. 

It immediately follows that this scheme cannot scale data to any set of accretion disc parameters for which $\Sigma/\Sigma_0 > 1$. In particular, since $\Sigma^{-1}$ is always directly proportional to the luminosity $L/L_{\rm Edd}$, we can never scale to smaller values of $L/L_{\rm Edd}$ unless they are offset by simultaneously scaling to, for example, smaller radii or greater $\Delta \epsilon$. We note that this scaling scheme is, therefore, useful for scaling lower Eddington ratio simulations to higher ones, as we do in this work, but not the other way around.

We showed in Section \ref{sec_dependence_1} that bulk Comptonization increases strongly with increasing $\Sigma^{-1}$. For the curve in Figure \ref{fig_omega_sigma} with the smallest value of $\Sigma^{-1}$, $\Sigma^{-1} = \Sigma_{\rm fid}^{-1} = 4 \Sigma_0^{-1}$, we see that $0.1 <y_{\rm p,C} < 0.3$. Therefore, 
the fact that we cannot scale to values of $\Sigma^{-1}$ smaller than $\Sigma_0^{-1}$ is not a significant limitation since it appears that bulk Comptonization is negligible for such values anyway. But this analysis assumes that the scaling scheme in K17 is valid over an arbitrarily large parameter range. If we want to scale to a regime with significantly different opacities, for example, then we really should use data from simulations with the relevant opacities included. For example, if the vertical structure is significantly different for sub-Eddington AGN because of changes in the opacities that occur in such regimes, then bulk Comptonization could be larger than we would infer from our analysis of the 110304a simulation data. On the other hand, this seems unlikely given that absorption opacities will substantially increase in this regime.

\subsection{Effect of bulk Comptonization on disc spectra}
\label{section_effect_spectra}
The effect of bulk Comptonization on disc spectra cannot only be to upscatter photons to higher energies because we also must take into account the back reaction on the disc vertical structure. Since energy conservation fixes the flux as a function of radius and the other accretion disc parameters, we expect that bulk Comptonization will be accompanied by a decrease in the gas temperature at the effective photosphere so that the total emitted flux will remain unchanged. For significant bulk Comptonization, the effect of this is to move the Wien tail to higher energy while moving the spectral peak to lower energy, broadening the spectrum. For moderate bulk Comptonization, the effect of lowering the gas temperature may not translate into a leftward shift of the spectral peak, but the spectrum will still be broadened in such a way that the total flux remains unchanged.

A decrease in the effective photosphere gas temperature is the simplest conceivable back reaction. This would occur if the only effect of bulk Comptonization on the gas is to remove kinetic energy from the turbulent cascade through radiation viscous dissipation \citep{kau16} so that less kinetic energy is dissipated and converted to gas internal energy. But to self-consistently model this phenomenon, bulk Comptonization must be implemented in the underlying radiation MHD simulations. The shearing box simulations used in this paper \citep{hir09}, for example, do not include bulk Comptonization since it is primarily a second order effect in velocity (see K17 Section 4.2), and the flux limited diffusion approximation does not capture second order effects \citep{kau16}.

\subsection{Effect of the horizontally averaged $z$ component of the velocity field on bulk Comptonization}
\begin{figure}
\includegraphics[width = 84mm]{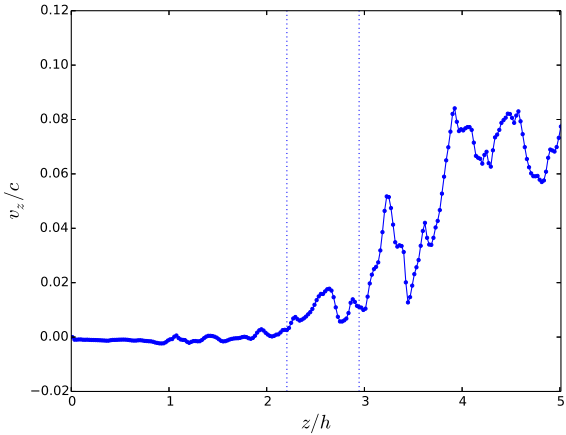}
\caption{Horizontally averaged profile at $r=14$ for the $M = 2 \times 10^6 M_\odot$,  $L/L_{\rm Edd} = 5$ parameter set (Table \ref{table_disc_param}) of the $z$ component of the velocity field. The dashed lines denote where $\tau_{\rm s} =$ $1$ and $\tau_{\rm s} =$ $10$.}
\label{fig_vz_profile}
\end{figure}

Bulk Comptonization includes effects that are both first and second order in the velocity field (K16) and the wave temperature defined in Section \ref{sec_step_map} captures only the second order effects. The first order effect is non-zero only for compressible modes and is negligible when the photon mean free path is large relative to the mode wavelength. As long as the mean photon energy is less than $4 k_{\rm B} \left(T_{\rm g} + T_{\rm w}\right)$, the second order effect always results in upscattering, analogous to thermal Comptonization. But the first order effect can result in either upscattering or downscattering depending on whether the velocity field is converging or diverging, respectively (K16). It follows that only long wavelength compressible modes should result in a non-negligible first order effect, since for shorter wavelength modes either the first order effect is negligible or upscattering in one region is offset by downscattering in another. Therefore, the variations with respect to $z$ of the density weighted, horizontal average of the $z$ component of the velocity field may result in a non-negligible first order effect. As in the case of the wave temperature profile (Section \ref{sec_justification}), density weighting is appropriate because photons scatter more times in higher density regions. We expect such long wavelength variations to exist since the vertical structure is stratified. For example, in Figure \ref{fig_vz_profile} we plot this profile at the 140 orbits timestep for the $M = 2 \times 10^6 M_\odot$,  $L/L_{\rm Edd} = 5$ parameter set (Table \ref{table_disc_param}) with $r=14$. In the remainder of this section we show that this effect is discernable but subdominant to the second order effect. We also show that once this effect is taken into account the slight discrepancy in Figure \ref{fig_spectra_Twave} between the spectra calculated directly with the turbulence and the spectra calculated by modeling the turbulence with the wave temperature vanishes. We therefore conclude that the wave temperature models the second order effect more accurately than we originally had reason to believe based on the preliminary analysis in Section \ref{sec_step_map}. All data in this section are scaled to the $M = 2 \times 10^6 M_\odot$,  $L/L_{\rm Edd} = 5$ parameter set (Table \ref{table_disc_param}). As in Section \ref{sec_justification}, all spectra and vertical structure profiles correspond to the 140 orbits timestep.

\begin{figure}
\includegraphics[width = 84mm]{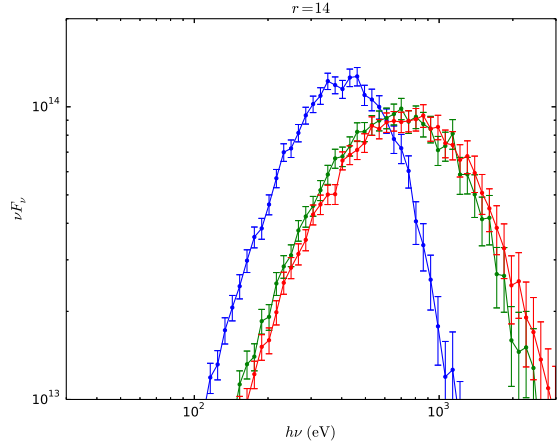}
\caption{Normalized accretion disc spectra at $r=14$ for the $M = 2 \times 10^6 M_\odot$,  $L/L_{\rm Edd} = 5$ parameter set (Table \ref{table_disc_param}) computed with (green) and without (blue) the velocities. For the red curve, the spectrum was computed with velocities but the horizontally averaged $z$ component of the velocity field was subtracted from the total $z$ component.}
\label{fig_spectra_no_vzavg}
\end{figure}

\begin{figure}
\includegraphics[width = 84mm]{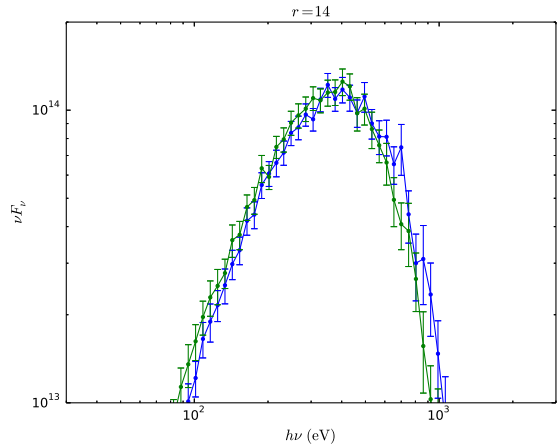} \\
\caption{Normalized accretion disc spectra at $r=14$ for the $M = 2 \times 10^6 M_\odot$,  $L/L_{\rm Edd} = 5$ parameter set (Table \ref{table_disc_param}), computed with data truncated at $\tau_{\rm s} = 20$. All gas temperatures were set to the horizontally averaged value at the base. For the blue curve, the velocities were not included, and for the green curve only the horizontally averaged $z$ component of the velocity field was included.}
\label{fig_spectra_no_turb}
\end{figure}

To begin, we calculate spectra with the original velocity field, both with and without subtracting off the horizontally averaged $z$ component, and plot the results for $r=14$ in Figure \ref{fig_spectra_no_vzavg}. We see that the spectrum computed with the horizontally averaged $z$ component included is shifted to slightly lower energies. The spectra at the other radii illustrate the same effect. Since any additional second order effect associated with this component can only \emph{increase} upscattering, this energy shift must either be due to the first order effect or vertical radiation advection. As explained in Section \ref{res_advection}, vertical radiation advection transports photons through the bulk Comptonization region faster, which decreases the number of photon scatterings in the region and may therefore reduce the overall second order effect. In order to show that the energy shift is predominantly due to the first order effect, not radiation advection, we calculate spectra with uniform temperature profiles both for the case of no velocities and for the case where only the horizontally averaged $z$ component is included, and plot the results for $r=14$ in Figure \ref{fig_spectra_no_turb}. The spectra at other radii illustrate the same effect. Since a uniform temperature profile with no velocity field has no effect on the base spectrum there is no second order effect, and so adding in the horizontally averaged $z$ component of the velocity field can shift the resulting spectra to lower energies only through the first order effect, not radiation advection. Since the spectrum in Figure \ref{fig_spectra_no_turb} is shifted by the same amount as in Figure \ref{fig_spectra_no_vzavg}, we conclude that the original shift is predominantly due to the first order effect, not vertical radiation advection. As a check on this analysis, we repeated the uniform temperature profile spectral calculations but included instead the negative of the horizontally averaged $z$ component of the velocity field, and found that the energy shifts were opposite in direction and equal in magnitude.

We also estimate the energy shift due to the first order effect heuristically and check that the result is consistent with our spectral calculations. The fractional energy change per scattering due to this effect is (K16)
\begin{align}
\frac{\Delta \epsilon}{\epsilon} = - \frac{\lambda_{\rm p} \nabla  \cdot {\bf v}}{3c}.
\end{align}
In this case, therefore, the region where this effect is greatest is near the photosphere (Figure \ref{fig_vz_profile}), where we conveniently just confirmed that vertical advection is dominated by diffusion. It follows that the average number of scatterings $dN$ in a region of optical depth $d \tau$ is approximately (Section \ref{res_reduce}) equal to $1.6(2\tau d \tau)$. The total approximate fractional energy change $f$ in this region is then 
\begin{align}
f &= -1 + \lim\limits_{\Delta \tau \rightarrow 0} \prod_{i} \left(1 - \frac{\lambda_{\rm p} \nabla  \cdot {\bf v}}{3c}\right)^{1.6\left(2 \tau_{i} \Delta \tau_{i} \right)} \\
&= -1 + \lim\limits_{\Delta \tau \rightarrow 0} \prod_{i} \exp \left(\ln \left( \left(1 - \frac{\lambda_{\rm p} \nabla  \cdot {\bf v}}{3c}\right)^{1.6\left(2 \tau_{i} \Delta \tau_{i} \right)} \right)\right)\\
&= -1 + \exp\left(\int 1.6\ln \left(1 - \frac{\lambda_{\rm p} \nabla  \cdot {\bf v}}{3c}\right)2\tau d \tau \right).
\end{align}
Since the fractional energy change per scattering is much smaller than unity,
\begin{align}
f \approx -1 + \exp\left(\int - 1.6\left(\frac{\lambda_{\rm p} \nabla  \cdot {\bf v}}{3c}\right)2\tau d \tau \right).
\end{align}
In this case we find that at all radii $f \approx -0.1$, consistent with the results in Figures \ref{fig_spectra_no_vzavg} and \ref{fig_spectra_no_turb}. We note that if the \emph{total} fractional energy change is also much less than unity, such as in this case, then
\begin{align}
f \approx \int - 1.6\left(\frac{\lambda_{\rm p} \nabla  \cdot {\bf v}}{3c}\right)2\tau d \tau.
\end{align} 

In order for Monte Carlo calculations to self-consistently capture the first order effect, one must take into account the time dependent nature of the problem (K16), either by performing time dependent simulations or by careful analysis of the results. This is because this effect can result in either upscattering or downscattering depending on whether a region is converging or diverging, and a diverging region will typically evolve into a converging one on the flow timescale, given by
\begin{align}
t_{\rm f} \sim \lambda_{\rm p} \tau /v_{\rm z}.
\end{align}
If the region is near the photosphere, such as in the case examined here, then the photons escape the region on the diffusion timescale, which is shorter, and the first order effect will on average broaden the spectrum. If the region is sufficiently deep inside the photosphere that the diffusion timescale,
\begin{align}
t_{\rm d} \sim \lambda_{\rm p}\tau^2 /c,
\end{align}
exceeds the flow timescale, then the flow will change significantly before photons can diffuse very far. This is the case for standing acoustic modes, for example \citep{bla11}. In this case, results from time independent Monte Carlo simulations can be trusted only if the spectrum is negligibly affected by the upscattering or downscattering in such regions.

\begin{figure}
\includegraphics[width = 84mm]{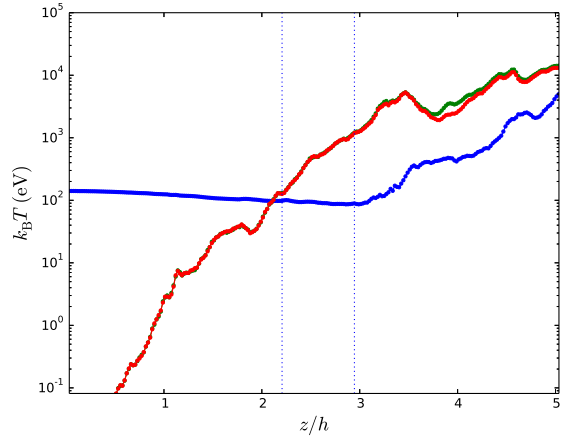}
\caption{Horizontally averaged profiles at $r=14$ for the $M = 2 \times 10^6 M_\odot$,  $L/L_{\rm Edd} = 5$ parameter set (Table \ref{table_disc_param}) for the gas temperature $T_{\rm g}$ (blue), wave temperature $T_{\rm w}$ (green), and wave temperature computed by first subtracting off the horizontally averaged $z$ component of the velocity field (red). The dashed lines denote where $\tau_{\rm s} =$ $1$ and $\tau_{\rm s} =$ $10$.}
\label{fig_T_profiles_mdotfactor_3_no_vzavg}
\end{figure}

\begin{figure}
\includegraphics[width = 84mm]{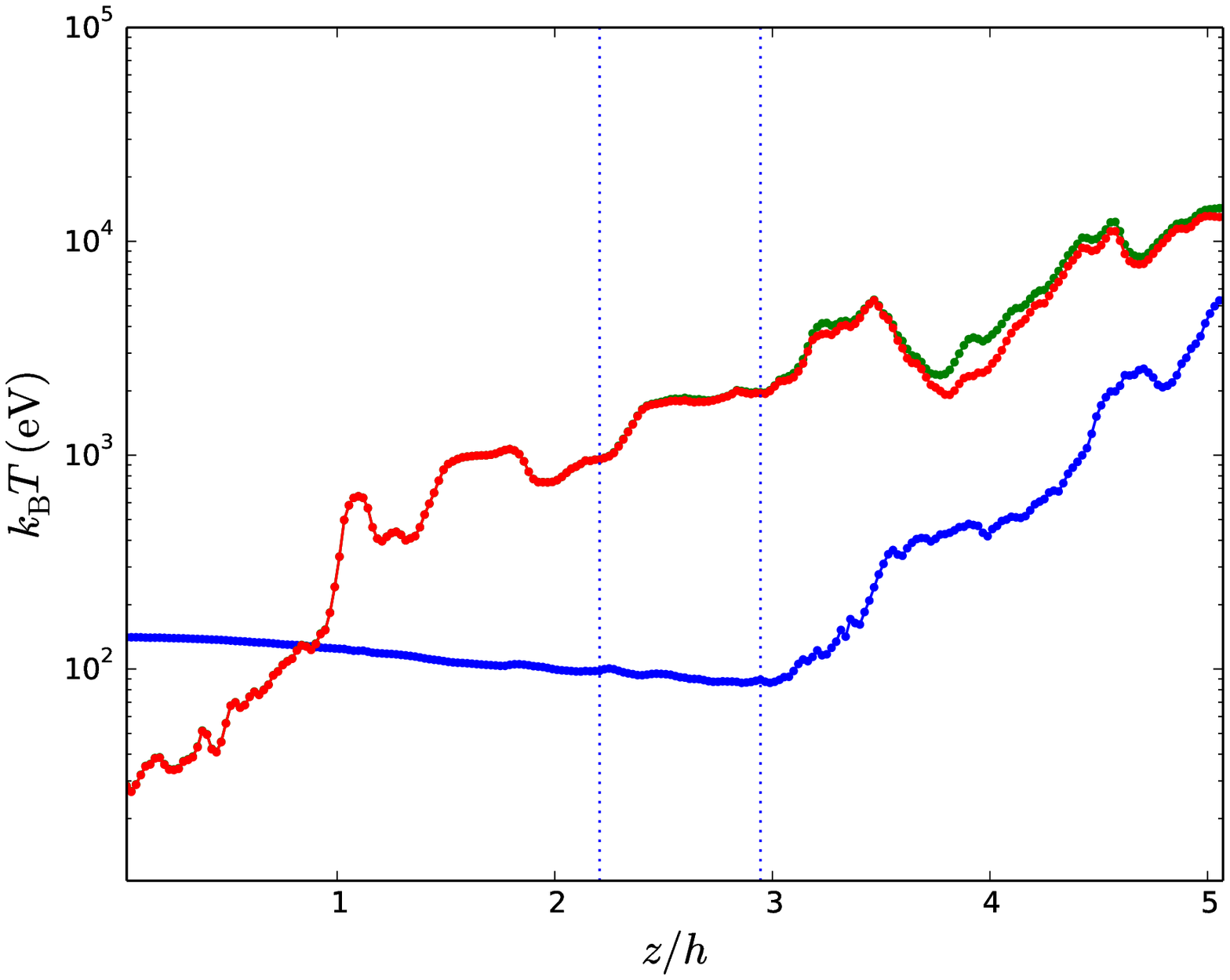}
\caption{Horizontally averaged profiles at $r=14$ for the $M = 2 \times 10^6 M_\odot$,  $L/L_{\rm Edd} = 5$ parameter set (Table \ref{table_disc_param}) for the gas temperature $T_{\rm g}$ (blue), bulk temperature $T_{\rm bulk}$ (green), and bulk temperature computed by first subtracting off the horizontally averaged $z$ component of the velocity field (red). The dashed lines denote where $\tau_{\rm s} =$ $1$ and $\tau_{\rm s} =$ $10$.}
\label{fig_T_profiles_mdotfactor_3_no_vzavg2}
\end{figure}

\begin{figure}
\includegraphics[width = 84mm]{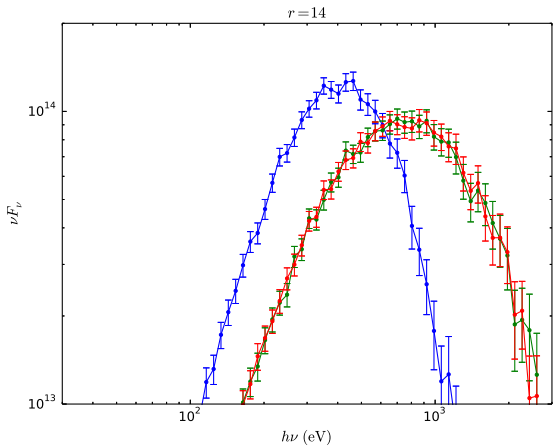} 
\caption{Normalized accretion disc spectra at $r=14$ for the $M = 2 \times 10^6 M_\odot$,  $L/L_{\rm Edd} = 5$ parameter set (Table \ref{table_disc_param}) computed without the velocities. For the green and red curves, the wave temperatures were added to the gas temperatures. For the red curve, the wave temperatures were computed by first subtracting off the horizontally averaged $z$ component of the velocity field.}
\label{fig_spectra_Twave_and_no_vzavg_Twave}
\end{figure}

In order to capture only second order effects in a Monte Carlo simulation, we can first subtract off the horizontally averaged $z$ component of the velocity field, but we can do this only if its second order effect is negligible. To investigate this, in Figure \ref{fig_T_profiles_mdotfactor_3_no_vzavg} for $r=14$ we plot the original wave temperature profile along with the wave temperature profile computed by first subtracting off the horizontally averaged $z$ component of the velocity field. We see that the resulting two curves are essentially identical except in a small region where bulk Comptonization is negligible since $\tau_{\rm s} \ll 1$. The horizontally averaged $v_z$ profile contributes negligibly to the wave temperature both because it contributes negligibly to the underlying bulk temperature profile, plotted in Figure \ref{fig_T_profiles_mdotfactor_3_no_vzavg2}, and because the wave temperature downweights long wavelength variations (Section \ref{sec_step_map}). To confirm that the contribution of the horizontally averaged $v_z$ profile to the second order effect is negligible, we calculate spectra in which we model the turbulence with wave temperatures calculated both with and without including the horizontally averaged $v_z$ and plot the results for $r=14$ in Figure \ref{fig_spectra_Twave_and_no_vzavg_Twave}. We see that there is no discrepancy between the respective curves, consistent with the wave temperature profiles in Figure \ref{fig_T_profiles_mdotfactor_3_no_vzavg}. The spectra at the other radii illustrate the same effect.

\begin{figure}
\includegraphics[width = 84mm]{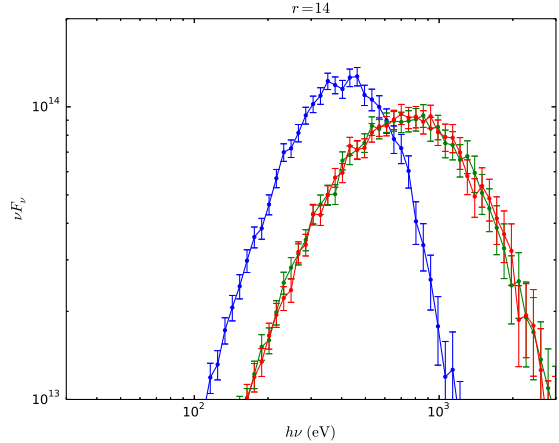}
\caption{Normalized accretion disc spectra at $r=14$ for the $M = 2 \times 10^6 M_\odot$,  $L/L_{\rm Edd} = 5$ parameter set (Table \ref{table_disc_param}) computed with (green) and without (blue) the velocities. For the spectrum computed with velocities, the horizontally averaged $z$ component of the velocity field was subtracted from the total $z$ component. For the red curve, the velocities were not included but the wave temperatures were added to the gas temperatures.}
\label{fig_spectra_no_vzavg_and_Twave}
\end{figure}

\begin{figure}
\includegraphics[width = 84mm]{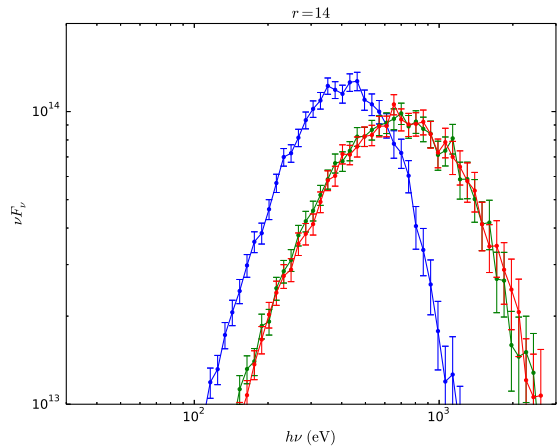}
\caption{Normalized accretion disc spectra at $r=14$ for the $M = 2 \times 10^6 M_\odot$,  $L/L_{\rm Edd} = 5$ parameter set (Table \ref{table_disc_param}) computed with (green) and without (blue) the velocities. For the red curve, the $z$ component of the velocity field was horizontally averaged, the $x$ and $y$ components were set to zero, and the wave temperatures were added to the gas temperatures.}
\label{fig_spectra_Twave_vzavg}
\end{figure}

In Section \ref{sec_step_map} we showed in Figure \ref{fig_spectra_Twave} that spectra computed with data in which the velocities were turned off and the wave temperatures were added to the gas temperatures approximated spectra computed with the velocities. Since the wave temperature captures only second order effects and since the horizontally averaged $z$ component of the velocity field results in a non-negligible first order effect, we expect that when this component is subtracted off the approximation will improve. We perform this comparison in Figure \ref{fig_spectra_no_vzavg_and_Twave} for $r = 14$. We see that in this case the approximation is so good that the respective spectra are indistinguishable from each other. The spectra at the other radii illustrate the same effect. This is not only consistent with our prediction but shows that the wave temperature captures second order effects even more accurately than we originally had reason to believe based on the preliminary analysis in Section \ref{sec_step_map}. Given these results, we also expect that spectra computed with the velocities turned off except for the horizontally averaged $z$ component and with the wave temperatures added to the gas temperatures will coincide with spectra computed with the velocities turned on, since both sets of spectra should capture both first and second order effects. We plot these spectra in Figure \ref{fig_spectra_Twave_vzavg} for $r=14$ and see that they agree with our prediction. The spectra at the other radii illustrate the same effect.

\section{Summary}
\label{sec_Conclusion}
We have simplified the bulk Comptonization model of K17 in order to explore a larger space of accretion disc parameters and develop greater physical insight into this phenomenon. Rather than fit the temperature and optical depth to spectra computed with Monte Carlo post processing simulations, we developed a procedure to calculate the Comptonization temperature and optical depth directly from the underlying vertical structure data (Section \ref{sec_modeling}). Using this, we plotted the dependence of the Comptonization parameters on the shearing box parameters and showed how these results can be understood in terms of the one dimensional temperature profiles (Sections \ref{res_overview} and \ref{res_part_1}). We then showed how we can analytically determine the dependence of bulk Comptonization on each accretion disc parameter individually (Sections \ref{res_part_2} and \ref{res_part_3}). Our principal results are as follows.

The primary independent variables in a shearing box are the surface density $\Sigma$, the vertical epicyclic frequency $\Omega_z$, and the strain rate, $\partial_x v_y$. We also allow $\alpha$, the ratio of the vertically integrated stress to the vertically integrated total pressure, to vary. For Kerr discs the scalings for the strain rate and vertical epicyclic frequency are always nearly equal (equation \ref{eq_strain_scaling}), which leaves three independent parameters. Using the velocity scalings (equations \ref{eq_v_turb}, \ref{eq_v_th_c}, and \ref{eq_v_th_ph}), we showed that the dependence of the Comptonization parameters on $\alpha$ can be subsumed into the other parameters (equations \ref{eq_tau}, \ref{eq_T}, and \ref{eq_yp}), which reduces the parameter space to two variables, $\Sigma$ and $\Omega_z$.

We plotted the dependence of the bulk Comptonization temperature, optical depth, and $y$ parameter on $\Sigma$ and $\Omega_z$ (Figure \ref{fig_omega_sigma}). We showed that these results can be understood from analyzing the one dimensional temperature profiles (Figure \ref{fig_T_profiles}) and the velocity scalings (equations \ref{eq_v_turb}, \ref{eq_v_th_c}, and \ref{eq_v_th_ph}). In particular, the Comptonization optical depth and $y$ parameter increase strongly with increasing $\Sigma^{-1}$ and weakly with increasing $\Omega_z^{-1}$. The Comptonization temperature also increases strongly with increasing $\Sigma^{-1}$, but decreases weakly with increasing $\Omega_z^{-1}$.

\begin{figure}
\includegraphics[width = 84mm]{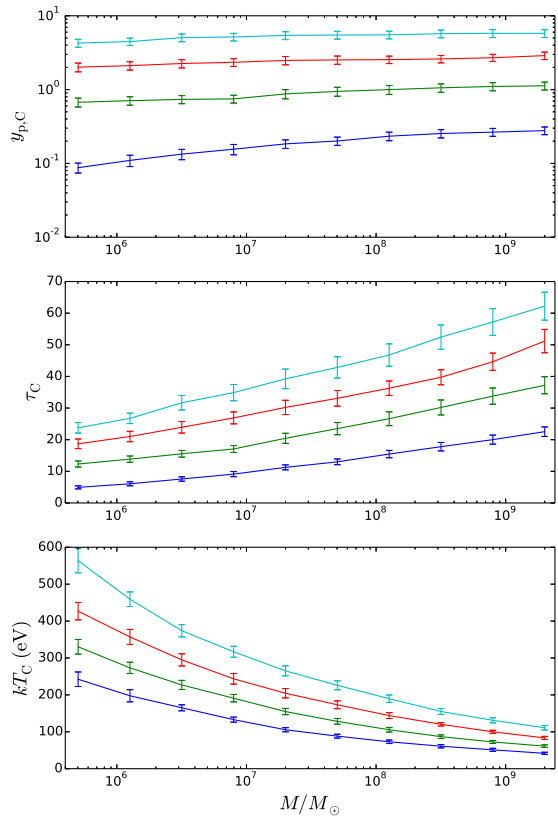}
\caption{Dependence of bulk Comptonization on mass. The blue, green, red, and cyan curves correspond to $L/L_{\rm fid} = $ $1$, $2$, $3.3$, and $5$, respectively, where $L_{\rm fid}/L_{\rm Edd} = 2.5$. The parameters held constant are $r=20$, $a=0$, $\Delta \epsilon = 0$, and $\alpha/\alpha_0 = 2$ (i.e. the other parameters in the $M = 2 \times 10^6 M_\odot$,  $L/L_{\rm Edd} = 2.5$ parameter set (Table \ref{table_disc_param}), with $r=20$). The only difference between this figure and Figure \ref{fig_omega_sigma} is the labeling of the axes.}
\label{fig_m_mdot}
\end{figure}

To determine the dependence of bulk Comptonization on accretion disc parameters, we write $\Sigma$ in terms of $F$ and then write the scalings for $F$ and $\Omega_z$ in terms of mass, luminosity, radius, spin, and inner boundary condition (Section \ref{res_part_2}). Since $\Omega_z^{-1}$ is directly proportional to mass, and $\Sigma$ is independent of mass, the dependence of bulk Comptonization on mass is identical to its dependence on $\Omega_z^{-1}$. Similarly, since $\Sigma^{-1}$ is directly proportional to luminosity, and $\Omega_z^{-1}$ is independent of luminosity, the dependence of bulk Comptonization on luminosity is identical to its dependence on $\Sigma^{-1}$. Therefore, Figure \ref{fig_omega_sigma} also summarizes the dependence of bulk Comptonization on mass and luminosity. Here, for clarity, we reproduce the plots from Figure \ref{fig_omega_sigma} in Figure \ref{fig_m_mdot} with the independent variables labeled as mass and luminosity.

The dependence of bulk Comptonization on the other accretion disc parameters is inferred by analyzing how they affect $\Sigma^{-1}$ since bulk Comptonization depends much more strongly on $\Sigma^{-1}$ than it does on $\Omega_z^{-1}$. Since $\Sigma^{-1}$ is proportional to the flux $F$ (equation \ref{eq_sigma_flux}), we showed that the dependence of bulk Comptonization on the other disc parameters can be understood intuitively in terms of how they effect $F$. In particular, at large radius (i.e. $r \gg r_{\rm in}$) bulk Comptonization always decreases with increasing radius. At small radius, whether bulk Comptonization increases or decreases with radius depends on the inner boundary condition parameter $\Delta \epsilon$. This parameter is defined to be the increase in the efficiency $\eta$ for a non-zero stress inner boundary condition, i.e. $L = \left(\eta + \Delta \epsilon \right) \dot{M}c^2$. Using the same line of reasoning, we showed that bulk Comptonization increases with both spin and the inner boundary condition parameter $\Delta \epsilon$ at small radius ($r \approx r_{\rm in}$), and decreases with those parameters at large radius. Finally, we showed that bulk Comptonization increases with $\alpha$, since once the accretion disc parameters are substituted in for $\Sigma$ and $\Omega_z$, $\Sigma^{-1}$ itself becomes proportional to $\alpha$ (equation \ref{eq_sigma_flux}) and this outweighs the dependence on $\alpha$ discussed earlier.

Next we studied bulk Comptonization for an entire accretion disc by examining how it varies when the radius is fixed to the region of maximum luminosity (Section \ref{res_part_3}). The dependence of bulk Comptonization on mass, luminosity, and $\alpha$ is unchanged from above since the radius of maximum luminosity does not vary with these parameters. But since this radius does depend on the spin and inner boundary condition parameter $\Delta \epsilon$, the dependence of bulk Comptonization on these parameters required a new treatment. We showed that in this case bulk Comptonization always increases with spin and $\Delta \epsilon$.

In Section \ref{res_advection} we showed that the effect of including vertical radiation advection at a fixed radius in an accretion disc is to decrease bulk Comptonizaton. We discussed how to include advection in our model more formally in Appendix \ref{sec_scalings}.

In Section \ref{sec_discussion_obs} we pointed out that our results broadly agree with the results of K17, which in turn agree with the analysis by D12 of the narrow-line Seyfert 1 REJ1034+396.

\begin{figure}
\includegraphics[width = 84mm]{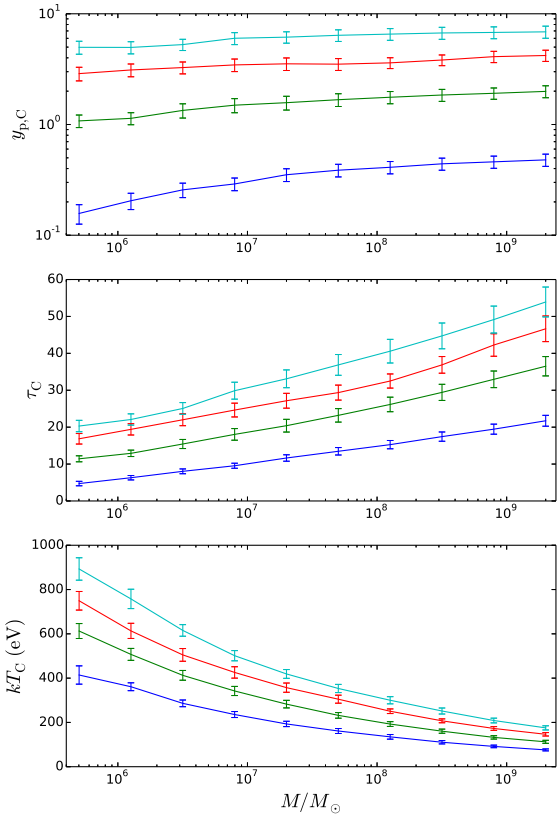}
\caption{Dependence of bulk Comptonization on mass for $\Delta \epsilon = 0.2$. The blue, green, red, and cyan curves correspond to $L/L_{\rm fid} = 0.25$, $0.5$, $0.75$, and $1$, respectively, where $L_{\rm fid}/L_{\rm Edd} = 2.5$. The parameters held constant are $r=6$, $a=0$, and $\alpha/\alpha_0 = 2$.}
\label{fig_m_mdot_deltaep}
\end{figure}

An important result of this work is that bulk Comptonization is strongly dependent on the disc inner boundary condition (Section \ref{section_inner}). In particular, the larger that $\Delta \epsilon$ is, the lower the luminosity can be without bulk Comptonization being negligible. Figure \ref{fig_m_mdot_deltaep} summarizes the dependence of bulk Comptonization on mass and luminosity for $\Delta \epsilon = 0.2$. In both Figures \ref{fig_m_mdot} and \ref{fig_m_mdot_deltaep} the radius is fixed to the region where the luminosity is greatest, but for $\Delta \epsilon = 0.2$ this radius corresponds to the innermost stable circular orbit. By comparing these two figures we see that for a given luminosity bulk Comptonization is significantly greater for $\Delta \epsilon = 0.2$. We also showed (Figure \ref{fig_m_deltaep_a0.5}) that bulk Comptonization is negligible in black hole X-ray binaries unless the disc inner boundary condition parameter is very large ($\Delta \epsilon \sim 0.1$) or the luminosity greatly exceeds Eddington. 

We expect that in a real disc bulk Comptonization at a given radius will be accompanied by a decrease in the gas temperature at the effective photosphere in order to leave the flux unchanged, which is required by energy conservation (Section \ref{section_effect_spectra}).

Because our model connects the bulk Comptonization parameters to the disc vertical structure one dimensional temperature profiles in a way that is physically intuitive, it provides a useful framework for understanding bulk Comptonization even in situations in which some of our specific results may not hold, such as shearing box or global radiation MHD simulations run in entirely different regimes.

Since our results outline how bulk Comptonization depends on fundamental accretion disc parameters, an observer who fits the soft X-ray excess with a warm Comptonization model can use them to distinguish contributions to the soft X-ray excess due to bulk Comptonization from those due to other physical mechanisms. In high Eddington sources, this can help provide a physical basis for and therefore constrain warm Comptonization models of the soft excess. Bulk Comptonization is likely insignificant in lower Eddington flows, on the other hand, even though the data show that in these flows the soft excess carries a more significant fraction of the power \citep{jin12,meh11,meh15}. Moreover, since bulk Comptonization depends on the properties of MRI turbulence through $\alpha$ (Sections \ref{res_part_2} and \ref{res_part_3}) and through the time variability of the temperature and density profiles (Section \ref{res_timestep}), our work indicates that observations of the soft X-ray excess may in turn advance our understanding of disc turbulence in the radiation pressure dominated regime.

\section*{Acknowledgements}
This work was supported by NASA Astrophysics Theory Program grant NNX13AG61G and the International Space Science Institute (ISSI) in Bern. SH was supported by Japan JSPS KAKENH 15K05040 and the joint research
project of ILE, Osaka University. The radiation MHD simulation was
partly carried out on the Cray XT4 at CfCA, National Astronomical
Observatory of Japan, and on SR16000 at YITP in Kyoto University.

\appendix
\section{Applying the wave temperature definition to simulation data}
\label{sec_implementation}
It is problematic to apply equation (\ref{eq_def_delta_v}) directly to simulation data for two reasons. First, a straightforward implementation runs too slowly for our purposes, since it requires computing an entire volume average for each grid cell. To speed up the computation for the applications in this work, we take the density in the entire region over which the spatial average is defined to be the density at position ${\bf r}$. With this approximation, equation (\ref{eq_def_delta_v}) can be implemented in Python without explicitly using ``for" loops to traverse the grid. In most of the simulation domain this approximation is sufficient since the probability that a photon scatters far from position ${\bf r}$ is negligible anyway. This approximation is less valid in the scattering photosphere, where the photon mean free path is large. But directly implementing equation (\ref{eq_def_delta_v}) in this region is problematic for an entirely different reason, which is that we do not have access to the velocity function above the top of the simulation domain. As a result, for values of ${\bf r}$ near the top of the simulation domain the spatial average underestimates $\langle (\Delta {\bf v})^2 \rangle_{{\bf r}}$ . To compensate for this as well as our original approximation, we define an additional parameter $\tau_{\rm break}$ as follows. At each pair of $x$ and $y$ coordinates we set $T_{\rm w} = T_{\rm bulk}$ for values of $z$ where $\tau_{\rm s} \leq \tau_{\rm break} < 1$, since $T_{\rm w}$ approaches $T_{\rm bulk}$ in the optically thin limit. We set $\tau_{\rm break} = 0.5$. However, because the number of photon scatterings in a given region scales with $\tau_{\rm s}^2$ (Section \ref{sec_step_TC}), it turns out that for our bulk Comptonization model the value of $T_{\rm w}$ does not matter for $\tau_{\rm s} < 1$ anyway, and so the approximations we make to define $T_{\rm w}$ in this region have no impact on our results. For example, we repeated the spectral calculations plotted in Figure \ref{fig_spectra_Twave} for $\tau_{\rm break} = 0$ and found that the results were unchanged.

\section{Including vertical radiation advection in shearing box scalings}
\label{sec_scalings}
\subsection{Derivation of shearing box scalings without assuming radiation diffusion}
Here we derive the shearing box scalings presented in K17 without assuming that the flux is carried by radiation diffusion. We give scalings for $\rho$, $T_{\rm g}$, $v_{\rm turb}$, and $v_{\rm s}$ in terms of $\Sigma$, $\Omega_z$, $\partial_x v_y$, $\alpha$, $\beta$, $f_{\rm col}$, and $h$. The result, therefore, of not assuming radiation diffusion is to leave the scale height $h$ as a free parameter.

To begin, the scaling for the density profile is still given by K17 equation (13), except that $h$ is now a free parameter:
\begin{align}
\rho\left(z\right) = \left(\frac{\Sigma}{\Sigma_0} \right) \left(\frac{h}{h_0} \right)^{-1} \rho_0\left(h_0 z/h\right).
\end{align}
The flux scaling is determined by K17 equations (2), (3), and (5), which give
\begin{align}
\label{eq_flux_scaling}
\left(\frac{F}{F_0}\right) = \left(\frac{\alpha}{\alpha_0}\right)\left(\frac{\Omega_{\rm z}}{\Omega_{\rm z,0}}\right)^2 \left(\frac{\partial_x v_y}{\partial_x v_{y,0}}\right)\left(\frac{\Sigma}{\Sigma_0}\right) \left(\frac{h}{h_0}\right)^2.
\end{align}
The scaling for the turbulent velocity profile is derived from K17 equations (2), (5), (10), (11), which give
\begin{align}
v(z) = \left(\frac{\alpha}{\alpha_0}\right)^{1/2} \left(\frac{\beta}{\beta_0}\right)^{1/2} \left(\frac{\Omega_{\rm z}}{\Omega_{\rm z,0}}\right) \left(\frac{h}{h_0}\right) v_0(h_0 z/h).
\end{align}
The scalings for the shear velocity, pressure and gas temperature profiles are unchanged, except that $h$ is now a free parameter. The scaling for the shear velocity profile is given by K17 equation (22):
\begin{align}
v_{\rm s}\left(x\right) = \left(\frac{\partial_x v_y}{\partial_x v_{y,0}}\right)\left(\frac{h}{h_0}\right) v_{\rm s,0}\left(h_0 x/h\right).
\label{eq_vshear_profile}
\end{align}
The pressure profile is given by K17 equation (17),
\begin{align}
P(z) =& P_{\rm ph,in} + \left(\frac{P_{\rm c}}{P_{\rm c,0}}\right)\left(P_0\left(h_0z/h\right) - P_0\left(h_0 z_{\rm ph}/h\right)\right),
\end{align}
where
\begin{align}
\left(\frac{P_{\rm c}}{P_{\rm c,0}}\right) = \left(\frac{\Omega_{\rm z}}{\Omega_{\rm z}}\right)^2 \left(\frac{\Sigma}{\Sigma_0}\right)\left(\frac{h}{h_0}\right)
\end{align}
and
\begin{align}
 P_{\rm ph,in} = \left(\frac{f_{\rm cor}}{f_{\rm cor,0}} \right)^4 \left(\frac{F}{F_0} \right) P_{\rm ph,in,0}.
\label{eq_pressure_ph}
\end{align}
The scaling for the gas temperature profile is given by K17 equations (19), (20), and (21):
\begin{align}
T_{\rm g,in}^4(z) =& T_{\rm g,ph}^4 + \left(\frac{P_{\rm c}}{P_{\rm c,0}}\right)\left(T_{\rm g,0}^4 \left(h_0z/h\right) - T_{\rm g,0}^4 \left(h_0 z_{\rm ph}/h\right)\right)
 \label{eq_Tgas_profile_1}
\end{align}
\begin{align}
T_{\rm g,ph}^4 = \left(\frac{P_{\rm ph,in}}{P_{\rm ph,in,0}}\right)T_{\rm g,ph,0}^4
\label{eq_Tgas_ph}
\end{align}
\begin{align}
T_{\text{g,out}}^4\left(z\right) =& \left(\frac{P_{\rm ph,in}}{P_{\rm ph,in,0}}\right) T_{\text{g},0}^4\left(z_{\text{ph},0} + h_0 (z-z_{\text{ph}})/h\right).
\label{eq_Tgas_profile_2}
\end{align}

\subsection{Modelling radiation advection with an effective $\kappa$}
The scaling for $h$ depends on how the radiation is vertically transported. For radiation diffusion, for example, the scaling for $h$ is given by K17 equation (7):
\begin{align}
\left(\frac{h}{h_0}\right) = \left(\frac{\alpha}{\alpha_0}\right)^{-1} \left(\frac{\kappa}{\kappa_0}\right)^{-1} \left(\frac{\partial_x v_y}{\partial_x v_{y,0}}\right)^{-1} \left(\frac{\Sigma}{\Sigma_0}\right)^{-1}.
\label{eq_h}
\end{align} 
In Section \ref{res_advection} we pointed out that including radiation advection is equivalent to simply decreasing $\kappa$ as far as the shearing box scalings are concerned. Now we also see that including radiation advection (at fixed surface density $\Sigma$) is therefore equivalent to simply \emph{increasing} the scale height relative to the value set by radiation diffusion alone.

We note that although including advection \emph{increases} the scale height in a shearing box, it has the opposite effect at a fixed radius in an accretion disc. This is because in a shearing box the surface density $\Sigma$ is fixed and the total flux $F$ is allowed to vary. In this case, including advection increases both the scale height and therefore the flux (equation \ref{eq_flux_scaling}). But if we substitute in equation (\ref{eq_flux_scaling}) everywhere for $\Sigma$, we can instead regard $F$ as a free parameter instead of $\Sigma$. Since $F$ is a function of accretion disc parameters such as the mass, mass accretion rate, radius, etc., this procedure gives the shearing box scalings in terms of accretion disc parameters rather than shearing box parameters. The scalings that result from this procedure, assuming radiation diffusion, are given in K17 Appendix C. The scaling for the scale height, for example, is given by
\begin{align}
\left(\frac{h}{h_0}\right) = \left(\frac{\kappa}{\kappa_0}\right) \left(\frac{\Omega_z}{\Omega_{z,0}}\right)^{-2}\left(\frac{F}{F_0}\right).
\end{align}
We see, therefore, that including advection \emph{decreases} the scale height at a fixed radius in the disc. In this process, the flux is held constant and the surface density, therefore, increases (equation \ref{eq_flux_scaling}). 

\section{Additional figures}
\label{sec_figures}
In this section we show the plots of spectra at multiple radii omitted from Section \ref{sec_modeling}.

\begin{figure*}
\begin{tabular}{ll}
\includegraphics[width = 84mm]{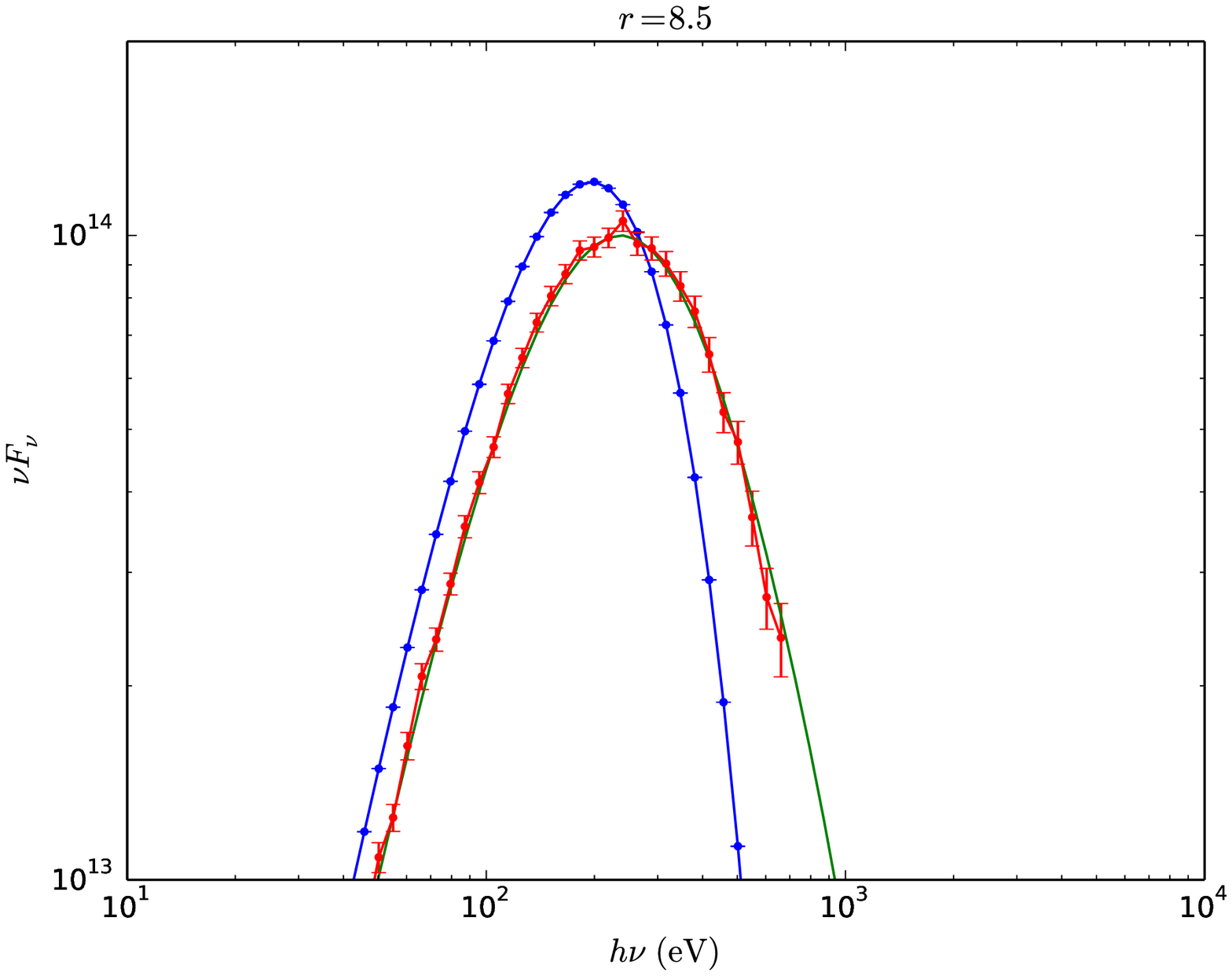} & \includegraphics[width = 84mm]{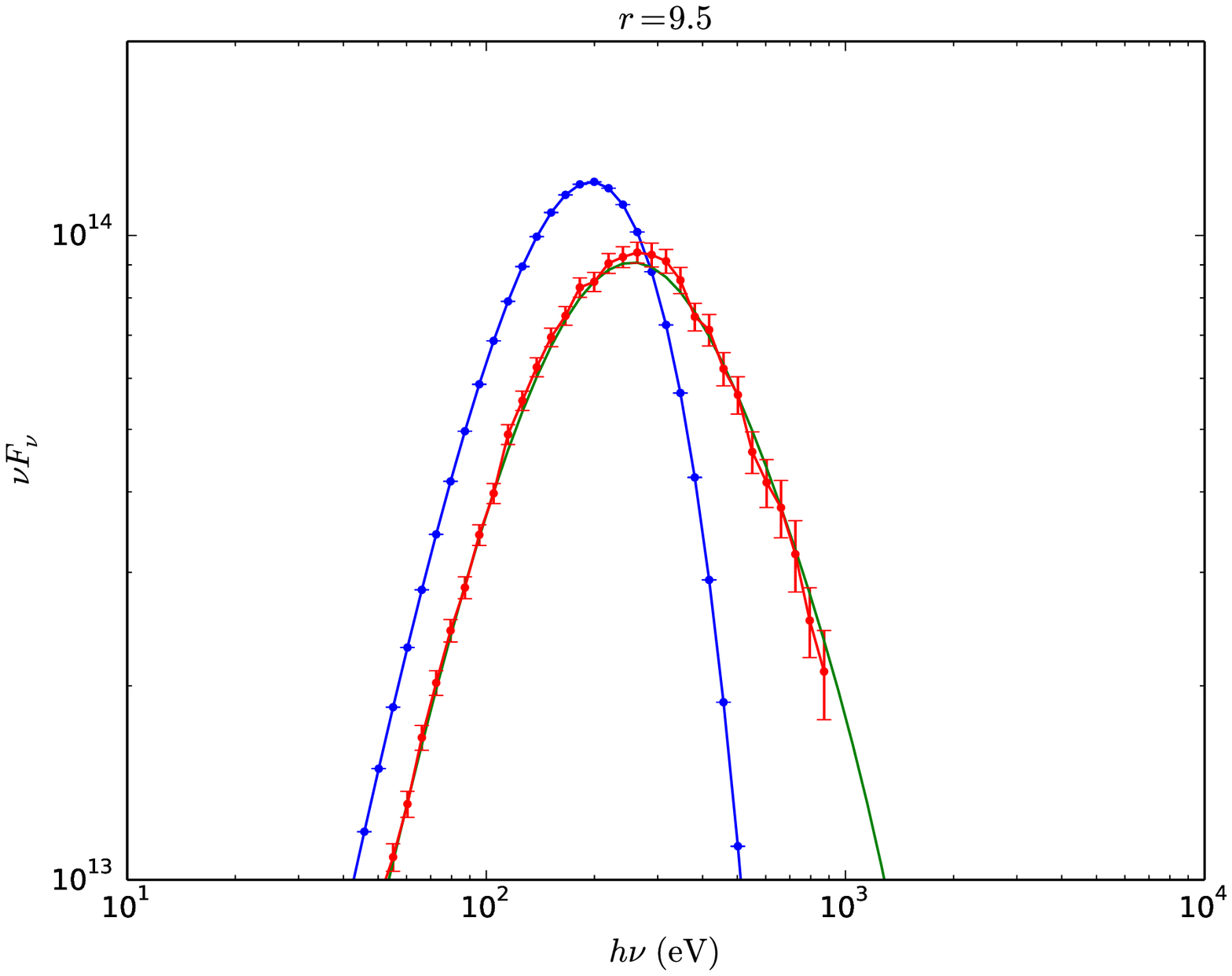} \\
\includegraphics[width = 84mm]{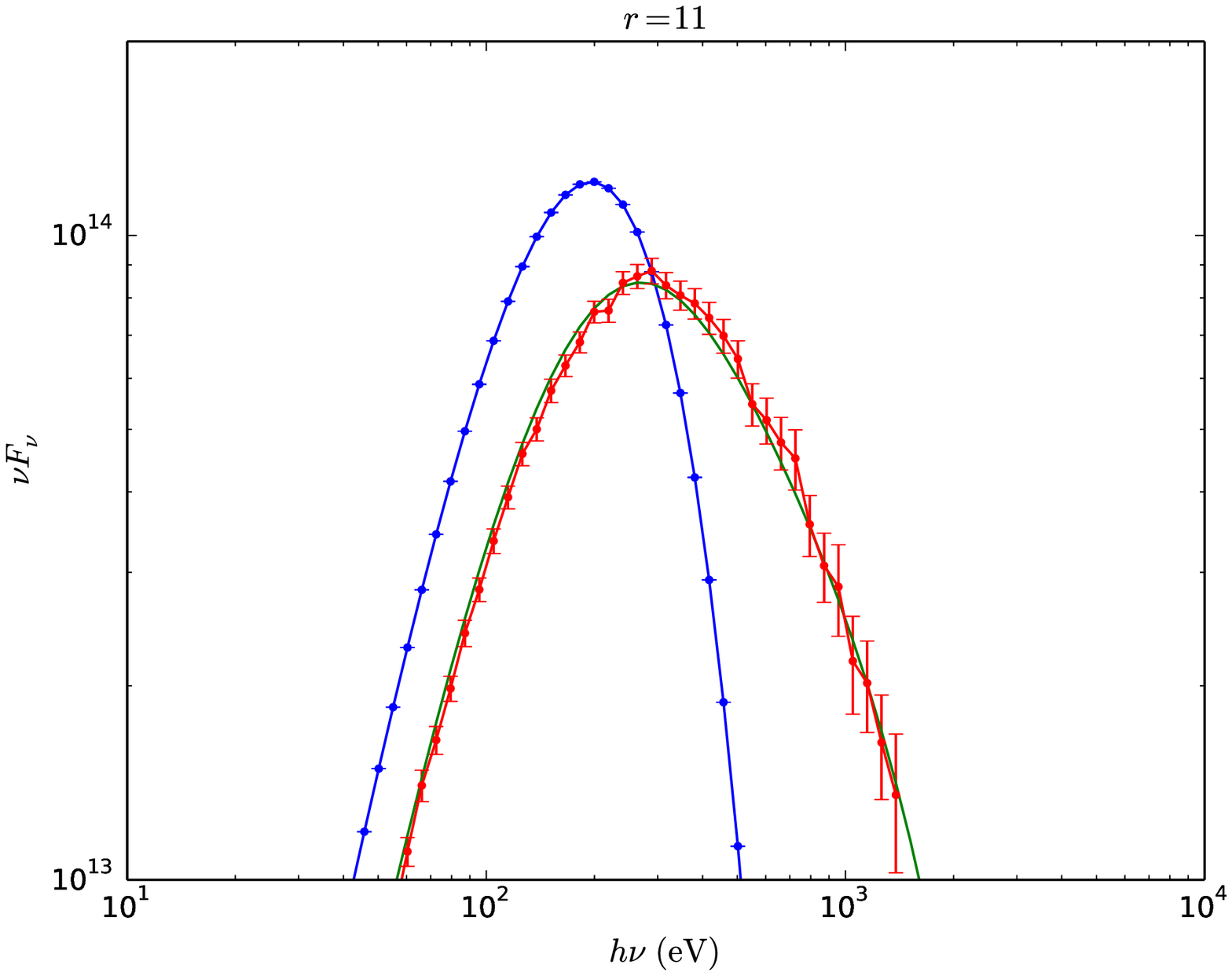} & \includegraphics[width = 84mm]{Comp_1D_r14.eps} \\
\includegraphics[width = 84mm]{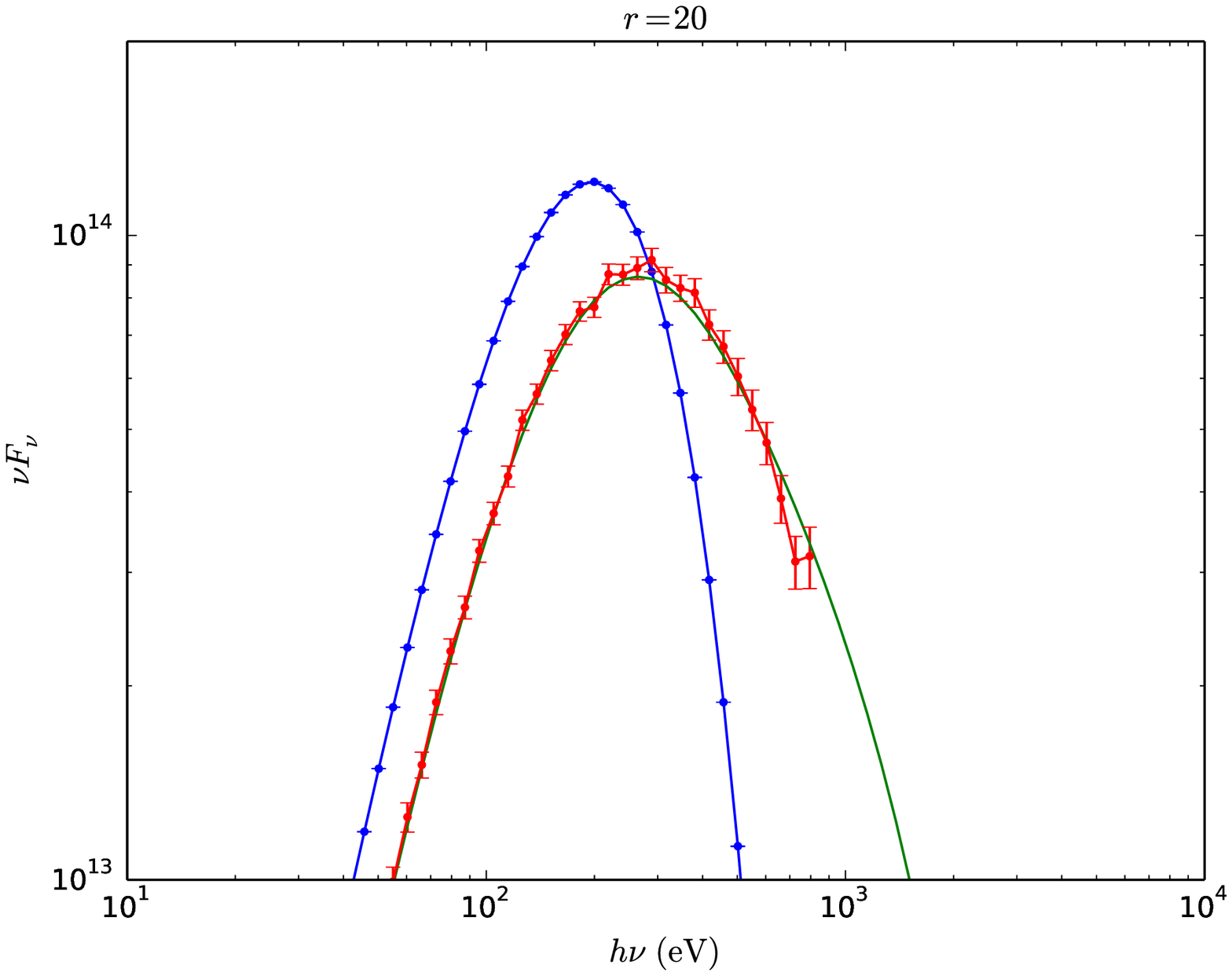} & \includegraphics[width = 84mm]{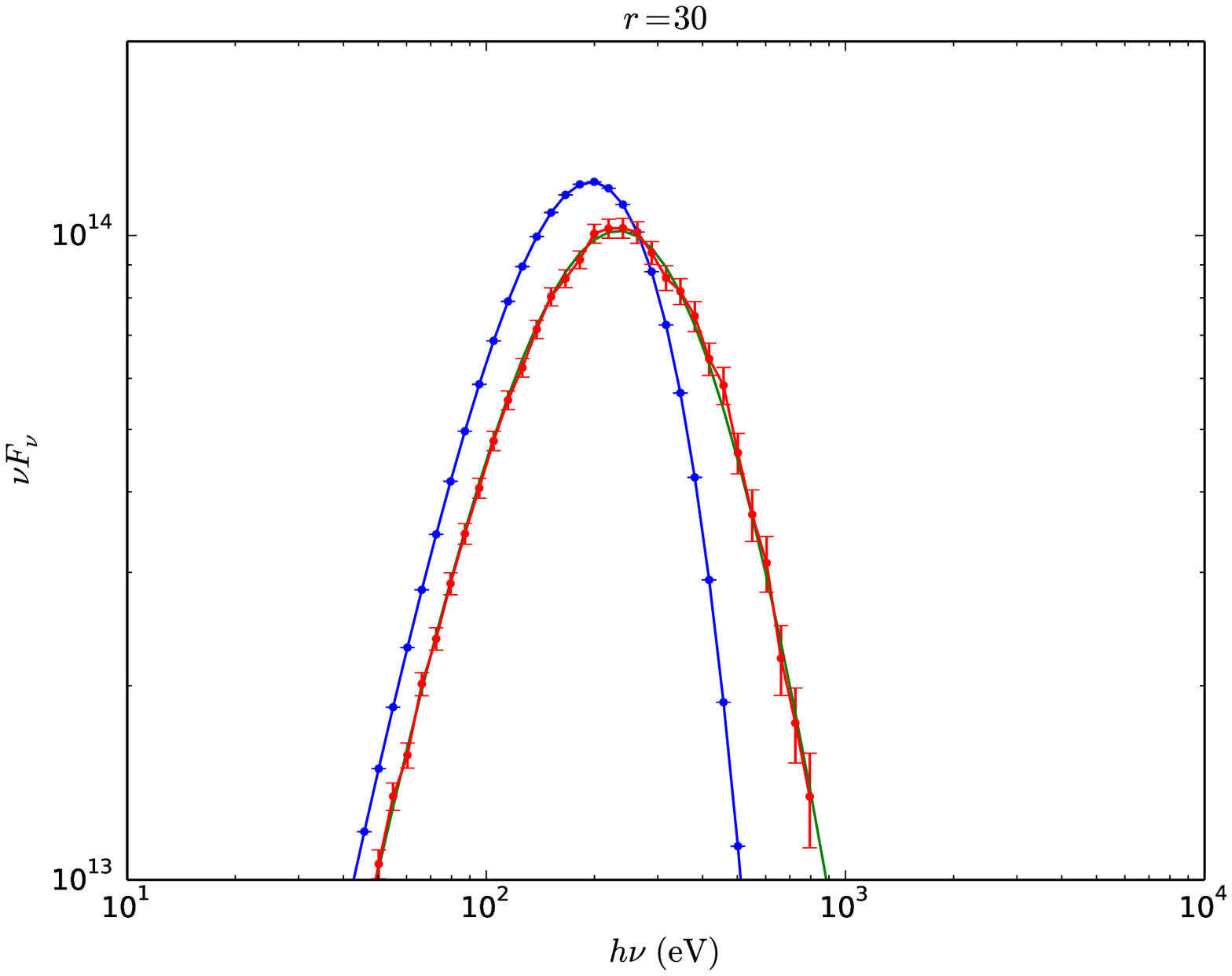} \\
\end{tabular}
\caption{Normalized spectra (red) computed by passing a $50{\rm eV}$ Planck source (blue) through vertical structure data truncated at $\tau_{\rm s} = 10$ at multiple radii for the $M = 2 \times 10^6 M_\odot$,  $L/L_{\rm Edd} = 5$ parameter set (Table \ref{table_disc_param}). In all cases the velocities are zeroed and the wave temperatures are added to the gas temperatures. The green curves are calculated by using the Kompaneets equation to pass the $50{\rm eV}$ Planck source through a homogeneous medium with temperature  $T_{\rm 1D}$, given in Table \ref{table_T_1D}. The spectra for only $r=14$ were originally plotted in Figure \ref{fig_Comp_1D}.}
\label{fig_Comp_1D_full}
\end{figure*}

\begin{figure*}
\begin{tabular}{ll}
\includegraphics[width = 84mm]{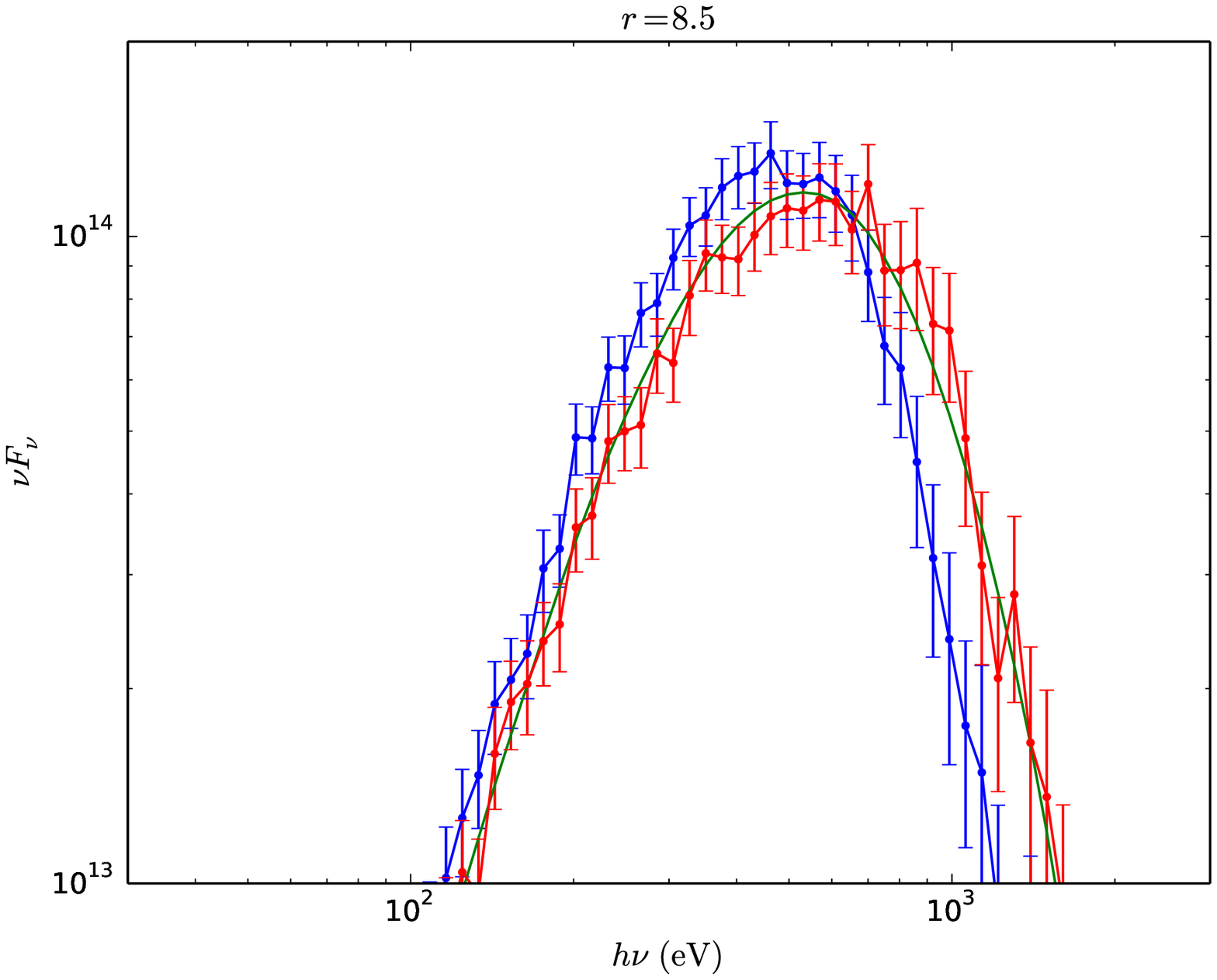} & \includegraphics[width = 84mm]{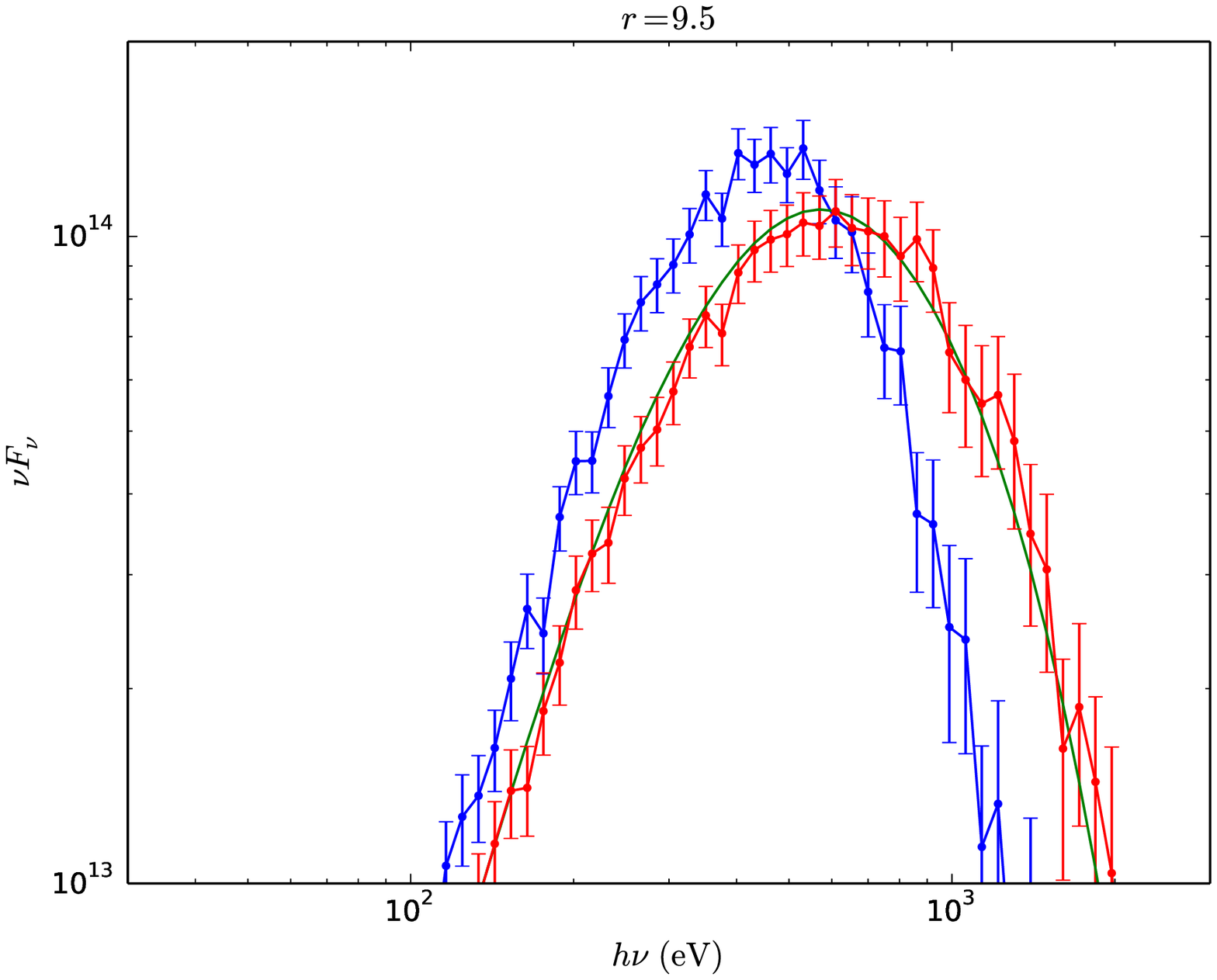} \\
\includegraphics[width = 84mm]{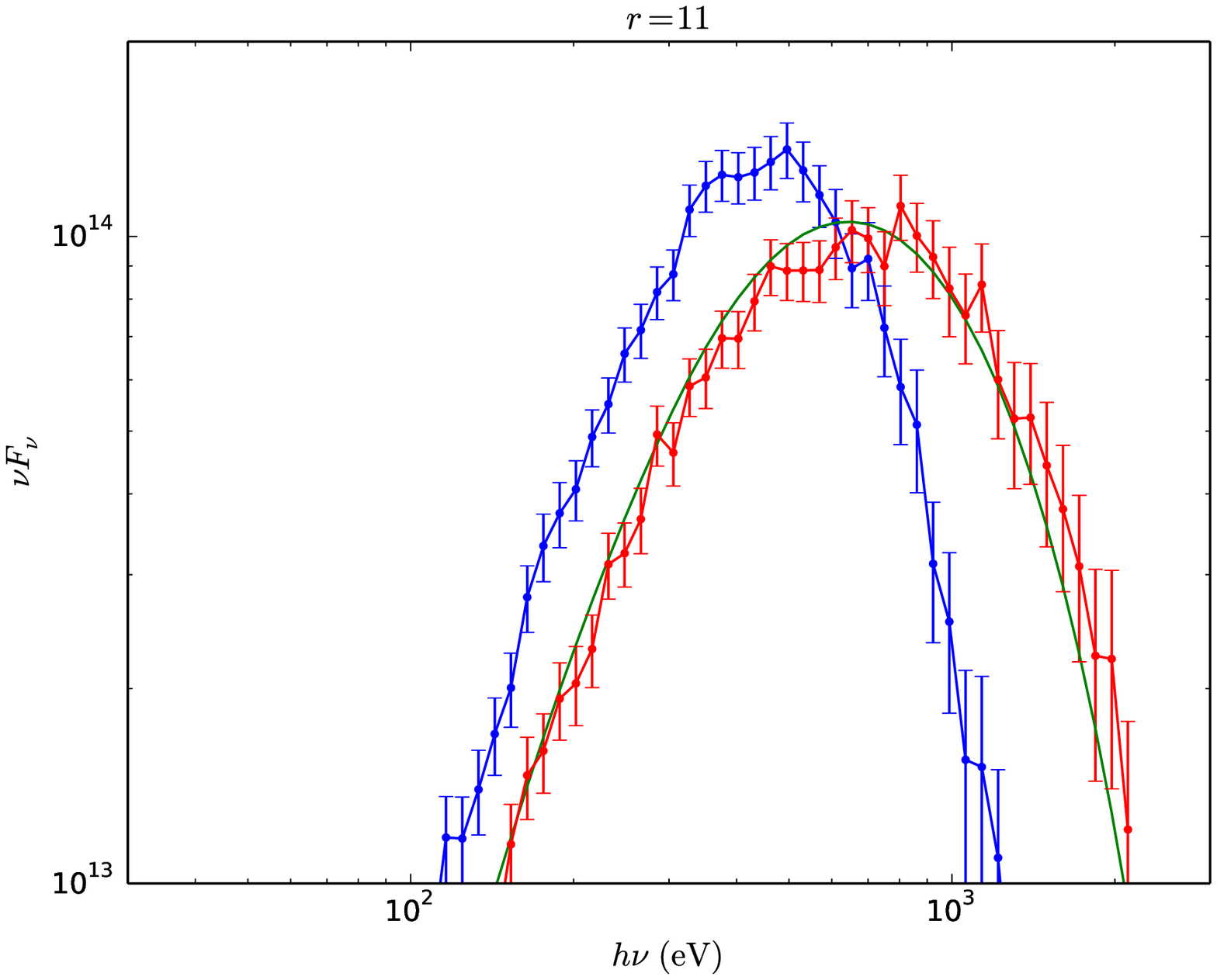} & \includegraphics[width = 84mm]{spectra_fit_set1_r14.eps} \\
\includegraphics[width = 84mm]{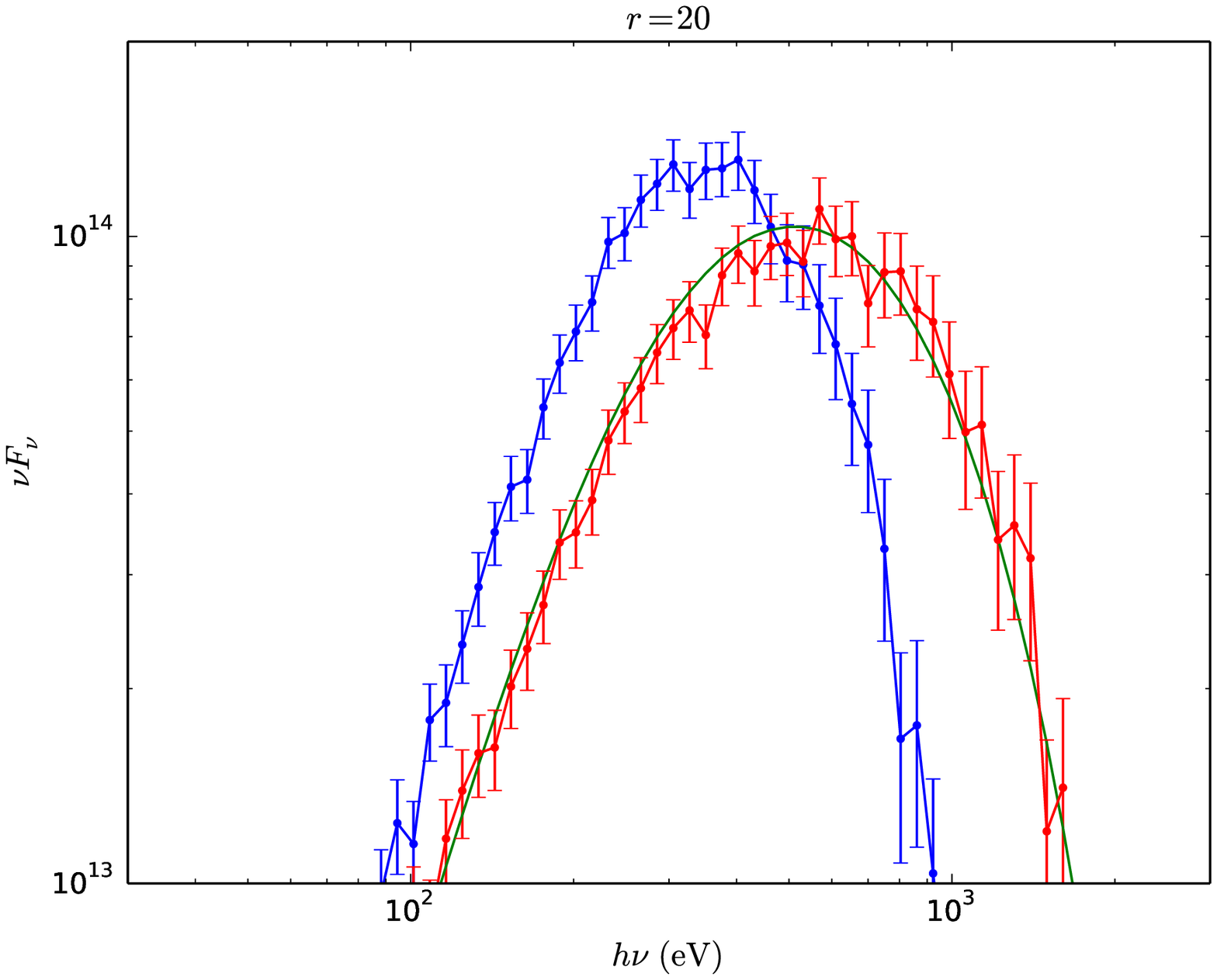} & \includegraphics[width = 84mm]{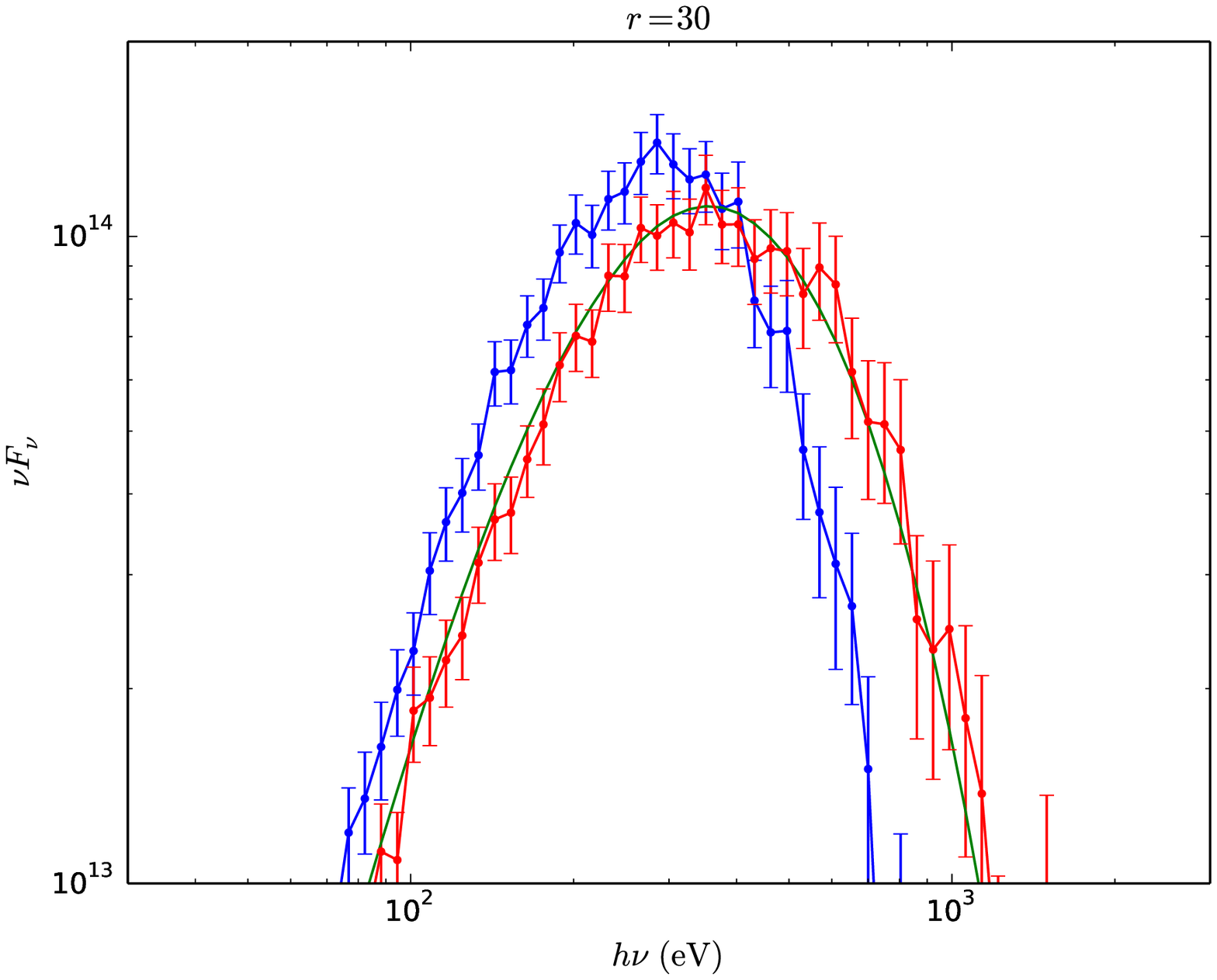} \\
\end{tabular}
\caption{Normalized spectra at multiple radii for the $M = 2 \times 10^6 M_\odot$,  $L/L_{\rm Edd} = 5$ parameter set (Table \ref{table_disc_param}) computed with (red) and without (blue) velocities. The green curves are calculated by using the Kompaneets equation to pass the blue curves through a homogeneous Comptonizing medium with parameters $T_{\rm C}$ and $\tau_{\rm C}$, given in Table \ref{table_comp_param_A}. The spectra for only $r=14$ were originally plotted in Figure \ref{fig_spectra_fit_set1}.}
\label{fig_spectra_fit_set1_full}
\end{figure*}

\begin{figure*}
\begin{tabular}{ll}
\includegraphics[width = 84mm]{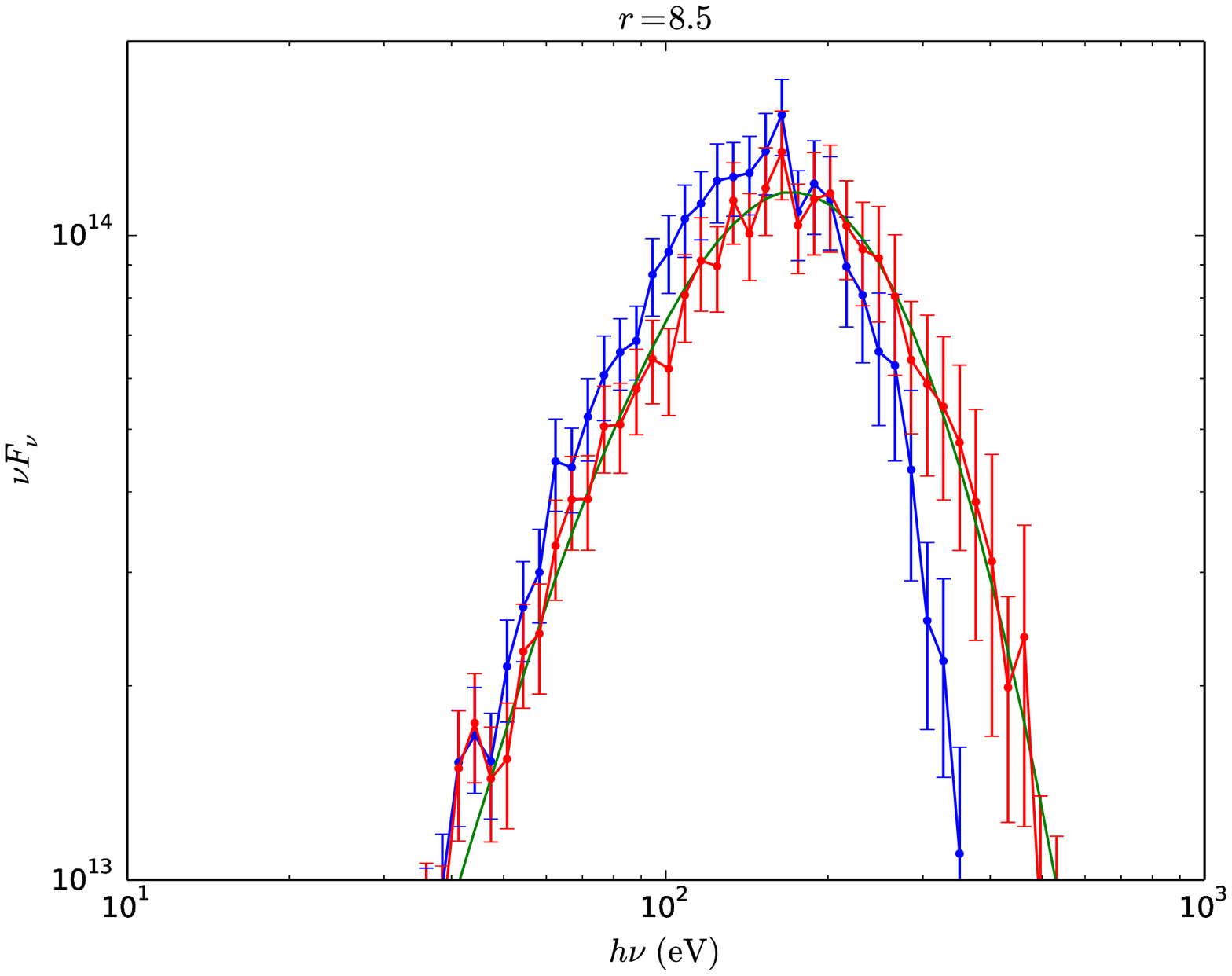} & \includegraphics[width = 84mm]{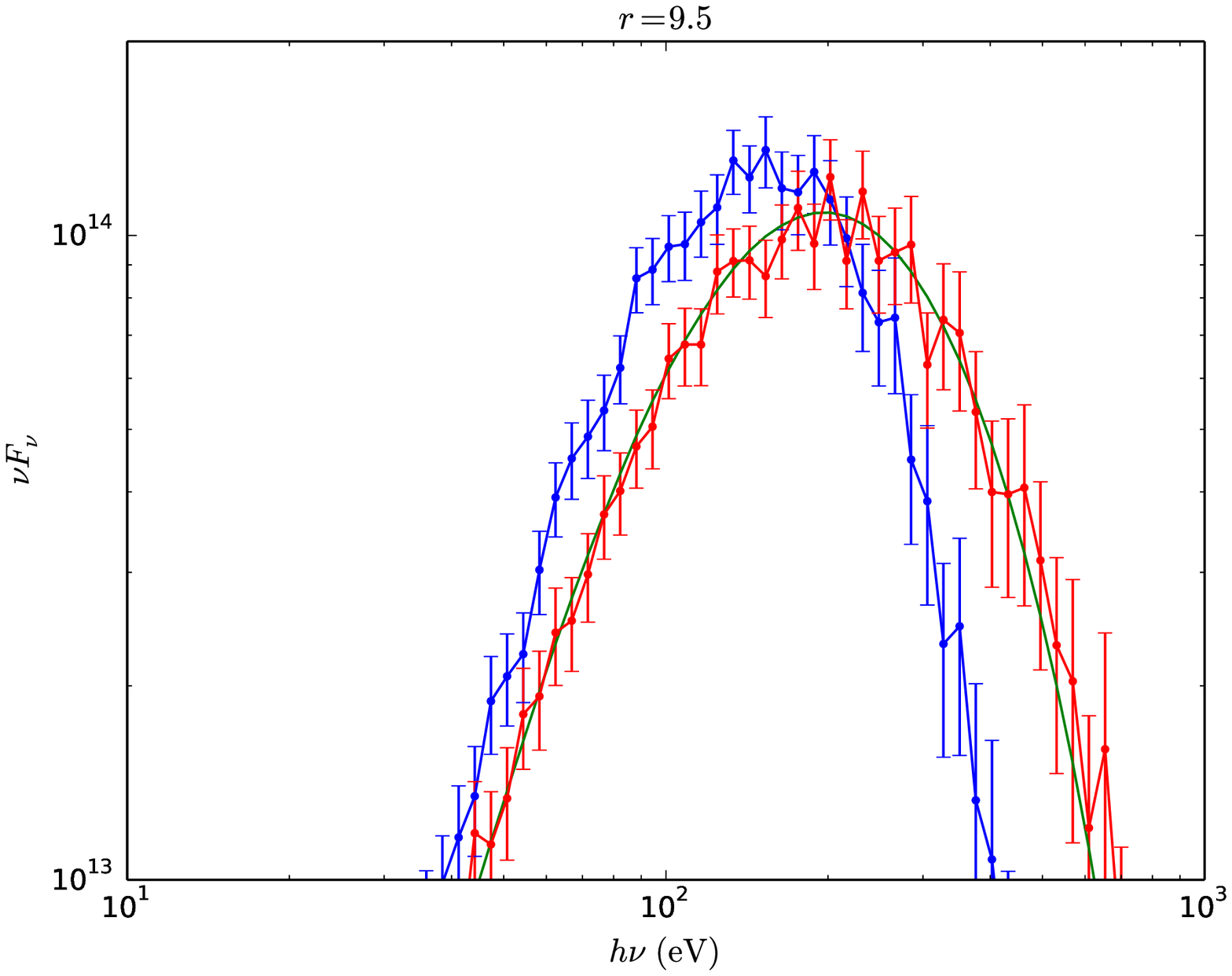} \\
\includegraphics[width = 84mm]{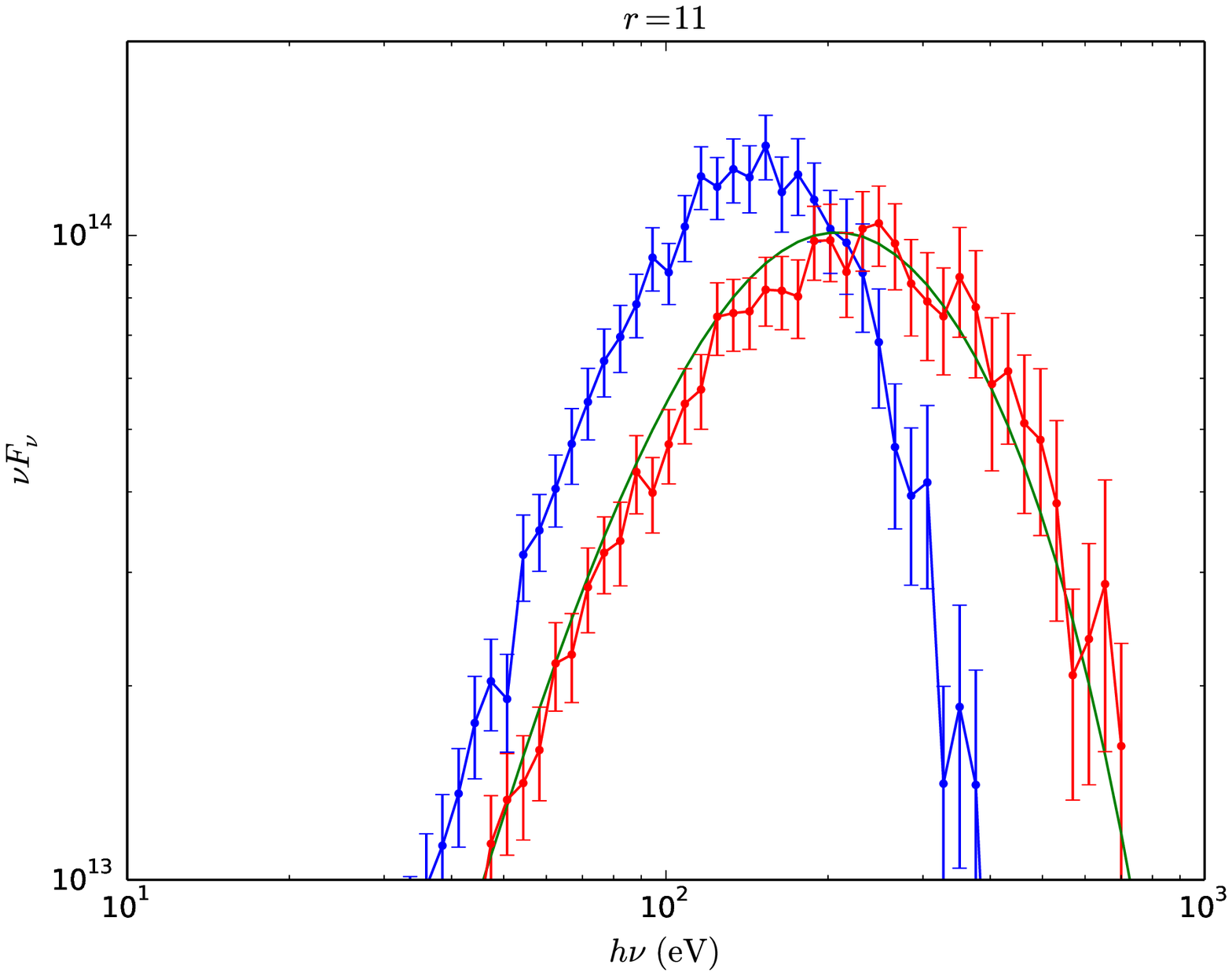} & \includegraphics[width = 84mm]{spectra_fit_set4_r14.eps} \\
\includegraphics[width = 84mm]{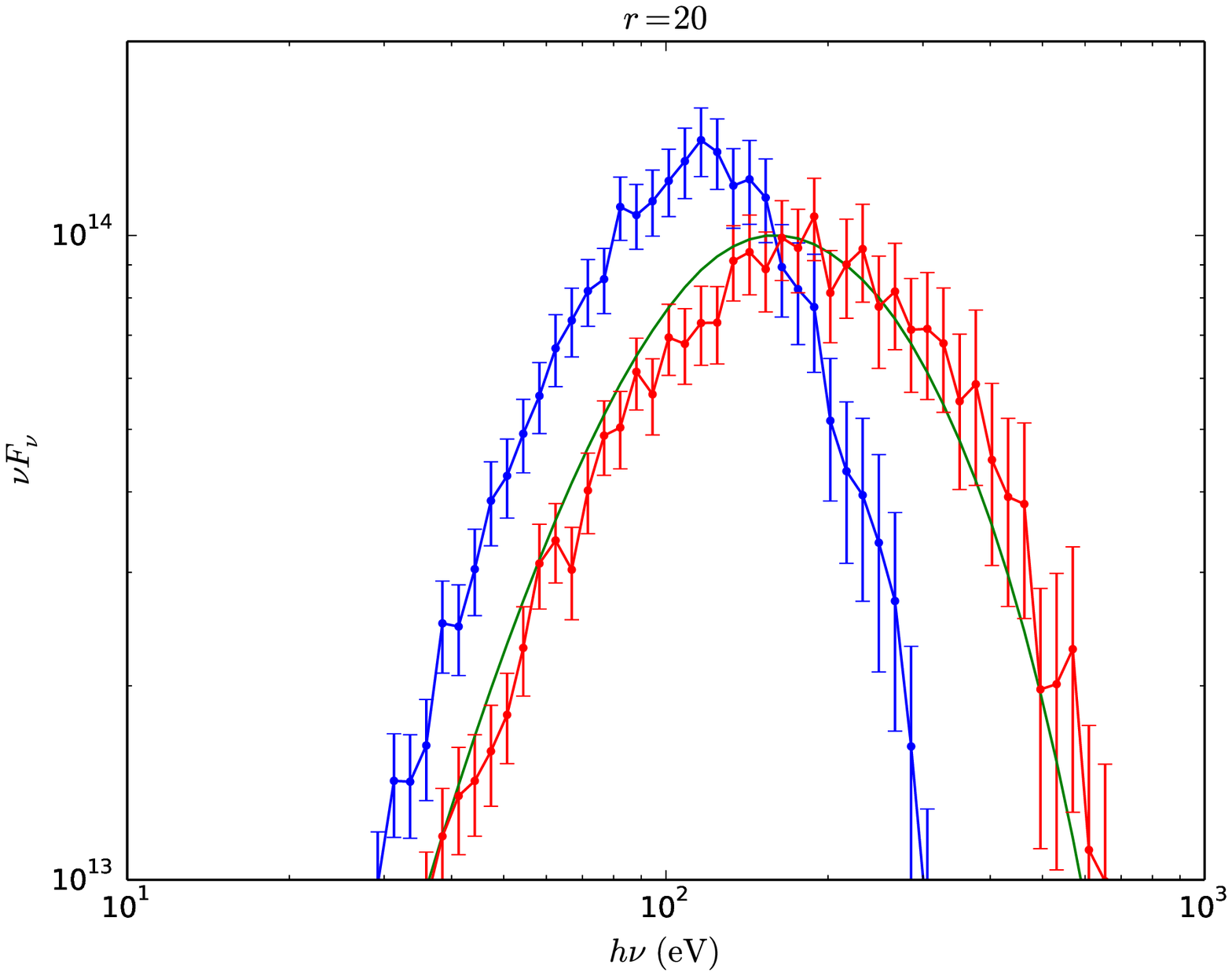} & \includegraphics[width = 84mm]{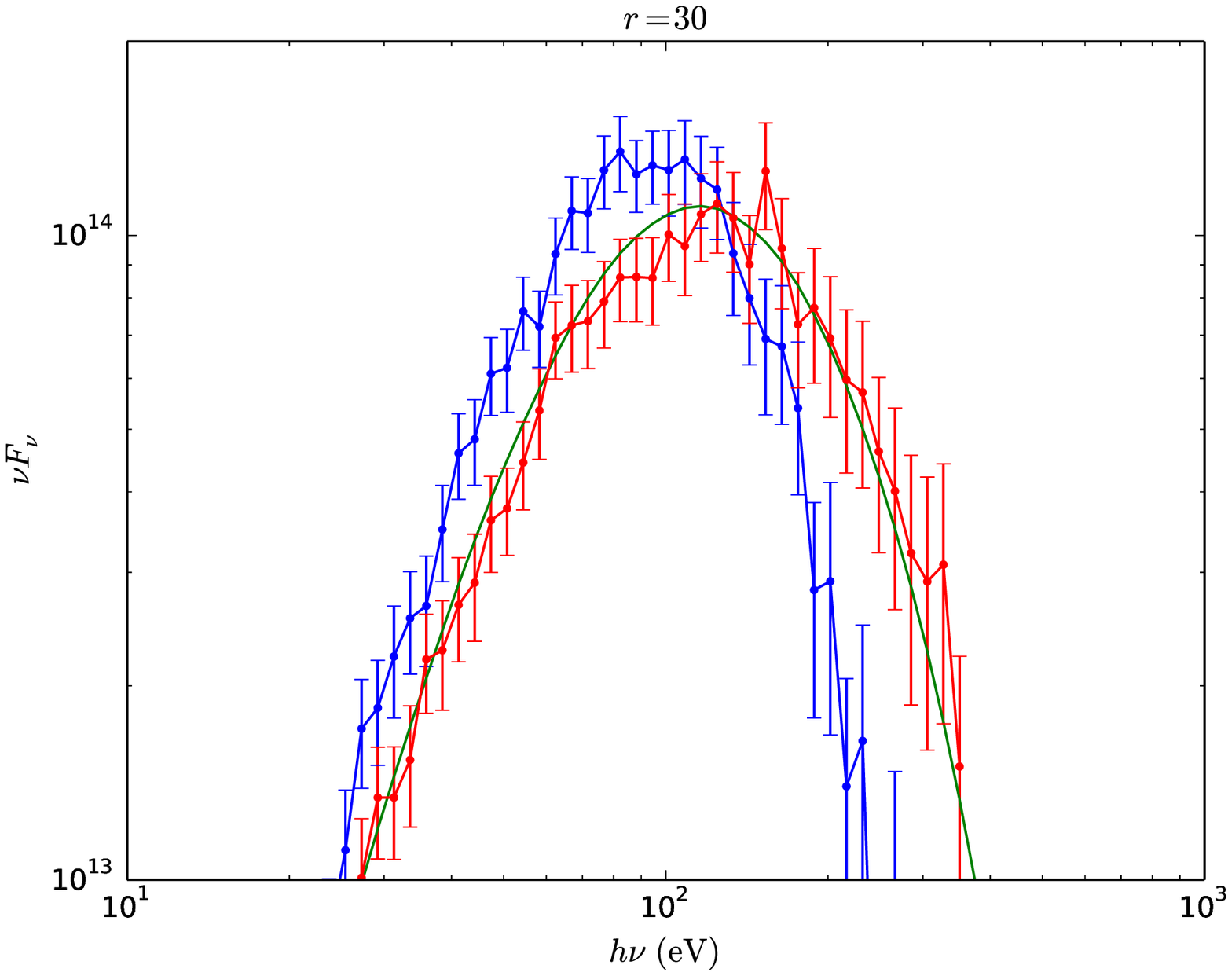} \\
\end{tabular}
\caption{Normalized spectra at multiple radii for the $M = 2 \times 10^8 M_\odot$,  $L/L_{\rm Edd} = 5$ parameter set (Table \ref{table_disc_param}) computed with (red) and without (blue) velocities. The green curves are calculated by using the Kompaneets equation to pass the blue curves through a homogeneous Comptonizing medium with parameters $T_{\rm C}$ and $\tau_{\rm C}$, given in Table \ref{table_comp_param_B}. The spectra for only $r=14$ were originally plotted in Figure \ref{fig_spectra_fit_set4}.}
\label{fig_spectra_fit_set4_full}
\end{figure*}

\begin{figure*}
\begin{tabular}{ll}
\includegraphics[width = 84mm]{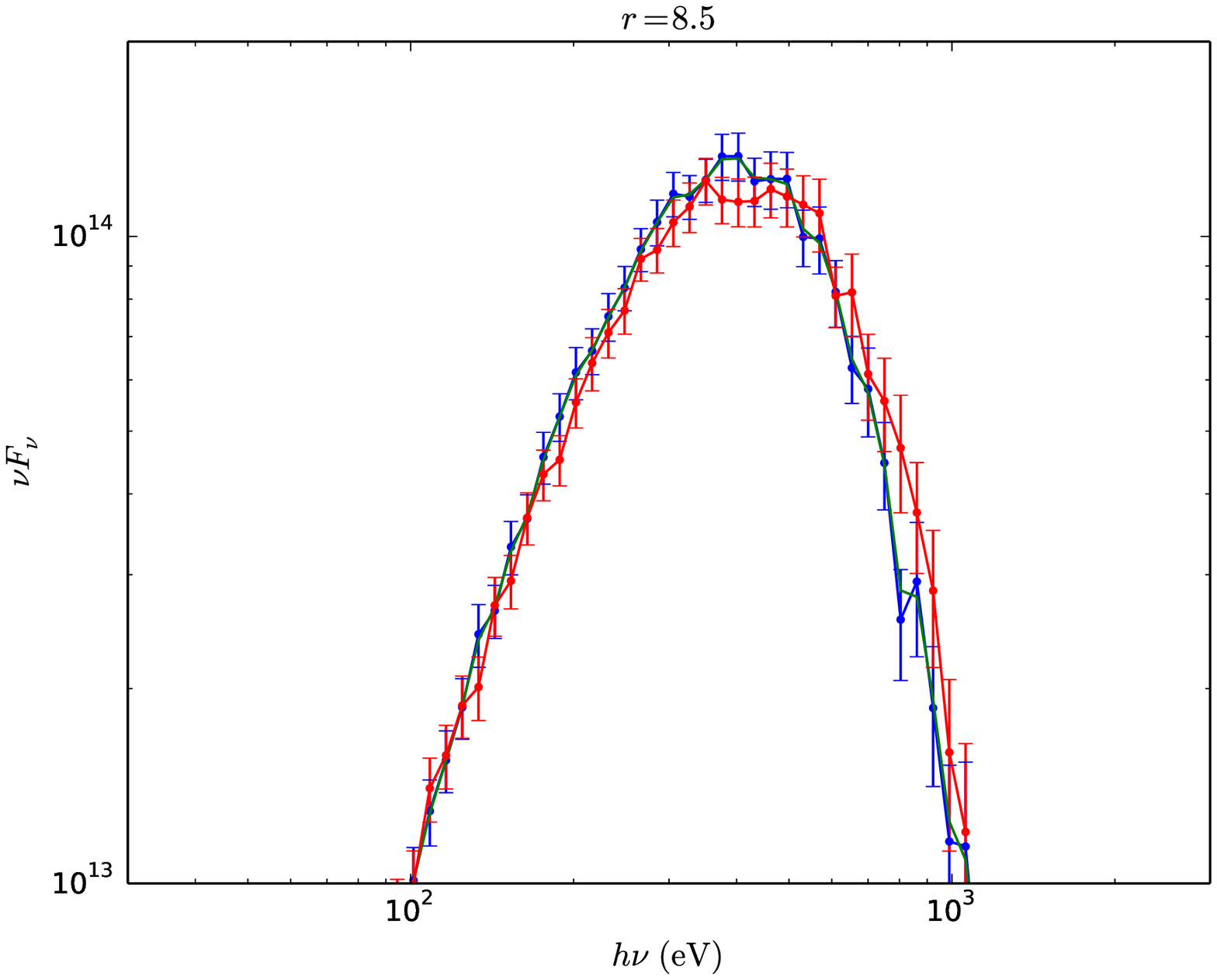} & \includegraphics[width = 84mm]{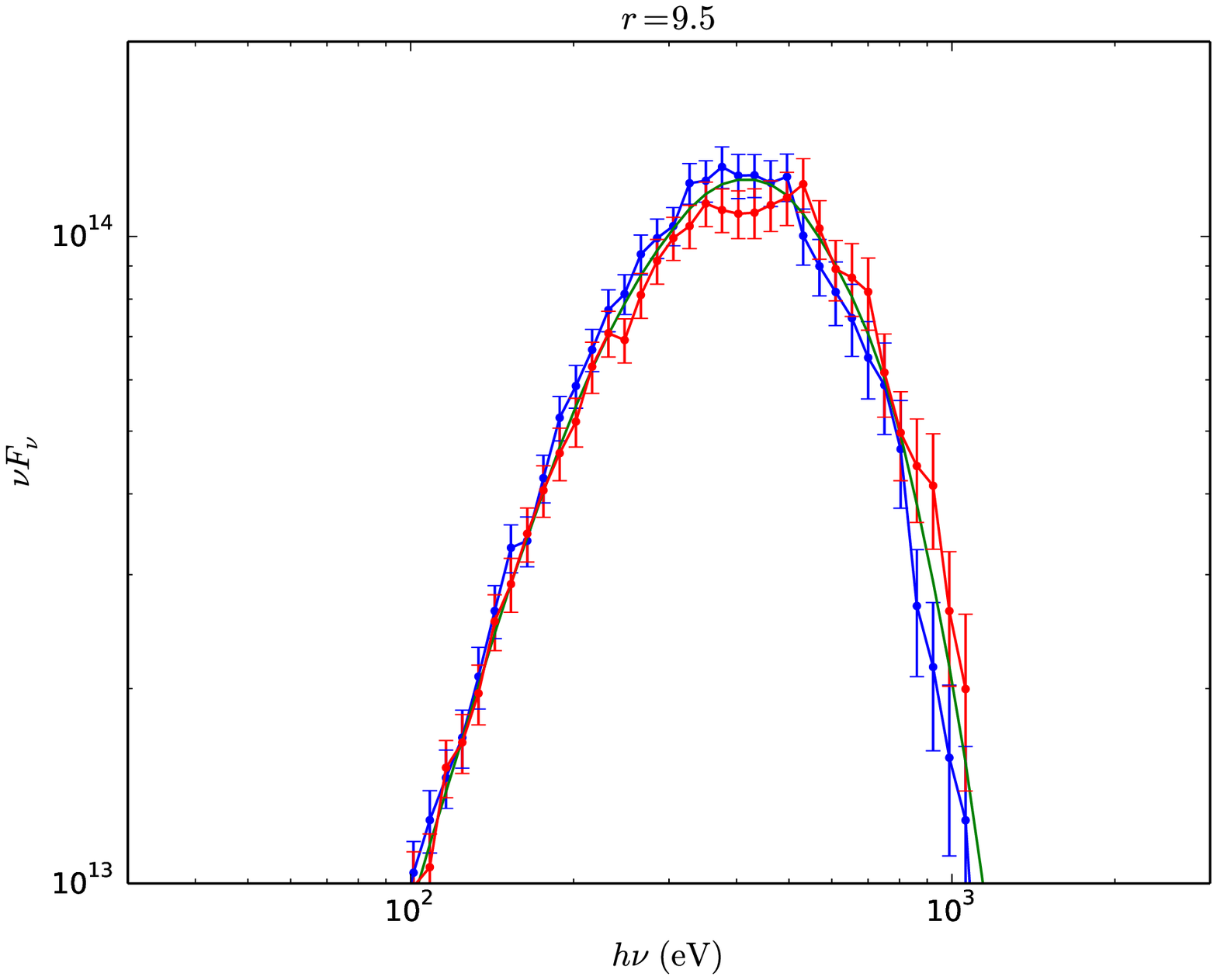} \\
\includegraphics[width = 84mm]{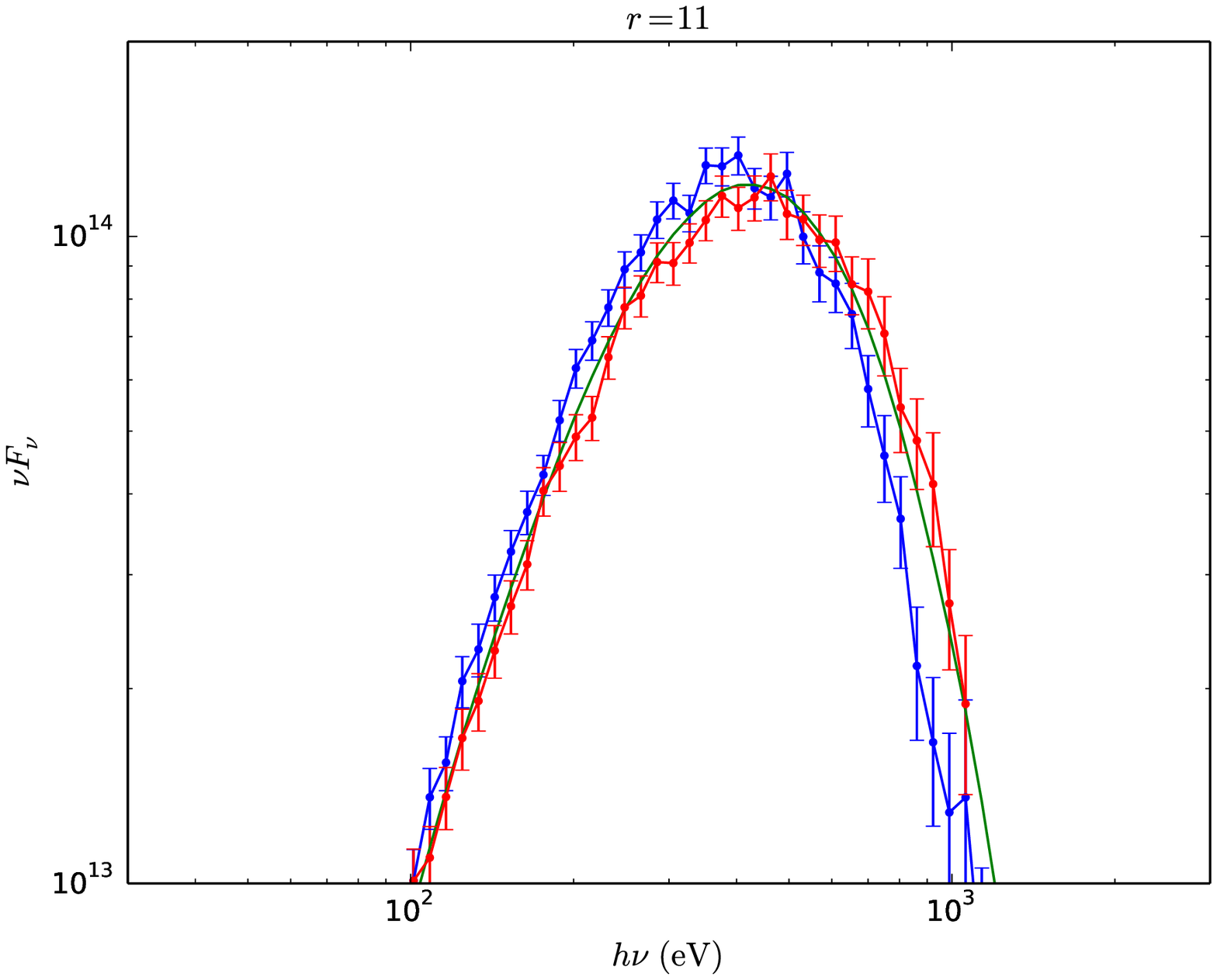} & \includegraphics[width = 84mm]{spectra_fit_set2_r14.eps} \\
\includegraphics[width = 84mm]{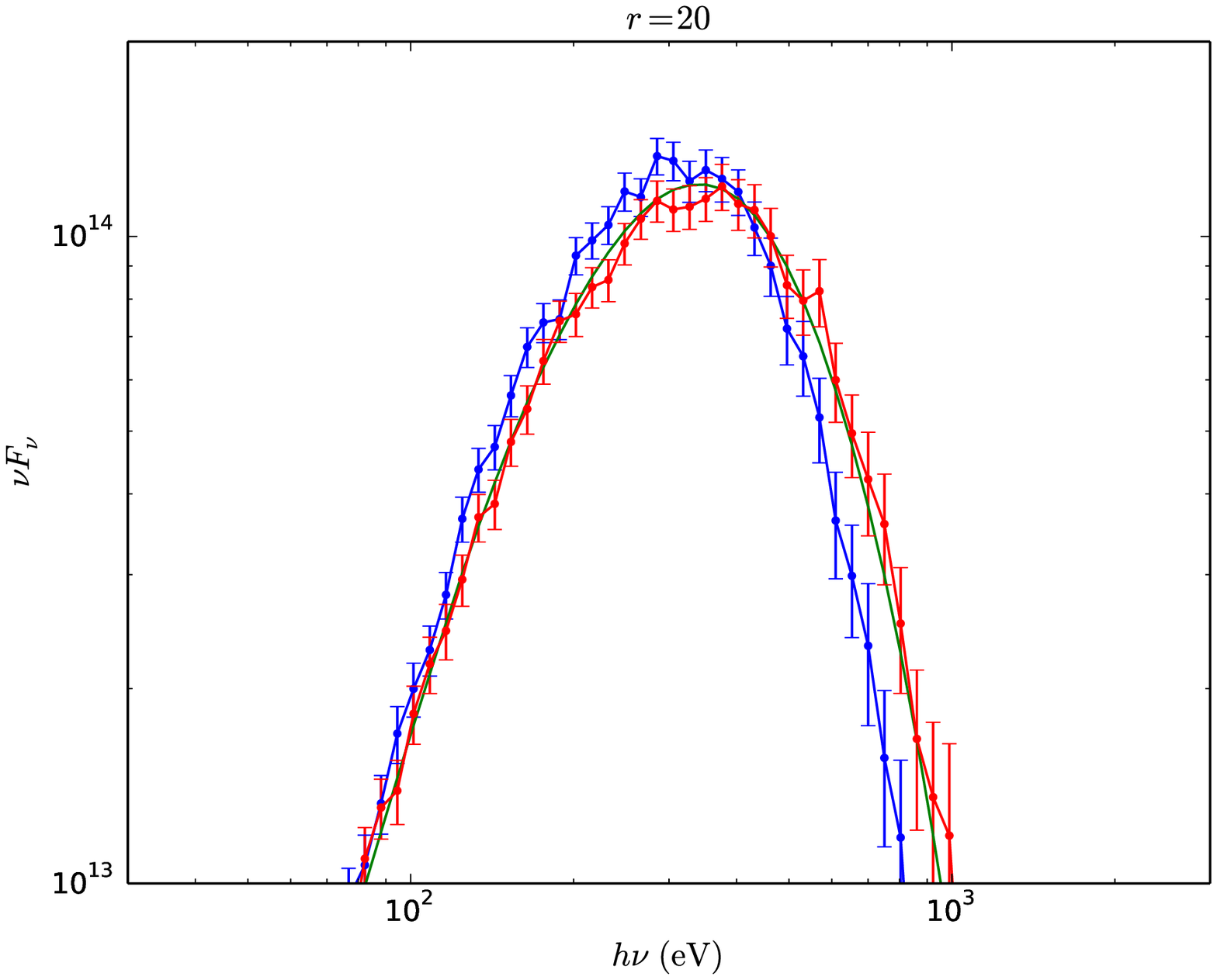} & \includegraphics[width = 84mm]{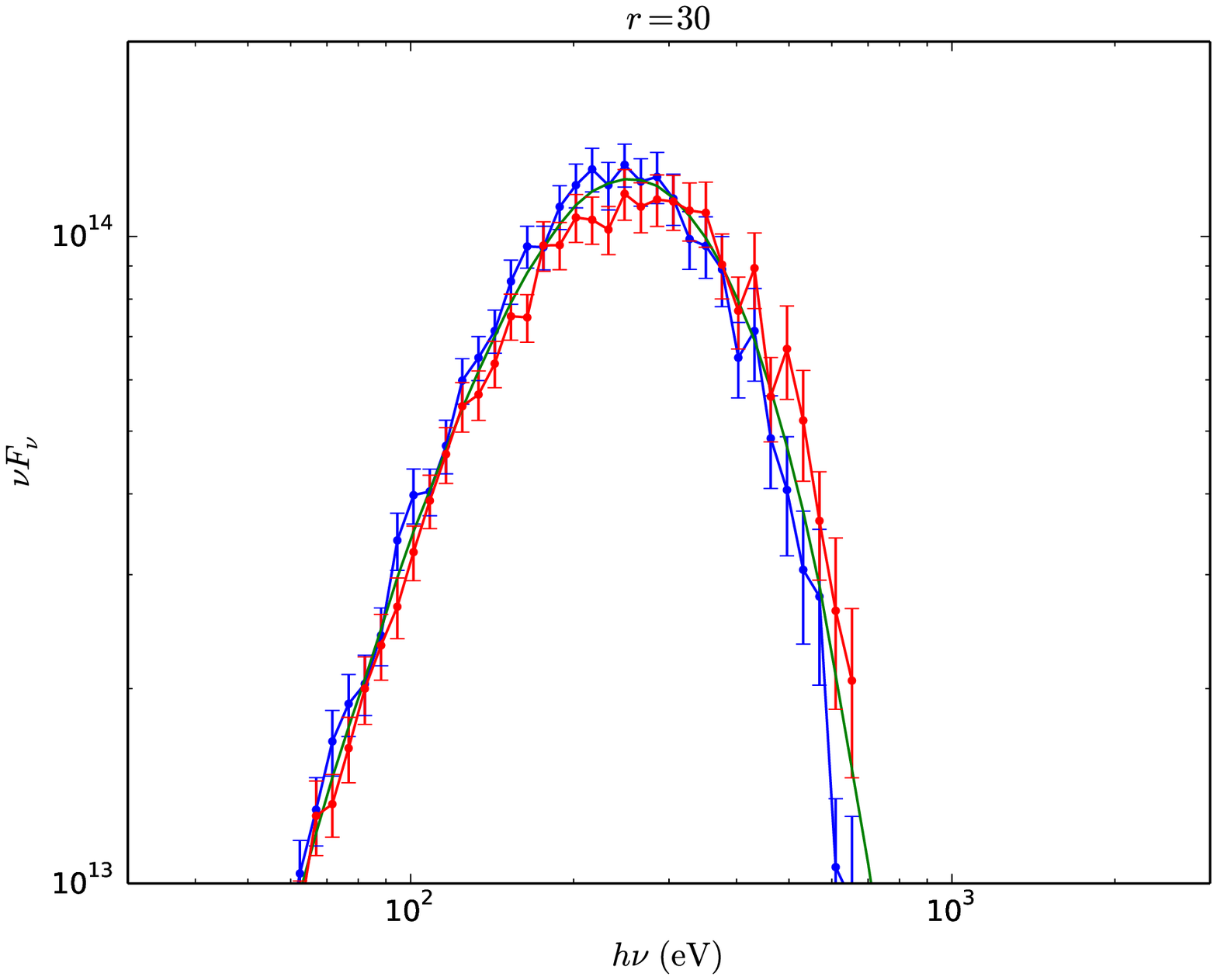} \\
\end{tabular}
\caption{Normalized spectra at multiple radii for the $M = 2 \times 10^6 M_\odot$,  $L/L_{\rm Edd} = 2.5$ parameter set (Table \ref{table_disc_param}) computed with (red) and without (blue) velocities. The green curves are calculated by using the Kompaneets equation to pass the blue curves through a homogeneous Comptonizing medium with parameters $T_{\rm C}$ and $\tau_{\rm C}$, given in Table \ref{table_comp_param_C}. The spectra for only $r=14$ were originally plotted in Figure \ref{fig_spectra_fit_set2}.}
\label{fig_spectra_fit_set2_full}
\end{figure*}
\end{document}